\documentclass[a4paper,11pt]{article}
\usepackage[margin=0.6in]{geometry}

\renewcommand{\title}[1]{\vbox{\center\LARGE{#1}}\vspace{5mm}}
\renewcommand{\author}[1]{\vbox{\center#1}\vspace{5mm}}
\newcommand{\address}[1]{\vbox{\center\em#1}}
\newcommand{\email}[1]{\vbox{\center\tt#1}\vspace{5mm}}

\renewcommand{\date}[1]{\vbox{\center#1}}

\usepackage{amssymb}
\usepackage{amsmath,bm}
\usepackage{amssymb}
\usepackage{graphicx}
\usepackage{amsfonts}         
\usepackage{fancybox}   

\usepackage{enumitem}

\usepackage{slashed}

\usepackage{epstopdf}
\usepackage[usenames,dvipsnames]{xcolor}%http://en.wikibooks.org/wiki/LaTeX/Colors
\usepackage[pagebackref, bookmarks={false}, pdfauthor={Diptarka Das}, pdftitle={IslandJAM}]{hyperref}
\hypersetup{colorlinks=true, linkcolor=darkblue, citecolor=red, filecolor=OliveGreen, urlcolor=blue, filebordercolor={.8 .8 1}, urlbordercolor={.8 .8 0}}%http://en.wikibooks.org/wiki/LaTeX/Hyperlinks
\usepackage{soul}%strike through text \st{} .  not in math mode.
\setstcolor{Red}

\usepackage{wrapfig}
\definecolor{jazzberryjam}{rgb}{0.65, 0.04, 0.37}
\definecolor{lust}{rgb}{0.9, 0.13, 0.13}
\definecolor{sandybrown}{rgb}{0.96, 0.64, 0.38}
\definecolor{mountainmeadow}{rgb}{0.19, 0.73, 0.56}
\definecolor{glaucous}{rgb}{0.38, 0.51, 0.71}
\definecolor{chromeyellow}{rgb}{1.0, 0.65, 0.0}
\definecolor{emerald}{rgb}{0.31, 0.78, 0.47}
\definecolor{deepsaffron}{rgb}{1.0, 0.6, 0.2}
\definecolor{darkgreen}{rgb}{0,0.4,0}
\definecolor{darkred}{rgb}{0.4,0,0}
\definecolor{darkblue}{rgb}{0,0,0.4}
\definecolor{lightblue}{rgb}{.6,.6,0.9}

\definecolor{uglybrown}{rgb}{0.8,  0.7,  0.5}

\definecolor{palatinatepurple}{rgb}{0.41, 0.16, 0.38}
\definecolor{celebrationcolor}{rgb}{0.75,  0.0,  0.9}

 \usepackage{mdframed}

\usepackage{framed}
\definecolor{shadecolor}{rgb}{0.90,0.90,0.90}

\usepackage{tkz-euclide}
\usepackage{makeidx}
\usepackage{cite}
\usepackage{bm}
\usepackage{geometry}
\usepackage{braket}
\usepackage[margin=20pt,small]{caption}

\usepackage[toc]{appendix}

\usepackage{tikz}
\usetikzlibrary{calc,decorations.markings,arrows.meta,shapes.misc,decorations.pathmorphing,calc,bending}

\usetikzlibrary{arrows.meta,shapes.misc,decorations.pathmorphing,calc,bending}

\tikzset{
  branch point/.style={cross out,draw=black,fill=none,minimum size=2*(#1-\pgflinewidth),inner sep=0pt,outer sep=0pt}, 
  branch point/.default=5
}
\tikzset{
  branch cut/.style={
    decorate,decoration=snake,
    to path={
      (\tikztostart) -- (\tikztotarget) \tikztonodes
    },
    }
  }

\input epsf

\newlength{\extraspace}
\newlength{\extraspaces}
\def\be{\begin{equation}}
\def\ee{\end{equation}}

\newcommand{\bea}{\begin{eqnarray}}
\newcommand{\eea}{\end{eqnarray}}

\def\Tr{{{\rm Tr~ }}}

\def\bra#1{{\langle}#1|}
\def\ket#1{|#1\rangle}

\def\II{\relax{I\kern-.10em I}}

\def\IB{\relax{\rm I\kern-.18em B}}

\def\ID{\relax{\rm I\kern-.18em D}}
\def\IE{\relax{\rm I\kern-.18em E}}
\def\IF{\relax{\rm I\kern-.18em F}}
\def\IG{\relax\hbox{$\inbar\kern-.3em{\rm G}$}}
\def\IGa{\relax\hbox{${\rm I}\kern-.18em\Gamma$}}
\def\IH{\relax{\rm I\kern-.18em H}}
\def\II{\relax{\rm I\kern-.18em I}}
\def\IK{\relax{\rm I\kern-.18em K}}
\def\inbar{\,\vrule height1.5ex width.4pt depth0pt}

\def\lp10{\ell_p^{10}}
\def\lp11{\ell_p^{11}}
\def\R11{R_{11}}

\def\frac#1#2{{#1 \over #2}}

\newdimen\tableauside\tableauside=1.0ex
\newdimen\tableaurule\tableaurule=0.4pt
\newdimen\tableaustep
\def\phantomhrule#1{\hbox{\vbox to0pt{\hrule height\tableaurule width#1\vss}}}
\def\phantomvrule#1{\vbox{\hbox to0pt{\vrule width\tableaurule height#1\hss}}}
\def\sqr{\vbox{%
  \phantomhrule\tableaustep
  \hbox{\phantomvrule\tableaustep\kern\tableaustep\phantomvrule\tableaustep}%
  \hbox{\vbox{\phantomhrule\tableauside}\kern-\tableaurule}}}
\def\squares#1{\hbox{\count0=#1\noindent\loop\sqr
  \advance\count0 by-1 \ifnum\count0>0\repeat}}
\def\tableau#1{\vcenter{\offinterlineskip
  \tableaustep=\tableauside\advance\tableaustep by-\tableaurule
  \kern\normallineskip\hbox
    {\kern\normallineskip\vbox
      {\gettableau#1 0 }%
     \kern\normallineskip\kern\tableaurule}%
  \kern\normallineskip\kern\tableaurule}}
\def\gettableau#1 {\ifnum#1=0\let\next=\null\else
  \squares{#1}\let\next=\gettableau\fi\next}

 \def\eqnn#1{\xdef #1{(\secsym\the\meqno)}\writedef{#1\leftbracket#1}%
 \global\advance\meqno by1\wrlabeL#1}
 \def\eqna#1{\xdef #1##1{\hbox{$(\secsym\the\meqno##1)$}}
 \writedef{#1\numbersign1\leftbracket#1{\numbersign1}}%
 \global\advance\meqno by1\wrlabeL{#1$\{\}$}}
 \def\eqn#1#2{\xdef #1{(\secsym\the\meqno)}\writedef{#1\leftbracket#1}%
 \global\advance\meqno by1$$#2\eqno#1\eqlabeL#1$$}

\global\newcount\itemno \global\itemno=0

\def\itemaut#1{\global\advance\itemno by1\noindent\item{\the\itemno.}#1}

\def\({\left(}
\def\){\right)}

\def\lsim{\mathrel{\mathstrut\smash{\ooalign{\raise2.5pt\hbox{$<$}\cr\lower2.5pt\hbox{$\sim$}}}}}
\def\gsim{\mathrel{\mathstrut\smash{\ooalign{\raise2.5pt\hbox{$>$}\cr\lower2.5pt\hbox{$\sim$}}}}}

\def\overleftrightarrow#1{\vbox{\ialign{##\crcr
     $\leftrightarrow$\crcr\noalign{\kern-0pt\nointerlineskip}
     $\hfil\displaystyle{#1}\hfil$\crcr}}}
     
     \def\overleftarrow#1{\vbox{\ialign{##\crcr
     $\leftarrow$\crcr\noalign{\kern-0pt\nointerlineskip}
     $\hfil\displaystyle{#1}\hfil$\crcr}}}

\usepackage[yyyymmdd,hhmmss]{datetime}
\newif{\ifeq}           % defines a new condition @eq tested by the conditional \ifeq
\eqtrue                 % if uncommented, declares @eq to be true
\newcounter{lecturecounter}
\usepackage[utf8]{inputenc}
\usepackage{comment}
\usepackage{float, xcolor}
\usepackage{physics}
\usepackage{ragged2e}
\usepackage{booktabs}
\usepackage{latexsym}
\usepackage{mathtools}

\usepackage{subcaption}
\usepackage{graphicx}
\usepackage{cleveref}
\usepackage{amsmath}

\begin{document}
	
	\title{Reflected Entropy for Communicating Black Holes II: Planck Braneworlds}
	
	\author{Mir Afrasiar${}^{\dagger}$, Jaydeep Kumar Basak${}^{\ddagger,\star,\dagger}$, Ashish Chandra${}^{\dagger}$ and Gautam Sengupta${}^{\dagger}$}
	
	\address{ \vspace{0.4cm}
		{\it $\ddagger$
			Department of Physics,\\ National Sun Yat-Sen University, \\ Kaohsiung 80424, Taiwan\\}
		{\it $\star$
			Center for Theoretical and Computational Physics, \\ Kaohsiung 80424, Taiwan\\}
		{\it $\dagger$
			Department of Physics,\\Indian Institute of Technology Kanpur, \\208016, India\\
	}}
	
	\date{}			
	
	\email{\href{mailto:afrasiar@iitk.ac.in}{afrasiar@iitk.ac.in}, \href{mailto:jkb.hep@gmail.com}{jkb.hep@gmail.com}, \href{mailto:achandra@iitk.ac.in}{achandra@iitk.ac.in},\\ \href{mailto:sengupta@iitk.ac.in}{sengupta@iitk.ac.in}}
	
\begin{comment}
	
	\author{Mir Afrasiar${}^{\dagger}$, Jaydeep Kumar Basak${}^{\ddagger,\star,\dagger}$, Ashish Chandra${}^{\dagger}$ and Gautam Sengupta${}^{\dagger}$}
	
%	\author[a]{Mir Afrasiar,}
%	\author[a]{Jaydeep Kumar Basak,}
%	\author[a]{Ashish Chandra}
%	\author[a]{and Gautam Sengupta}

\address{ \vspace{0.4cm}
	{\it $\ddagger$
		Department of Physics,\\ National Sun Yat-Sen University, \\ Kaohsiung 80424, Taiwan\\}
	{\it $\star$
		Center for Theoretical and Computational Physics, \\ Kaohsiung 80424, Taiwan\\}
	{\it $\dagger$
		Department of Physics,\\Indian Institute of Technology Kanpur, \\208016, India\\
}}
	
%	\affiliation[a]{
%		Department of Physics,\\
%		Indian Institute of Technology,\\
%		Kanpur 208 016, India
%		\bigskip
%	}
	
	\emailAdd{afrasiar@iitk.ac.in}
	\emailAdd{jkb.hep@gmail.com}
	\emailAdd{achandra@iitk.ac.in}
	\emailAdd{sengupta@iitk.ac.in}
\end{comment}
\abstract{ 
We obtain the reflected entropy for bipartite mixed state configurations of two adjacent and disjoint subsystems at a finite temperature in finite-sized non-gravitating reservoirs described by $CFT_2$s each coupled to two quantum dots at their boundaries in the large central charge limit through a replica technique. These field theory results are substantiated through a holographic computation involving the entanglement wedge cross section in the dual bulk BTZ black hole geometry truncated by two Planck branes. The two Planck branes are the holographic duals of the quantum dots described by $AdS_2$ slices with JT black holes. Our results reproduce the holographic duality between the reflected entropy and the bulk entanglement wedge cross section in the context of the $AdS_3/CFT_2$ correspondence. Subsequently we analyze the behaviour of the holographic Markov gap between the reflected entropy and the mutual information for different scenarios involving the subsystem sizes and time.

}

\newpage
\tableofcontents

\vfill\eject

\section{Introduction}
Understanding the black hole information problem \cite{Hawking:1975vcx,Hawking:1976ra} has led to various key insights into the issue of a quantum theory of gravity. The intriguing aspect of this problem involves the fine-grained entanglement entropy of the Hawking radiation from an evaporating black hole dominating the coarse-grained thermodynamic entropy at late times which leads to a violation of unitarity. For a unitary evolution the entanglement entropy of the Hawking radiation is expected to follow the Page curve \cite{Page:1993wv}. Recently, this puzzle has been investigated for toy models of quantum field theories coupled to semiclassical gravity, and a possible resolution to this issue was proposed through the ``island" ( quantum extremal surface) formula for the fine-grained generalized entanglement entropy of the Hawking radiation. Specifically the authors in \cite{Almheiri:2019psf,Penington:2019npb} proposed the quantum extremal surface (QES) formula motivated by the quantum corrected Ryu-Takayanagi (RT) prescription described in \cite{Ryu:2006bv,Ryu:2006ef,Hubeny:2007xt,Faulkner:2013ana,Engelhardt:2014gca,Almheiri:2019hni,Almheiri:2020cfm,Almheiri:2019yqk}. The ``island" formula emphasizes that the fine-grained generalized entanglement entropy of a subregion in quantum field theories coupled to semiclassical gravity receives contribution 
from regions termed entanglement island at late times\footnote{Recently, there has been a rich development in this directions which can be found in \cite{Almheiri:2019psy,Anderson:2020vwi,Chen:2019iro,Balasubramanian:2020hfs,Chen:2020wiq,Gautason:2020tmk,Bhattacharya:2020ymw,Anegawa:2020ezn,Hashimoto:2020cas,Hartman:2020swn,Krishnan:2020oun,Alishahiha:2020qza,Geng:2020qvw,Li:2020ceg,Chandrasekaran:2020qtn,Bak:2020enw,Krishnan:2020fer,Karlsson:2020uga,Hartman:2020khs,Balasubramanian:2020coy,Balasubramanian:2020xqf,Sybesma:2020fxg,Chen:2020hmv,Ling:2020laa,Hernandez:2020nem,Marolf:2020rpm,Matsuo:2020ypv,Akal:2020twv,Caceres:2020jcn,Raju:2020smc,Deng:2020ent,Anous:2022wqh,Bousso:2022gth,Hu:2022ymx,Grimaldi:2022suv,Akers:2022max,Yu:2021rfg,Geng:2021mic,Chou:2021boq,Hollowood:2021lsw,He:2021mst,Arefeva:2021kfx,Ling:2021vxe,Bhattacharya:2021dnd,Azarnia:2021uch,Saha:2021ohr,Hollowood:2021wkw,Sun:2021dfl,Li:2021dmf,Aguilar-Gutierrez:2021bns,Ahn:2021chg,Yu:2021cgi,Lu:2021gmv,Caceres:2021fuw,Akal:2021foz,Arefeva:2022cam,Arefeva:2022guf,Bousso:2022ntt,Krishnan:2021ffb,Zeng:2021kyb,Teresi:2021qff,Okuyama:2021bqg,Chen:2021jzx,Pedraza:2021ssc,Guo:2021blh,Kibe:2021gtw,Renner:2021qbe,Dong:2021oad,Raju:2021lwh,Nam:2021bml,Kames-King:2021etp,Chen:2021lnq,Sato:2021ftf,Kudler-Flam:2021alo,Geng:2021iyq,Wang:2021afl,Ageev:2021ipd,Buoninfante:2021ijy,Cadoni:2021ypx,Marolf:2021ghr,Chu:2021gdb,Urbach:2021zil,Li:2021lfo,Neuenfeld:2021bsb,Aalsma:2021bit,Ghosh:2021axl,Bhattacharya:2021jrn,Geng:2021wcq,Krishnan:2021faa,Verheijden:2021yrb,Bousso:2021sji,Karananas:2020fwx,Goto:2020wnk,Bhattacharya:2020uun,Chen:2020jvn,Agon:2020fqs,Laddha:2020kvp,Akers:2019nfi,Chen:2019uhq,Basu:2022reu,Uhlemann:2021nhu,Uhlemann:2021itz,Germani:2022rac,Yadav:2022fmo,Omidi:2021opl} and the references therein.}. It was shown in \cite{Almheiri:2019hni} that the entanglement island appears in the bulk entanglement wedge for the subregion in the QFT at late times in the context of the AdS/CFT scenario. The corresponding generalized entanglement entropy of a subregion $\mathcal{R}$ in the radiation flux of an evaporating black hole is given by
\begin{align}\label{IsformEE}
	S[\mathcal{R}]=\min \left\{\operatorname{ext}_{Is(\mathcal{R})}\left[\frac{\operatorname{Area}[\partial Is(\mathcal{R})]}{4 G_{N}}+S_{eff}[\operatorname{\mathcal{R}} \cup Is(\mathcal{R})]\right]\right\}.
\end{align} 
In the above equation, $Is(\mathcal{R})$ is the island region in the black hole geometry corresponding to the subregion 
$\mathcal{R}$ \footnote{Higher dimensional generalization of the island construction for the entanglement entropy has been studied in some recent papers \cite{Almheiri:2019psy,Chen:2020uac,Chen:2020hmv,Ling:2020laa,Krishnan:2020fer}.}. The proof of the above ``island" formula was provided in \cite{Penington:2019kki,Almheiri:2019qdq} through a gravitational path integral in the context of two dimensional Jackiw-Teitelboim (JT) gravity \cite{Jackiw:1984je,Teitelboim:1983ux} involving saddle points described by replica wormhole configurations.

Earlier, the ``island" formalism was explored in toy models of a semi-infinite non-gravitating radiation reservoir coupled to a quantum mechanical system (quantum dot) at the boundary \cite{Almheiri:2019hni,Almheiri:2019yqk}. The holographic dual of the quantum dot is a Planck brane containing JT gravity in a $AdS_2$ slice of a bulk $AdS_3$ geometry. The  non-gravitating reservoir and the Planck branes are described by a $CFT_2$ matter field with transparent boundary conditions implemented at the junction\cite{Almheiri:2019hni,Almheiri:2019yqk}. At a finite temperature two such copies of semi-infinite $CFT_2$ reservoirs coupled to quantum dots at their boundaries constitute a thermo-field double (TFD) state. In this construction, the Page curve for the entanglement entropy of the radiation reservoirs was reproduced utilizing the ``island" prescription.  

In a generalization of the above construction, the authors in \cite{Balasubramanian:2020coy} considered a finite-sized non-gravitating reservoir described by $CFT_2$ matter field coupled to two quantum dots at their boundaries. At a finite temperature this model involves two such copies of finite sized non-gravitating radiation reservoirs characterized by $CFT_2$s each with quantum dots located at their boundaries. The bulk dual geometry corresponding to these quantum dots are described by two Planck branes truncating the $AdS_3$ space time which are $AdS_2$ slices involving eternal JT black holes. Transparent boundary conditions were implemented at the junctions of the Planck branes with the non gravitating radiation reservoirs. Subsequently, the authors computed the generalized entanglement entropy of a finite segment in both the radiation reservoirs which involves the communication of entanglement between the two eternal JT black holes on the two Planck branes. 

In a different context, it is known from quantum information theory that the entanglement entropy serves as a valid entanglement measure for pure states. However the entanglement entropy is not suitable to characterize mixed state entanglement as it receives contributions from irrelevant classical and quantum correlations. This requires the introduction of other measures in quantum information theory to characterize the entanglement for mixed states. In this connection
various entanglement and correlation measures were proposed to describe mixed state entanglement such as the reflected entropy, entanglement negativity, entanglement of purification amongst others in the context of quantum information theory \cite{Vidal:2002zz,Terhal:2002,Plenio:2007zz,Horodecki:2009zz,Dutta:2019gen}. Some of these measures could be explicitly computed for $CFT_{2}$ through field theory replica techniques in \cite{Dutta:2019gen,Calabrese:2012ew,Calabrese:2012nk,Calabrese:2014yza,Hirai:2018jwy,Caputa:2018xuf} and were substantiated through bulk holographic computations described in \cite{Rangamani:2014ywa,Chaturvedi:2016rcn,Chaturvedi:2016rft,Jain:2017aqk,Takayanagi:2017knl,Malvimat:2018txq,Malvimat:2018ood,Kudler-Flam:2018qjo,Kusuki:2019zsp,Dutta:2019gen,KumarBasak:2020eia,KumarBasak:2021lwm,Dong:2021clv,Basu:2021axf,Basu:2021awn,Basu:2022nds} as well. In \cite{Afrasiar:2022ebi}, the holographic entanglement negativity for various bipartite mixed states of two adjacent and disjoint intervals in the communicating black holes model described in \cite{Balasubramanian:2021xcm} were computed through the ``island" prescription. Subsequently, analogue of the Page curves for the entanglement negativity were reproduced for different scenarios involving the subsystem sizes and time.

For the past few years the mixed state correlation measure termed the reflected entropy $S_R$ \cite{Dutta:2019gen} has gained significant attention due to its holographic duality with the bulk entanglement wedge cross-section (EWCS). The
reflected entropy in various models of evaporating black holes was also investigated in the context of the island construction in \cite{Chandrasekaran:2020qtn,Li:2020ceg,Ling:2021vxe,Akers:2022max} and analogue of the corresponding Page curve were reproduced. In this article, we compute the reflected entropy for bipartite mixed states of two adjacent and disjoint subsystems in the model of two JT black holes communicating through finite-sized non-gravitating reservoirs \cite{Balasubramanian:2021xcm} described earlier. Subsequently, we also obtain the holographic mutual information $I$ for the mixed states under consideration and compare their profiles with the corresponding profiles for the holographic reflected entropy for these mixed states for various subsystem sizes and the time. The difference between these two measures characterized an important feature of  multipartite entanglement termed the holographic Markov gap \cite{Hayden:2021gno,Bueno:2020fle,Camargo:2021aiq,Bueno:2020vnx} in this scenario.

This article is organized as follows. In \cref{review}, we briefly review some earlier works related to our article. In sub\cref{subsec: two comm} we review the model discussed in \cite{Balasubramanian:2021xcm}. Subsequently, we provide a short review of the reflected entropy for bipartite mixed states in sub\cref{subsec: srreview} in the context of the $AdS/CFT$ scenario. Finally in the sub\cref{subsec: markovreview}, we describe the issue of Markov gap in quantum information theory. Next, in \cref{subsec:srcomputation}, a detailed computation of the reflected entropy is described for bipartite mixed states of two adjacent and disjoint subsystems in finite-sized reservoirs each coupled to two quantum dots. Subsequently, we substantiate these field theory results from the explicit holographic computations of the bulk EWCS in the dual eternal BTZ black hole geometry. In \cref{subsec:markovcomputaion}, we discuss the holographic Markov gap for different scenarios of the adjacent and disjoint subsystems by comparing the corresponding profiles of the reflected entropy and the mutual information. Finally in \cref{discussion}, we discuss and summarize our results with some future open issues.

%%%%%%%%%%%%%%%%%%%%%%%%%%%%%%%%%%%%%%%%%%%%%%%%%%%%%%%%%%%%%%%%%%%%%%%%%%%%%%%%%%%%

\section{Review of earlier results}\label{review}
In this section, we begin with a brief review of the configuration of two communicating eternal JT black holes with two finite sized non-gravitating reservoirs in two dimensions coupled to two quantum dots at their boundaries as described in \cite{Balasubramanian:2021xcm}. The bulk dual of this configuration is described by $(2+1)$-dimensional eternal BTZ black hole geometry truncated by two Planck branes with $AdS_2$ geometries. The two dimensional communicating eternal JT black holes in this case are located on these Planck branes and the entire system of the
black holes and the reservoirs is described by a matter $CFT_2$ with transparent boundary conditions at the junctions \cite{Almheiri:2019yqk,Almheiri:2019hni}. Subsequently, we will also review the field theory replica technique for the computation of the reflected entropy and the corresponding bulk entanglement wedge cross section in the $AdS_{3}/CFT_{2}$ scenario. Finally, we will discuss the emergence of the holographic Markov gap between the reflected entropy and the corresponding mutual information.

\subsection{Communicating black holes}\label{subsec: two comm} 

\begin{figure}[H]
	\centering
	\begin{subfigure}[b]{0.45\textwidth}
		\centering
		\includegraphics[width=\textwidth]{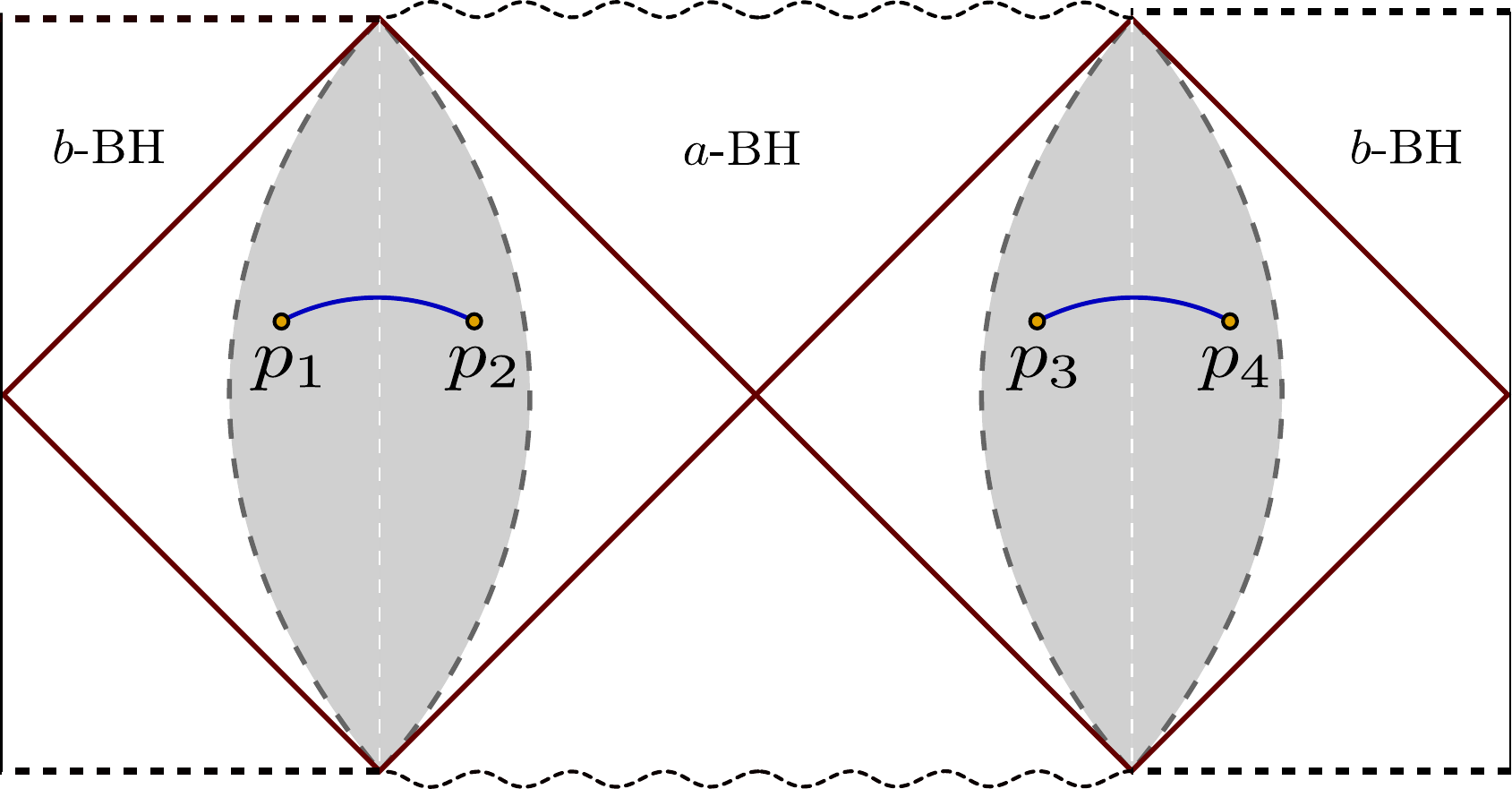}
		\caption{}
		\label{a}
	\end{subfigure}
	\hspace{.5cm}
	\begin{subfigure}[b]{0.45\textwidth}
		\centering
		\includegraphics[width=\textwidth]{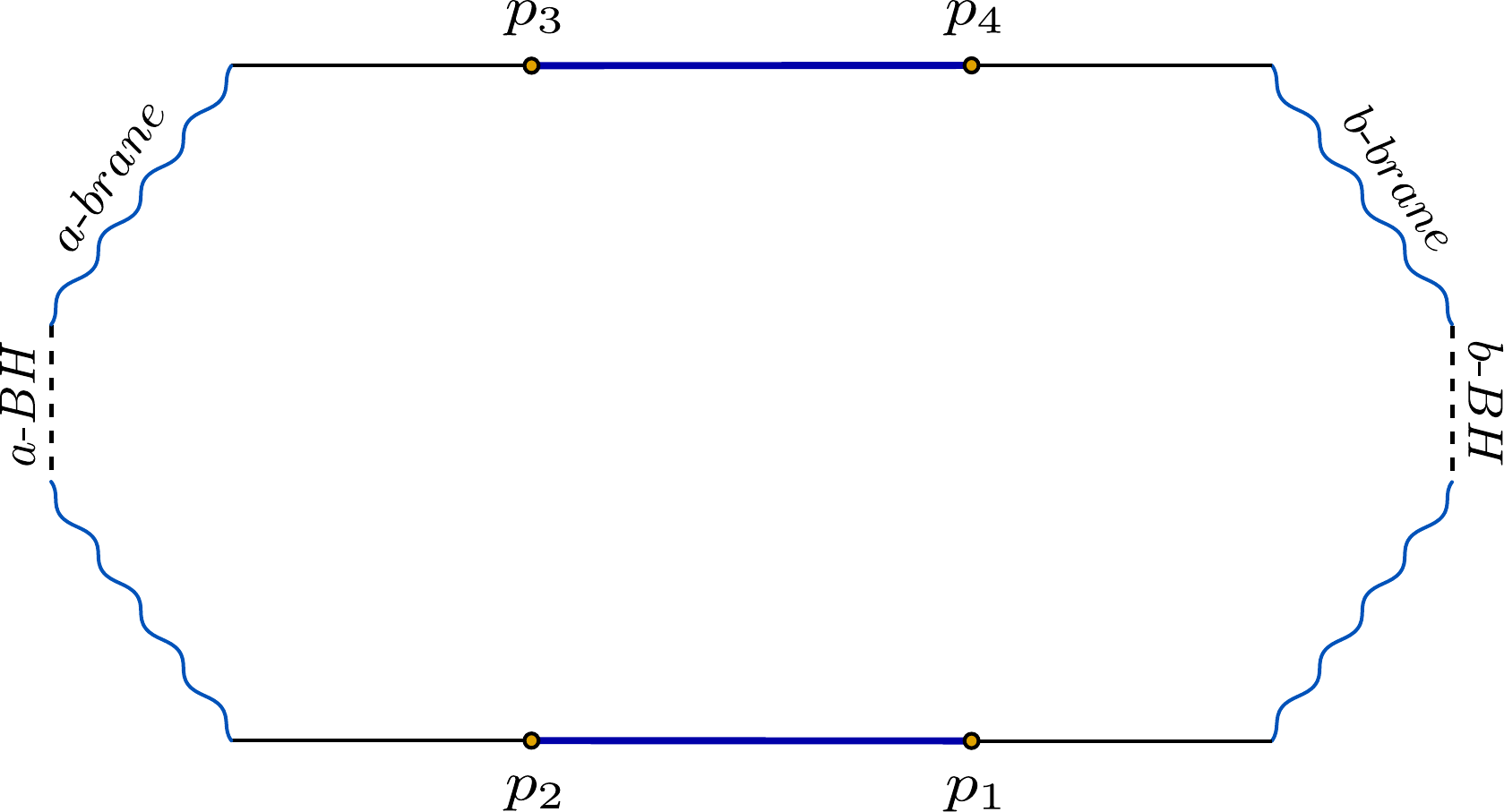}
		\caption{}
		\label{b}
	\end{subfigure}
	\caption{Schematics depict the maximal extension of the Penrose diagram for two $AdS_2$ eternal black holes. The union of segments considered in the radiation reservoirs  (shaded regions) describes a subsystem with the endpoints labeled as  $p_1$, $p_2$, $p_3$, $p_4$. The eternal black holes located on the JT branes are denoted as $a$ and $b$. (Figure modified from \cite{Balasubramanian:2021xcm})  }\label{penrose2}
\end{figure}

In this subsection, we describe the model of \cite{Balasubramanian:2021xcm} and consider two eternal JT black holes at the same temperature. For this case, the bulk computation of the generalized entanglement entropy for a subsystem $A$ described by the union of two identical intervals in the two reservoirs has been described in \cite{Afrasiar:2022ebi}.
The Penrose diagram of the eternal JT black holes\footnote{ The two eternal JT black holes together with the two Planck branes are labeled as $a$ and $b$.} located on the Planck branes is described in \cref{penrose2} which are coupled to each other through the shared reservoirs.

For this configuration, the metric describing the exterior regions of the two eternal JT black holes may be expressed as
\begin{eqnarray}
	ds_1^2 &=& \frac{4\pi^2}{\beta^2} \frac{-dt^2+d\xi^2}{\sinh^2 \frac{2\pi \xi}{\beta}} , \quad \xi \in \left(-\infty, - \epsilon\right]\,, \label{MetricPhysL1b1}\\
	ds_2^2 &=& \frac{4\pi^2}{\beta^2} \frac{-dt^2+d\xi^2}{\sinh^2 \frac{2\pi }{\beta}(\xi - L)},\quad \xi \in \left[L + \epsilon, +\infty\right) \label{MetricPhysL2b2}\,.
\end{eqnarray} 
The corresponding radiation reservoirs are defined by the flat metric
\begin{equation}\label{MetricBath}
	ds_{R}^2 = \frac{-d t^2 + d\xi^2}{\epsilon^2}\,, \quad \xi \in \left[- \epsilon, L+\epsilon\right]\,, 
\end{equation}
where the reservoir is glued continuously to the surfaces $\xi=-\epsilon$ and $L+\epsilon$. The dilaton profiles for the two eternal JT black holes on the Planck branes are then given as follows
\begin{equation}\label{dilaton2BH}
\begin{aligned}
	\phi_{a}(\xi) =&\Phi_0+ \frac{2\pi\phi_r}{\beta}  \coth \frac{2\pi \xi}{\beta}\,, \\
	\phi_{b}(\xi) =& \Phi_0+\frac{2\pi\phi_r}{\beta}  \coth \frac{2\pi}{\beta}\left(\xi-L\right). 
\end{aligned}
\end{equation}
Note that here the three dimensional bulk geometry consists of an eternal BTZ black hole truncated by the two Planck branes which is described by the following metric \cite{Grimaldi:2022suv,Almheiri:2019hni,Balasubramanian:2021xcm}
\begin{equation}\label{BTZmetric}
	ds^2= -\frac{(1-\frac{z^2}{z_h^2})}{z^2}dt^2+\frac{1}{z^2 (1-\frac{z^2}{z_h^2})}dz^2+\frac{1}{z^2}dx^2\,,
\end{equation}  
where the horizon at $z_h$ is related to the inverse temperature as $\beta=2\pi z_h$. 

%%%%%%%%%%%%%%%%%%%%%%%%%%%%%%%%%%%%%%%%%%%%%%%%%%%%%%%%%%%%%%%%%%%%%%%%%%%%%%%%%

\subsection{Reflected entropy} \label{subsec: srreview}

In this subsection we provide a brief review of the reflected entropy for the case of two disjoint subsystems as described in \cite{Dutta:2019gen}. The authors in this work first proposed the mixed state correlation measure of the reflected entropy which involved a canonical purification of a bipartite mixed state. In this regards, consider a bipartite mixed state $\rho_{AB}$ of disjoint subsystems $A$ and $B$. The canonically purified state $\ket{\sqrt{\rho_{AB}}}$ in a doubled Hilbert space $\mathcal{H}_A\otimes\mathcal{H}_B\otimes\mathcal{H}_{A^\star}
\otimes\mathcal{H}_{B^\star}$ may be constructed to purify\footnote{ For more details about the construction of the canonically purified state $\ket{\sqrt{\rho_{AB}}}$ see also \cite{Dutta:2019gen,Jeong:2019xdr}} the given mixed state $\rho_{AB}$ where $A^\star$ and $B^\star$ are the CPT conjugate copies of the subsystems $A$ and $B$ respectively. The reflected entropy $S_R(A: B)$ for the bipartite subsystems may be defined as \cite{Dutta:2019gen}   
\begin{align}\label{RE}
	S^R(A: B) = S_{vN}(\rho_{AA^\star})_{\sqrt{\rho_{AB}}}
\end{align}
where $S_{vN}$ denotes the von Neumann entropy.
In \cref{RE}, the reduced density matrix  $\rho_{AA^{\star}}$ is given by
\begin{equation}
	\rho_{AA^{\star}}=\Tr_{\mathcal{H}_B\otimes\mathcal{H}_{B^\star}}\ket{\sqrt{\rho_{AB}}}\bra{\sqrt{\rho_{AB}}}.
\end{equation} 
The authors of \cite{Dutta:2019gen} developed a replica technique to compute the reflected entropy for bipartite mixed states in $CFT_2$s. For the bipartite state $|\rho^{m/2}_{AB}\rangle \equiv \ket{\psi_{m}}$, the replica manifold involves $m$-replication of the original manifold with $m \in 2\mathbb{Z^{+}}$. The corresponding
reduced density matrix after tracing over the subsystems $BB^*$ is then described as
\begin{equation}
	\rho^{(m)}_{AA^{*}}= \text{Tr}_{\mathcal{H}_{B} \otimes \mathcal{H}_{B^{*}}} \ket{\psi_{m}} \bra{\psi_{m}}.
\end{equation}
Using this reduced density matrix, the R\'enyi reflected entropy may now be defined  as $S_n(\rho^{(m)}_{AA^{*}})_{\psi_m}$ which involves an $nm$-sheeted replica manifold with $n \in 2\mathbb{Z}$. The reflected entropy may then be obtained by implementing the replica limits $n \to 1$ and $m \to 1$ as \footnote{The two replica limits $n \to 1$ and $m \to 1$ do not commute with each other as discussed in \cite{ Dutta:2019gen, Akers:2022max}. In this work, we first consider $n \to 1$ and subsequently $m \to 1$ limit to compute the reflected entropy as suggested in \cite{Dutta:2019gen, Akers:2022max}.}
\begin{equation}\label{SR-def}
	S^{R}(A:B)=\lim_{n,m \to1} S_{n}(AA^{*})_{\psi_{m}}.
\end{equation}
We now consider the case of the mixed state configuration of two disjoint subsystems$A \equiv [z_1,z_2]$ and $B \equiv [z_3,z_4]$ in a $CFT_2$.  As described in \cite{Dutta:2019gen} the reflected entropy involving four point twist field correlators may be obtained through the replica technique above as follows

\begin{equation}
	\begin{aligned}
		S^R(A:B) &=\lim_{ n \to 1 }\lim_{m \to 1 }\,S_n\left(AA^*\right)_{\psi_m}\\ 
		&= \lim_{ n \to 1 }\lim_{m \to 1 } \frac{1}{1-n}\log\frac{\left<\sigma_{g^{}_A}(z_1)\sigma_{g_A^{-1}}(z_2)\sigma_{g^{}_B}(z_3)\sigma_{g_B^{-1}}(z_4)\right>_{\mathrm{CFT}^{\bigotimes mn}}}{\left(\left<\sigma_{g^{}_m}(z_1)\sigma_{g_m^{-1}}(z_2)\sigma_{g^{}_m}(z_3)\sigma_{g_m^{-1}}(z_4)\right>_{\mathrm{CFT}^{\bigotimes m}}\right)^n}.\label{Renyi-reflected}
	\end{aligned}
\end{equation}
Here the twist operators $\sigma_{g_A}$ and $\sigma_{g_B}$ are located at the endpoints of the corresponding subsystems $A$ and $B$. 
The conformal dimensions for these twist operators are given by
\begin{equation}\label{reflected-twist-field}
	h_{g_A^{-1}}=h_{g^{}_B}=\frac{n \,c}{24}\left(m-\frac{1}{m}\right),
	\quad h_{{g^{}_B} g_A^{-1}}=\frac{2 \,c}{24}\left(n-\frac{1}{n}\right), \quad h_{g_m}=\frac{c}{24}\left(m-\frac{1}{m}\right).
\end{equation}

Following this, the authors of \cite{Chandrasekaran:2020qtn} developed a generalized version for the reflected entropy of a bipartite system $AB$ in a holographic semi-infinite $CFT_2$ coupled to a semi-classical gravity which is given as follows
\begin{align} \label{srew}
	S^R_{\rm gen}(A:B)=S^R_{\text{eff}}(A\cup \text{Is}_R(A):B\cup \text{Is}_R(B))
	+\frac{\text{Area}[Q']}{2G^{(2)}_{N}}~,
\end{align}
where $Q'=\partial \text{Is}_R(A)\cap \partial \text{Is}_R(B)$. In the above expression, the first term can be computed utilizing the formula described in \cref{SR-def} and the second term is given by the value of the dilaton field at the island point $Q'$ on the JT brane. In this context, since the $CFT_2$ located on the brane and the flat non-gravitating region is considered to be holographic, the term $S_R^{(\text{eff})}$ can be obtained from the doubly holographic perspective by computing the area of the dual EWCS in the bulk geometry as
\begin{align} \label{srew2}
	S^R_{\rm gen}(A:B)&=\frac{2 \rm Area(EWCS)}{4 G_N}
	+\frac{\text{Area}[Q']}{2G^{(2)}_{N}}\notag\\
	&=2 E_{W}(A: B)
	+\frac{\text{Area}[Q']}{2G^{(2)}_{N}}~,
\end{align}
where we have defined $E_{W}(A: B)=\frac{2 \rm Area(EWCS)}{4 G_N}$ in the above expression. In the following subsection, we describe the holographic Markov gap \cite{Hayden:2021gno} as the difference between the reflected entropy and the mutual information in the framework of  the $AdS_3/CFT_2$ correspondence.

\subsection{Markov gap} \label{subsec: markovreview}
In quantum information theory, the Markov gap may be described through the Markov recovery process which is defined as a recovery of the quantum state $\rho_{A B C}$ from the bipartite mixed state $\rho_{AB}$. Now if we define a quantum channel ${\cal R}_{B\to BC}$ whose action on the bipartite mixed state $\rho_{AB}$ produces tripartite state $\tilde{\rho}_{A B C}$ as
\begin{align}
	\tilde{\rho}_{A B C}=\mathcal{R}_{B \rightarrow B C}\left(\rho_{A B}\right)\,,
\end{align}
where the quantum channel $\mathcal{R}$ acts on the subsystem $B$ only. If this tripartite state ($\tilde{\rho}_{A B C}$) is equal to the quantum state $\rho_{A B C}$ then the Markov recovery process is perfect, and the corresponding state is said to be a quantum Markov chain with the ordering $A\to B\to C$. As discussed in \cite{Petz:1986tvy}, the conditional mutual information vanishes for the perfect Markov recovery process. In this context, the authors of \cite{Fawzi_2015} further studied the conditional mutual information and proposed a bound in terms of the quantum Fidelity $F\left(\rho_{A B C}, \mathcal{R}_{B \rightarrow B C}\left(\rho_{A B}\right)\right)$ as 
\begin{align}\label{CMI}
	I(A: C \mid B) \geq-\max _{\mathcal{R}_{B \rightarrow B C}} \log F\left(\rho_{A B C}, \mathcal{R}_{B \rightarrow B C}\left(\rho_{A B}\right)\right)\,.
\end{align}
Here the quantum Fidelity becomes 1 when the quantum recovery process is perfect and it becomes zero when the corresponding density matrices have support on the orthogonal subspaces. The authors of \cite{Fawzi_2015} also analyzed the bound in \cref{CMI} for the Markov recovery process of the reduced density matrix $\rho_{A B B^*}$ in the context of canonical purification of the subsystem B. This led to the following constraint on the conditional mutual information 
\begin{align}\label{bound}
	I(A: B \mid B^*) &\geq-\max _{\mathcal{R}_{B \rightarrow B B^*}} \log F\left(\rho_{A B B^*}, \mathcal{R}_{B \rightarrow B B^*}\left(\rho_{A B}\right)\right)\,,\\
	S^R(A:B)-I(A:B)&\geq-\max _{\mathcal{R}_{B \rightarrow B B^*}} \log F\left(\rho_{A B B^*}, \mathcal{R}_{B \rightarrow B B^*}\left(\rho_{A B}\right)\right)\,,
\end{align}
where in the last inequality, the conditional mutual information is expressed as the difference between reflected entropy and the mutual information. This difference is called as Markov gap as described in \cite{Hayden:2021gno}. Subsequently, the authors also demonstrate that the bound in \cref{bound} may be expressed geometrically in the framework of $AdS_3/CFT_2$ as
\begin{align}\label{MG}
	S^{R}(A: B)-I(A: B) \geq \frac{\log (2) \ell_{\text {AdS }}}{2 G_{N}} \times(\# \text { of boundaries of  EWCS })+\mathcal{O}\left(\frac{1}{G_{N}}\right)\,,
\end{align}
where $\#$ represents number of non-trivial boundaries of the EWCS in the bulk of $AdS_3$ geometry. Note that the endpoints of the EWCS located at spatial infinity does not contribute in the \cref{MG}.
%%%%%%%%%%%%%%%%%%%%%%%%%%%%%%%%%%%%%%%%%%%%%%%%%%%%%%%%%%%%%%%%%%%%%%%%%%%%%%%%%%%%%%

\section{Reflected entropy and EWCS in two Communicating black holes }\label{subsec:srcomputation}

In this section we compute the reflected entropy for various bipartite mixed states described by two adjacent and disjoint subsystems in the radiation reservoirs for the two communicating black holes configuration \cite{Balasubramanian:2021xcm} reviewed in sub\cref{subsec: two comm}. In this context we first consider the adjacent subsystems in the radiation reservoirs and obtain the reflected entropy for different channels of the corresponding twist correlator in the large central charge limit. We then follow a similar analysis for the computation of the reflected entropy for two disjoint subsystems. Finally, we substantiate these field theory results from a holographic computation of the EWCS in the dual bulk geometry for these cases.  Note that in the configurations (6),(7), (8) and (9) for two adjacent and disjoint subsystems, we receive an additional contribution from the dilaton term \cref{dilaton2BH} in the generalized reflected entropy formula described in the \cref{srew,srew2}. However in the following section, this additional term becomes zero for other contributions since there is no QES point located on the JT brane.

\subsection{Reflected entropy} 
\subsubsection{Adjacent subsystems}\label{sradjacent}
We start our discussion with the computation of the reflected entropy for two adjacent subsystems $A\equiv[p_1,p_2]\cup[p_5,p_6]$ and $B\equiv[p_2,p_3]\cup[p_4,p_5]$ located in the radiation reservoirs as depicted 
in the figures below. To this end we utilize the replica technique developed in \cite{Dutta:2019gen} to obtain the reflected entropy for the above bipartite mixed state for different subsystem sizes. It is observed that the reflected entropy receives contributions from various dominant channels of the corresponding twist correlators in the large central charge limit.
In what follows we compute these contributions to the reflected entropy for different configurations described by the relative subsystem sizes for the communicating black hole setup as shown in \cref{penrose2}.

\subsubsection*{Configuration-1}
We begin with the computation of the reflected entropy for the contribution as described in \cref{taka} where the twist operators are located at the endpoints of the two adjacent subsystems and the Planck branes where the entire black hole/reservoir setup is described by the same $CFT_2$. The R\'enyi reflected entropy for this configuration may be obtained from the following twist correlator

\begin{align}\label{sr-1}
	S^{R}_{n,m}(A:B) = 2\frac{1}{1-n}\log \frac{\left< \sigma_{g^{-1}_B}(a_1)\sigma_{g_A}(p_1)\sigma_{g_Bg_A^{-1}}(p_2)\sigma_{g_B^{-1}}(p_3)\sigma_{g_B}(b_1) \right>_{\mathrm{CFT}^{\bigotimes mn}}}{\left<\sigma_{g^{-1}_m}(a_1)\sigma_{g_m}(p_1)\sigma_{g_m^{-1}}(p_3)\sigma_{g_m}(b_1) \right>^n_{\mathrm{CFT}^{\bigotimes m}}}\,,	
\end{align}
where the points $a_1$ and $b_1$ are located on the $a$ and $b$-branes respectively. In the above equation, the factor two correspond to the contribution of the reflected entropy from the TFD copy of the radiation reservoir. Note that the correlator in \cref{sr-1} factorizes into the respective contractions in the large central charge limit as follows \cite{Chandrasekaran:2020qtn}  
\begin{align}	
	S^{R}_{n,m}(A:B) = 2\frac{1}{1-n}\log\frac{\left< \sigma_{g^{-1}_B}(a_1)\sigma_{g_A}(p_1)\sigma_{g_Bg_A^{-1}}(p_2)\right>_{\mathrm{CFT}^{\bigotimes mn}}\left<\sigma_{g_B^{-1}}(p_3)\sigma_{g_B}(b_1)\right>_{\mathrm{CFT}^{\bigotimes mn}}}{\left(\left<\sigma_{g^{-1}_m}(a_1)\sigma_{g_m}(p_1)\right>_{\mathrm{CFT}^{\bigotimes m}}\left<\sigma_{g_m^{-1}}(p_3)\sigma_{g_m}(b_1) \right>_{\mathrm{CFT}^{\bigotimes m}}\right)^n} \,.
\end{align}
In this case the dominant contribution to the R\'enyi reflected entropy arises from the correlator
\begin{align}	
	S^{R}_{n,m}(A:B) = 2\frac{1}{1-n}\log\frac{\left< \sigma_{g^{-1}_B}(a_1)\sigma_{g_A}(p_1)\sigma_{g_Bg_A^{-1}}(p_2)\right>_{\mathrm{CFT}^{\bigotimes mn}}}{\left(\left<\sigma_{g^{-1}_m}(a_1)\sigma_{g_m}(p_1)\right>_{\mathrm{CFT}^{\bigotimes m}}\right)^n} \label{srcase1}\,
\end{align}
which may be obtained following the analysis described in  \cite{Chandrasekaran:2020qtn}. In the
replica limit ($n\rightarrow1$ and $m\rightarrow1$) the reflected entropy is then given from the above equation as follows
\begin{align}\label{reflectedtaka}	
	S^{R}_{\text{eff}}(A:B) = \frac{2c}{3}\log\left[\frac{\sinh\frac{2\pi(p_2-p_1)}{\beta}\sinh\frac{2\pi(a_1+p_2)}{\beta}}{\sinh\frac{2\pi(a_1+p_1)}{\beta}\sinh\frac{2\pi(\epsilon)}{\beta}}\right]\,,
\end{align}
where $\epsilon$ is the UV cutoff in the $CFT_2$.
%%%%%%%%%%%%%%%%%%%%%%%%%%%%%%%%%%%%%%%%%%%%%%%%%%%%%%%%%%%%%%%%%%%%%%%%%%%%%%%%%

\subsubsection*{Configuration-2}
\begin{figure}[H]
	\centering
	\begin{subfigure}[b]{0.45\textwidth}
		\centering
		\includegraphics[width=\textwidth]{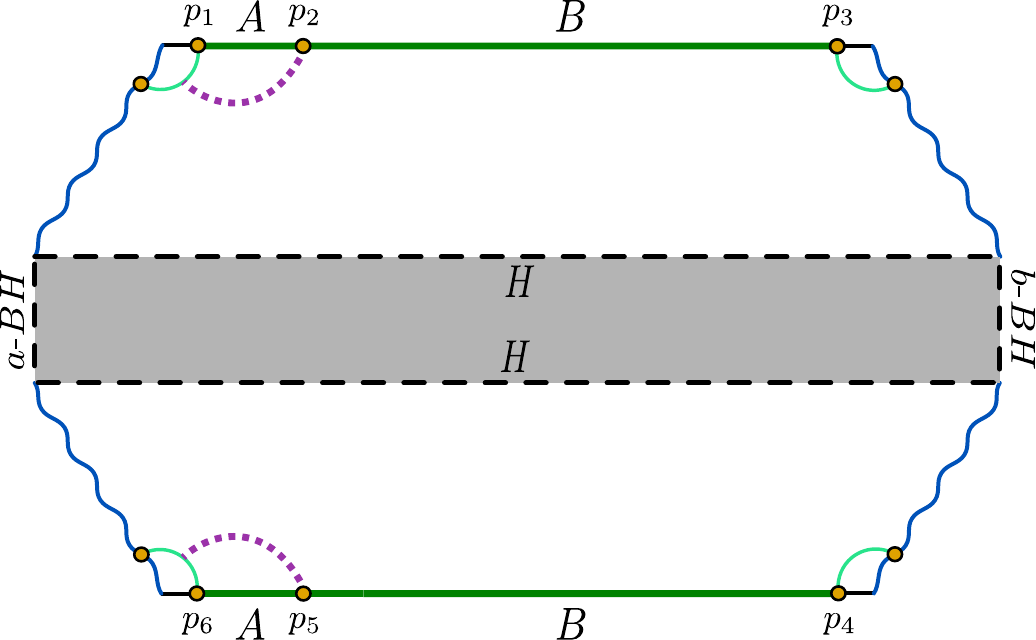}
		\caption{Configuration-1}
		\label{taka}
	\end{subfigure}
	\hspace{.5cm}
	\begin{subfigure}[b]{0.45\textwidth}
		\centering
		\includegraphics[width=\textwidth]{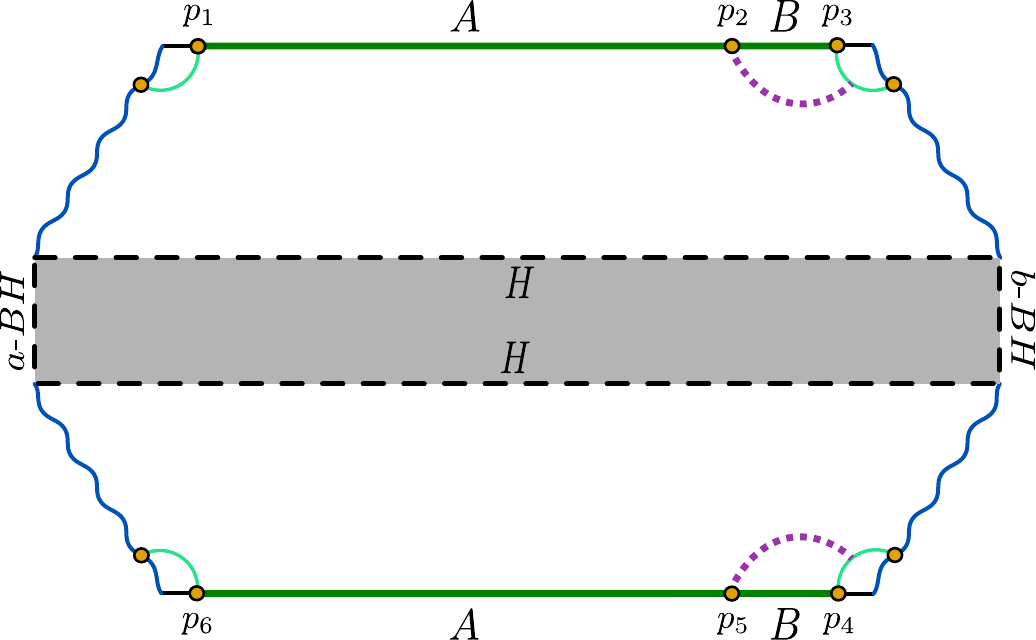}
		\caption{Configuration-2}
		\label{taka2}
	\end{subfigure}
	\hspace{.5cm}
	\begin{subfigure}[b]{0.45\textwidth}
		\centering
		\includegraphics[width=\textwidth]{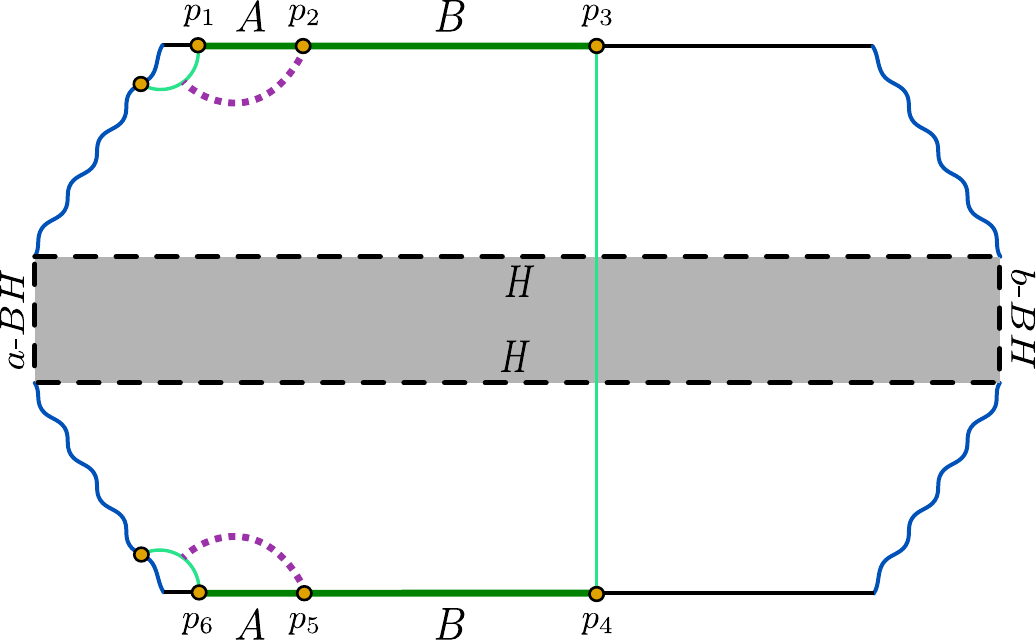}
		\caption{Configuration-3}
		\label{takabulk}
	\end{subfigure}
	\hspace{.5cm}
	\begin{subfigure}[b]{0.45\textwidth}
		\centering
		\includegraphics[width=\textwidth]{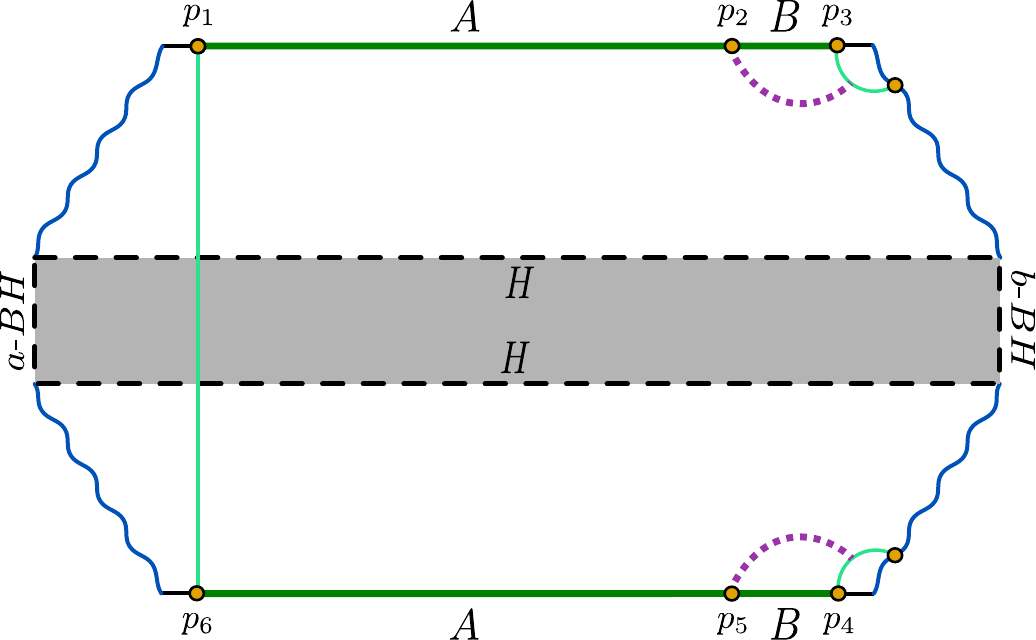}
		\caption{Configuration-4}
		\label{takabulk2}
	\end{subfigure}
	\vspace{.5cm}
	\begin{subfigure}[b]{0.45\textwidth}
		\centering
		\includegraphics[width=\textwidth]{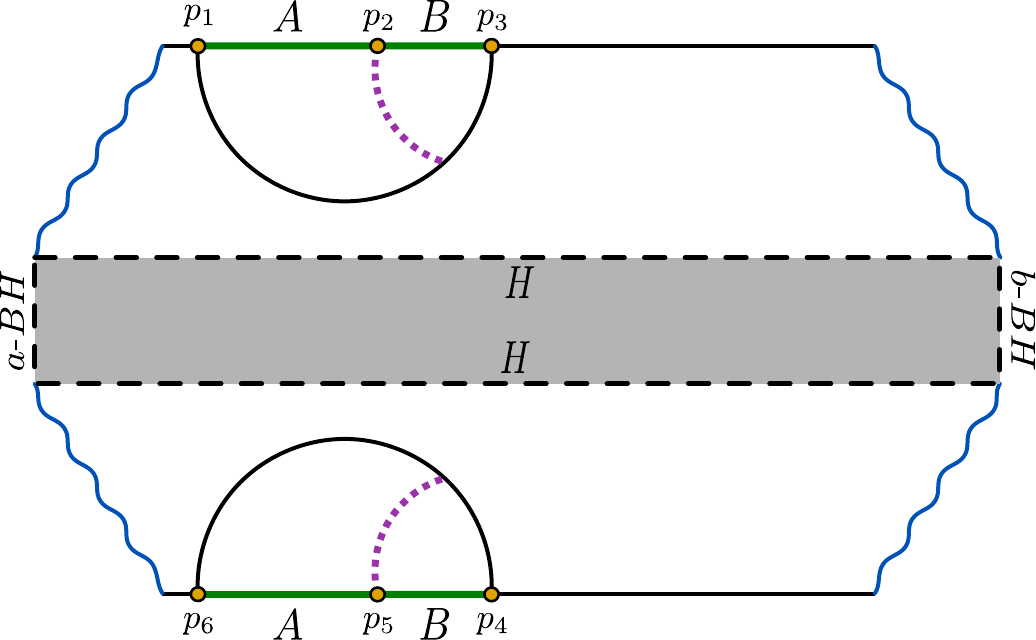}
		\caption{Configuration-5}
		\label{RTending}
	\end{subfigure}
	\caption{Schematics depicts all the possible contributions to the reflected entropy for the case of two adjacent subsystems $A$ and $B$ where we get connected phase of entanglement island for initial four phases only. }\label{takaresults}
\end{figure}

In this case, the computation of the reflected entropy for the two adjacent subsystems is similar to configuration-1, however, the factorization of the twist correlator in \cref{sr-1} in the large central charge limit is different
due to the change in the size of the subsystems $A$ and $B$ as shown in \cref{taka2} which alters the location of the relevant twist operators. In this case the R\'enyi reflected entropy is given by the following expression
\begin{align}	
	S^{R}_{n,m}(A:B) = 2\frac{1}{1-n}\log\frac{\left< \sigma_{g^{-1}_B}(a_1)\sigma_{g_A}(p_1)\right>_{\mathrm{CFT}^{\bigotimes mn}}\left<\sigma_{g_Bg_A^{-1}}(p_2)\sigma_{g_B^{-1}}(p_3)\sigma_{g_B}(b_1)\right>_{\mathrm{CFT}^{\bigotimes mn}}}{\left(\left<\sigma_{g^{-1}_m}(a_1)\sigma_{g_m}(p_1)\right>_{\mathrm{CFT}^{\bigotimes m}}\left<\sigma_{g_m^{-1}}(p_3)\sigma_{g_m}(b_1) \right>_{\mathrm{CFT}^{\bigotimes m}}\right)^n} \,,
\end{align}
where the factor two in the above expression arises from the reservoir copy of the double TFD states. Finally we obtain the dominant contribution to the R\'enyi reflected entropy in this configuration as follows
\begin{align}	
	S^{R}_{n,m}(A:B) = 2\frac{1}{1-n}\log\frac{\left< \sigma_{g_Bg_A^{-1}}(p_2)\sigma_{g_B^{-1}}(p_3)\sigma_{g_B}(b_1)\right>_{\mathrm{CFT}^{\bigotimes mn}}}{\left(\left<\sigma_{g_m^{-1}}(p_3)\sigma_{g_m}(b_1)\right>_{\mathrm{CFT}^{\bigotimes m}}\right)^n} \label{srcase2}\,,
\end{align}
which involves the points $p_2$, $p_3$ and $b_1$ after the factorization in the large central charge limit. The final expression for the reflected entropy in the replica limit following \cite{Dutta:2019gen,Chandrasekaran:2020qtn} is then given as
\begin{align}\label{reflectedtaka2}	
	S^{R}_{\text{eff}}(A:B) = \frac{2c}{3}\log\left[\frac{\sinh\frac{2\pi(p_3-p_2)}{\beta}\sinh\frac{2\pi(b_1+p_2)}{\beta}}{\sinh\frac{2\pi(b_1+p_3)}{\beta}\sinh\frac{2\pi(\epsilon)}{\beta}}\right]\,,
\end{align}
where $\epsilon$ is the $CFT_2$ UV cutoff.

\subsubsection*{Configuration- 3 and 4}
These configurations are depicted in the \cref{takabulk,takabulk2} and the computation of the reflected entropy for these cases is similar to the configurations 1 and 2. Here, the dominant contributions for the configuration-3 and 4 are given by the \cref{srcase1,srcase2} in the large central charge limit. Note that the only difference in these configurations arise due to the locations of the twist operators at the points $p_3$ and $p_6$ for the configuration-3 while the other configuration involves the points $p_1$ and $p_4$. However, these twist correlators cancel from the numerator and the denominator in the computation of the reflected entropy in the replica limit. Hence the reflected entropy for the configuration- 3 is given by \cref{reflectedtaka} and similarly \cref{reflectedtaka2} provides the result for the configuration- 4.
\subsubsection*{Configuration- 5}
The computation of the reflected entropy for this configuration as depicted in \cref{RTending} is trivial since it reduces to the analysis for the reflected entropy for two adjacent subsystems in a standard $CFT_2$. Hence the reflected entropy may be obtained for this case as follows
\begin{align}\label{srRTending}	
	S^{R}(A:B) = \frac{2c}{3}\log\left[\frac{\sinh\frac{2\pi(p_3-p_2)}{\beta}\sinh\frac{2\pi(p_3-p_2)}{\beta}}{\sinh\frac{2\pi(p_3-p_1)}{\beta}\sinh\frac{2\pi(\epsilon)}{\beta}}\right].
\end{align}
\subsubsection*{Configuration-6}
In this configuration the nontrivial contribution to the reflected entropy arises from the twist correlators involving
the QES point $a_2$ on the $a$-brane and the point $p_2$ in the reservoir as depicted in \cref{braneending}. There is also an additional contribution from the Weyl factor involving the point $a_2$ while the
Weyl factors associated with the coincident points $a_1$ and $b_1$ cancels from the numerator and the denominator in the expression for the reflected entropy. Hence the R\'enyi reflected entropy for this configuration 
is given as follows
    \begin{align}\label{sr-2}
	S^{R}_{n,m}(A:B) = 2\frac{1}{1-n}\log \frac{\left< \sigma_{g_B^{-1}g_A}(a_2)\sigma_{g_A^{-1}}(a_1)\sigma_{g_A}(p_1)\sigma_{g_Bg_A^{-1}}(p_2)\sigma_{g_B^{-1}}(p_3)\sigma_{g_B}(b_1) \right>_{\mathrm{CFT}^{\bigotimes mn}}}{\left<\sigma_{g^{-1}_m}(a_1)\sigma_{g_m}(p_1)\sigma_{g_m^{-1}}(p_3)\sigma_{g_m}(b_1) \right>^n_{\mathrm{CFT}^{\bigotimes m}}}\,,
\end{align}
The correlator in \cref{sr-2}  factorizes in the large central charge limit as follows
\begin{align}\label{Rsr6}
	S^{R}_{n,m}(A:B)= 2\frac{1}{1-n}\log \frac{\Big< \sigma_{g_A^{-1}}(a_1)\sigma_{g_A}(p_1)\Big>\left<\sigma_{g_B^{-1}g_A}(a_2)\sigma_{g_Bg_A^{-1}}(p_2)\right>\left<\sigma_{g_B^{-1}}(p_3)\sigma_{g_B}(b_1) \right>}{\left(\Big<\sigma_{g^{-1}_m}(a_1)\sigma_{g_m}(p_1)\Big>\left<\sigma_{g_m^{-1}}(p_3)\sigma_{g_m}(b_1) \right>\right) ^n}\,.
\end{align}
Note that the correlators in the numerator and denominator of \cref{Rsr6} are defined in the $mn$ and $m$ replicated sheets respectively. The dominant contribution to the R\'enyi reflected entropy in this case then arises from the
correlator involving the points $a_2$ and $p_2$ in the large central charge limit as follows
\begin{align}\label{sr6}
	S^{R}_{n,m}(A:B)= 2\frac{1}{1-n}\log \left<\sigma_{g_B^{-1}g_A}(a_2)\sigma_{g_Bg_A^{-1}}(p_2)\right>_{\mathrm{CFT}^{\bigotimes mn}}\,. 
\end{align}
Finally the reflected entropy for this configuration may be obtained in the replica limit as
\begin{align}\label{srconfig6}
	S^{R}_{\text{eff}}(A:B)= \frac{2c}{3} \log\left[\frac{\beta}{\pi}\frac{ \cosh \left(\frac{2 \pi  \left(a_2-p_2\right)}{\beta }\right)-1}{\sinh\left(\frac{2 \pi  a_2}{\beta }\right)}\right]\,.
\end{align}
Now we may obtain the generalized reflected entropy utilizing the island formula in \cref{srew} as follows 
\begin{align}\label{srbraneending}
	S^{R} _{\text{gen}}(A:B)= 2\Phi_0+\frac{4 \pi  \Phi_r}{\beta } \coth \left(\frac{2 \pi a_2}{\beta }\right)+\frac{2c}{3} \log\left[\frac{\beta}{\pi}\frac{ \cosh \left(\frac{2 \pi  \left(a_2-p_2\right)}{\beta }\right)-1}{\sinh\left(\frac{2 \pi  a_2}{\beta }\right)}\right]\,.
\end{align}
\subsubsection*{Configuration-7}
\begin{figure}[H]
	\centering
	\begin{subfigure}[b]{0.45\textwidth}
		\centering
		\includegraphics[width=\textwidth]{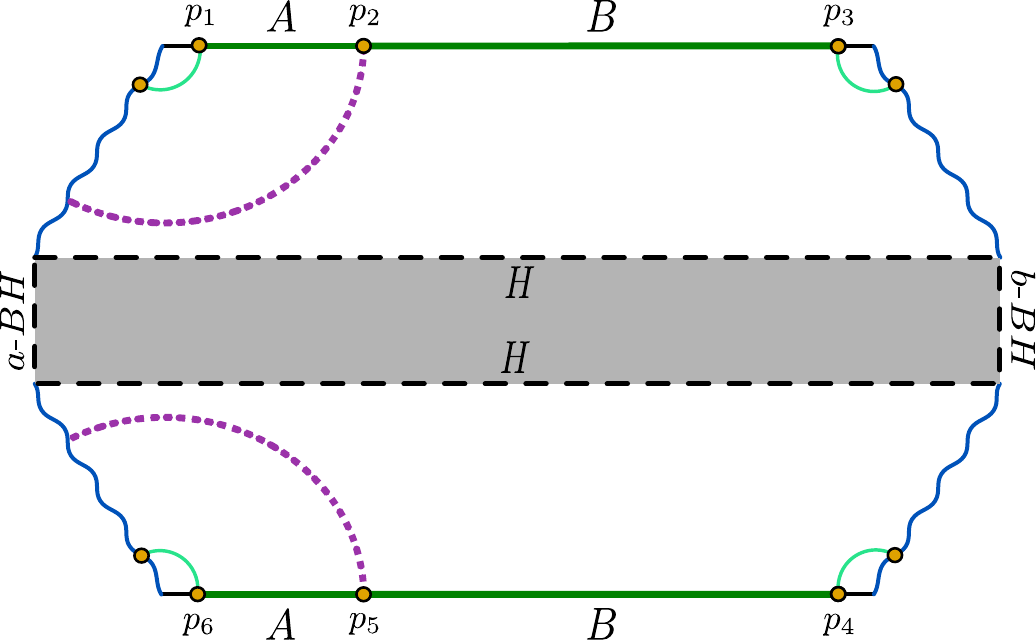}
		\caption{Configuration-6}
		\label{braneending}
	\end{subfigure}
	\hspace{.5cm}
	\begin{subfigure}[b]{0.45\textwidth}
		\centering
		\includegraphics[width=\textwidth]{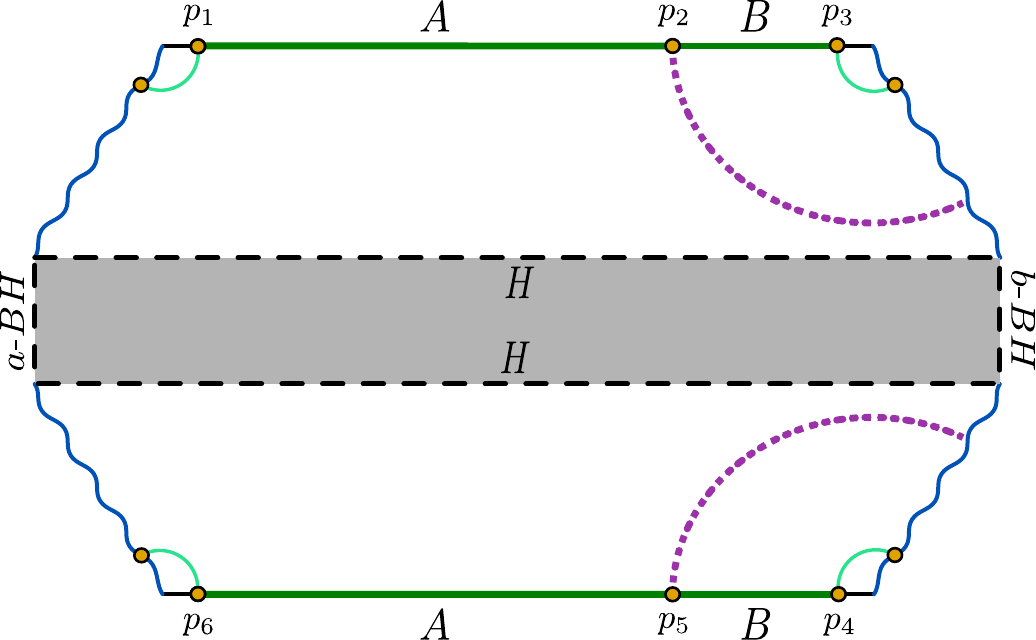}
		\caption{Configuration-7}
		\label{braneending2}
	\end{subfigure}
	\hspace{.5cm}
	\begin{subfigure}[b]{0.45\textwidth}
		\centering
		\includegraphics[width=\textwidth]{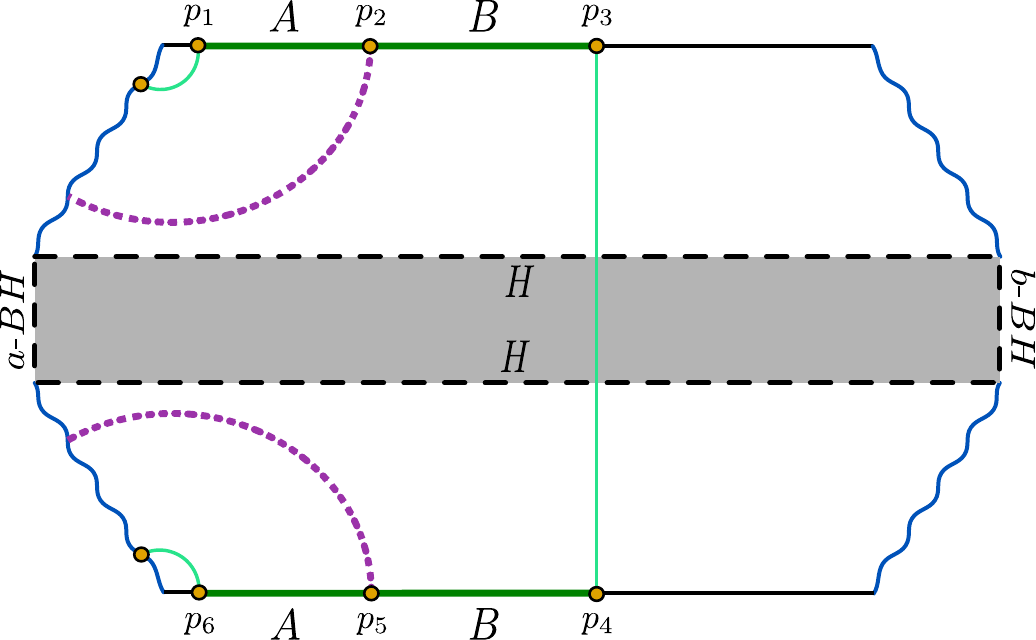}
		\caption{Configuration-8}
		\label{braneendingbulk}
	\end{subfigure}
	\hspace{.5cm}
	\begin{subfigure}[b]{0.45\textwidth}
		\centering
		\includegraphics[width=\textwidth]{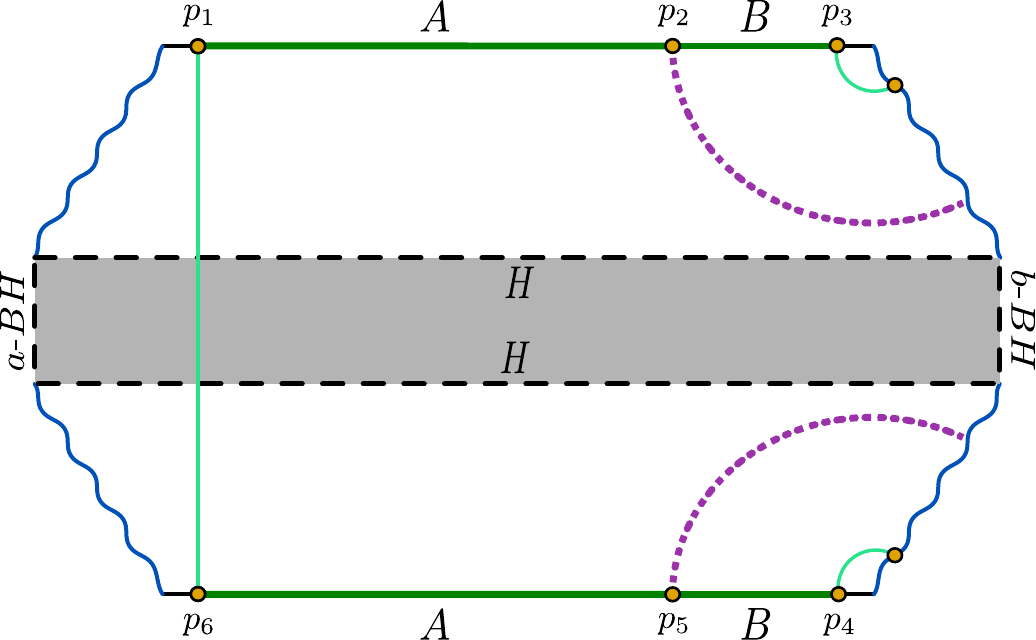}
		\caption{Configuration-9}
		\label{braneendingbulk2}
	\end{subfigure}
	\caption{ The non-trivial cross section in each of the following channels for the reflected entropy is shown with a connected phase of the entanglement island. }\label{braneendingresults}
\end{figure}

This configuration is similar to the configuration-6, however, the reflected entropy in this case receives contribution from the QES point $b_2$ located on the $b$-brane. Hence, the twist correlator for the R\'enyi reflected entropy factorizes to the following contraction in the large central large limit
\begin{align}\label{Rsr7}
	S^{R}_{n,m}(A:B)= 2\frac{1}{1-n}\log \frac{\Big< \sigma_{g_A^{-1}}(a_1)\sigma_{g_A}(p_1)\Big>\left<\sigma_{g_B^{-1}g_A}(b_2)\sigma_{g_Bg_A^{-1}}(p_2)\right>\left<\sigma_{g_B^{-1}}(p_3)\sigma_{g_B}(b_1) \right>}{\left(\Big<\sigma_{g^{-1}_m}(a_1)\sigma_{g_m}(p_1)\Big>\left<\sigma_{g_m^{-1}}(p_3)\sigma_{g_m}(b_1) \right>\right)^n}\,, 
\end{align}
 where the dominant contribution in the above equation involves twist operators located at the points $b_2$ and $p_2$ as shown in \cref{braneending2}. Thus the R\'enyi reflected entropy is given by the following equation
 \begin{align}\label{sr7}
 	S^{R}_{n,m}(A:B)= 2\frac{1}{1-n}\log \left<\sigma_{g_B^{-1}g_A}(b_2)\sigma_{g_Bg_A^{-1}}(p_2)\right>_{\mathrm{CFT}^{\bigotimes mn}}\,. 
 \end{align}
  Finally, the expression of the reflected entropy may be obtained in the replica limit as follows
\begin{align}\label{srconfig7}
	S^{R}_{\text{eff}}(A:B)= \frac{2c}{3} \log\left[\frac{\beta}{\pi}\frac{ \cosh \left(\frac{2 \pi  \left(b_2-p_2\right)}{\beta }\right)-1}{\sinh\left(\frac{2 \pi  b_2}{\beta }\right)}\right]\,.
\end{align}
For this configuration, we may now obtain the generalized reflected entropy by using the island formula in \cref{srew} as
\begin{align}\label{srbraneending2}
	S^{R} _{\text{gen}}(A:B)= 2\Phi_0+\frac{4 \pi  \Phi_r}{\beta } \coth \left(\frac{2 \pi (L-b_2)}{\beta }\right)+\frac{2c}{3} \log\left[\frac{\beta}{\pi}\frac{ \cosh \left(\frac{2 \pi  \left(b_2-p_2\right)}{\beta }\right)-1}{\sinh\left(\frac{2 \pi  b_2}{\beta }\right)}\right]\,.
\end{align}
%%%%%%%%%%%%%%%%%%%%%%%%%%%%%%%%%%%%%%%%%%%%%%%%%%%%%%%%%%%%%%%%%%%%%%%%%%%%%%%

\subsubsection*{Configuration-8 and 9}
The computation of the reflected entropy in this cases follow the analysis described for the configurations-6 and 7. 
The dominant contributions in these cases are then given by \cref{sr6,sr7} in the large central charge limit. The only difference between these two configurations arise from the location of the twist operators
at points $p_3$ and $p_6$ as depicted in \cref{braneendingbulk} for the first while the other configuration \ref{braneendingbulk2} incorporate twist operators located at the points $p_1$ and $p_4$. However, the contribution from these correlators cancel in the replica limit. Therefore the reflected entropy for the configuration- 8 is given by the \cref{srbraneending} and similarly the \cref{srbraneending2} describes the result for the configuration- 9.
%%%%%%%%%%%%%%%%%%%%%%%%%%%%%%%%%%%%%%%%%%%%%%%%%%%%%%%%%%%%%%%%%%%%%%%%%%%%%%%

\subsubsection*{Configuration-10}
Now we describe the next configuration where the reflected entropy involves contribution from the points located on both the radiation reservoirs as depicted in the \cref{bulkewcs}. In this case, the Weyl factors cancel from the numerator and denominator in the reflected entropy expression in the replica limit. Hence the 
R\'enyi reflected entropy for this configuration is given by
\begin{equation}
\begin{aligned}\label{sr-3}
	S^{R}_{n,m}(A:B) =\frac{1}{1-n}\log &\frac{\Big{<} \sigma_{g_A^{-1}}(a_1)\sigma_{g_A}(p_1)\sigma_{g_Bg_A^{-1}}(p_2)\sigma_{g_B^{-1}}(p_3)\sigma_{g_B}(b_1) }{\Big{<}\sigma_{g^{-1}_m}(a_1)\sigma_{g_m}(p_1)\sigma_{g_m^{-1}}(p_3)\sigma_{g_m}(b_1) }\\ 
	&\times\frac{\sigma_{g_A^{-1}}(a_2)\sigma_{g_A}(p_4)\sigma_{g_Bg_A^{-1}}(p_5)\sigma_{g_B^{-1}}(p_6)\sigma_{g_B}(b_2)\Big{>}_{\mathrm{CFT}^{\bigotimes mn}}}{\sigma_{g^{-1}_m}(a_2)\sigma_{g_m}(p_4)\sigma_{g_m^{-1}}(p_5)\sigma_{g_m}(b_2)\Big{>}^n_{\mathrm{CFT}^{\bigotimes m}}}\,.
\end{aligned}
\end{equation}
In the large central charge limit, the above expression factorizes to the following contractions 
\begin{equation}
\begin{aligned}\label{sr3next}
	S^{R}_{n,m}(A:B) =\frac{1}{1-n}\log &\frac{\Big{<} \sigma_{g_A^{-1}}(a_1)\sigma_{g_A}(p_1)\Big{>}\Big{<}\sigma_{g_Bg_A^{-1}}(p_2)\sigma_{g_Bg_A^{-1}}(p_5)\Big{>}\Big{<}\sigma_{g_B^{-1}}(p_3)\sigma_{g_B}(b_1)\Big{>} }{\left(\Big{<}\sigma_{g^{-1}_m}(a_1)\sigma_{g_m}(p_1)\Big{>}\Big{<}\sigma_{g_m^{-1}}(p_3)\sigma_{g_m}(b_1)\Big{>}\right)^n}\\ 
	&\times\frac{\Big{<}\sigma_{g_A^{-1}}(a_2)\sigma_{g_A}(p_4)\Big{>}\Big{<}\sigma_{g_B^{-1}}(p_6)\sigma_{g_B}(b_2)\Big{>}}{\left(\Big{<}\sigma_{g^{-1}_m}(a_2)\sigma_{g_m}(p_4)\Big{>}\Big{<}\sigma_{g_m^{-1}}(p_6)\sigma_{g_m}(b_2)\Big{>}\right)^n}\,.
\end{aligned}
\end{equation}
Now the reflected entropy in this configuration may be obtained from the dominant correlator involving the twist operators located at the points $p_2$ and $p_5$ since the other correlators in \cref{sr3next} cancel from the numerator and denominator in the replica limit. Hence, the R\'enyi reflected entropy reduced to the following expression
\begin{align}\label{dominant10}
	S^{R}_{n,m}(A:B) =\frac{1}{1-n}\log &\Big{<}\sigma_{g_Bg_A^{-1}}(p_2)\sigma_{g_Bg_A^{-1}}(p_5)\Big{>}\,.
\end{align}
 Note that the correlator involving twist operators from the TFD copies of the $CFT_2$s are also studied in the article \cite{Hartman:2013qma}. Finally, we may obtain the reflected entropy as follows 
\begin{align}\label{srbulk}
	S^{R}_{\text{eff}}(A:B) =\frac{2c}{3} \log \left[\frac{\beta}{\pi}
	\cosh \left(\frac{2 \pi t}{\beta }\right)\right]\,.
\end{align}
\subsubsection*{Configuration-11 and 12}
\begin{figure}[H]
	\centering
	\begin{subfigure}[b]{0.45\textwidth}
		\centering
		\includegraphics[width=\textwidth]{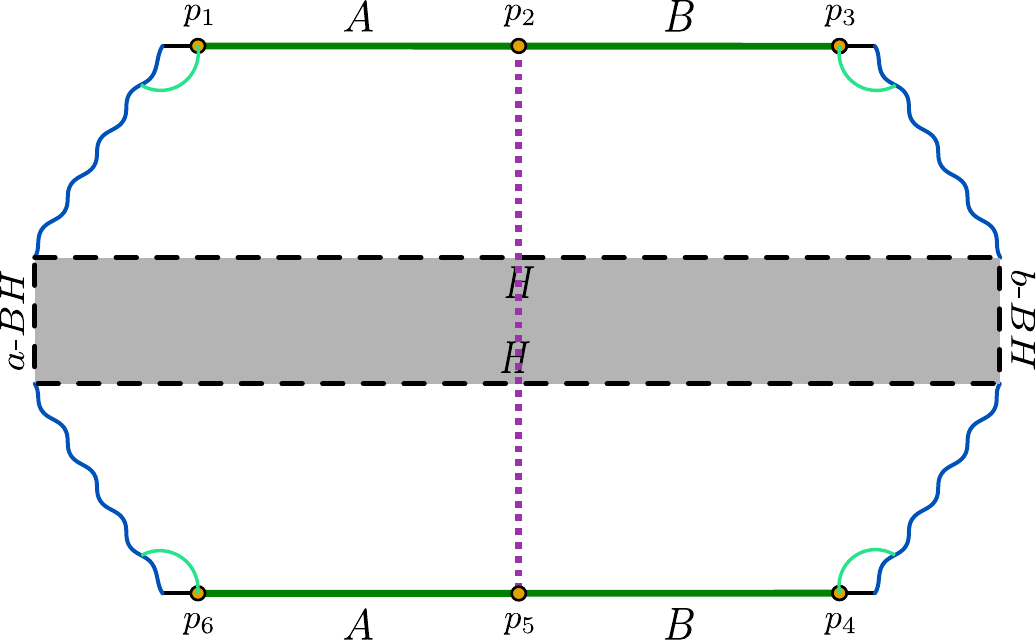}
		\caption{Configuration-10}
		\label{bulkewcs}
	\end{subfigure}
	\hspace{.5cm}
	\begin{subfigure}[b]{0.45\textwidth}
		\centering
		\includegraphics[width=\textwidth]{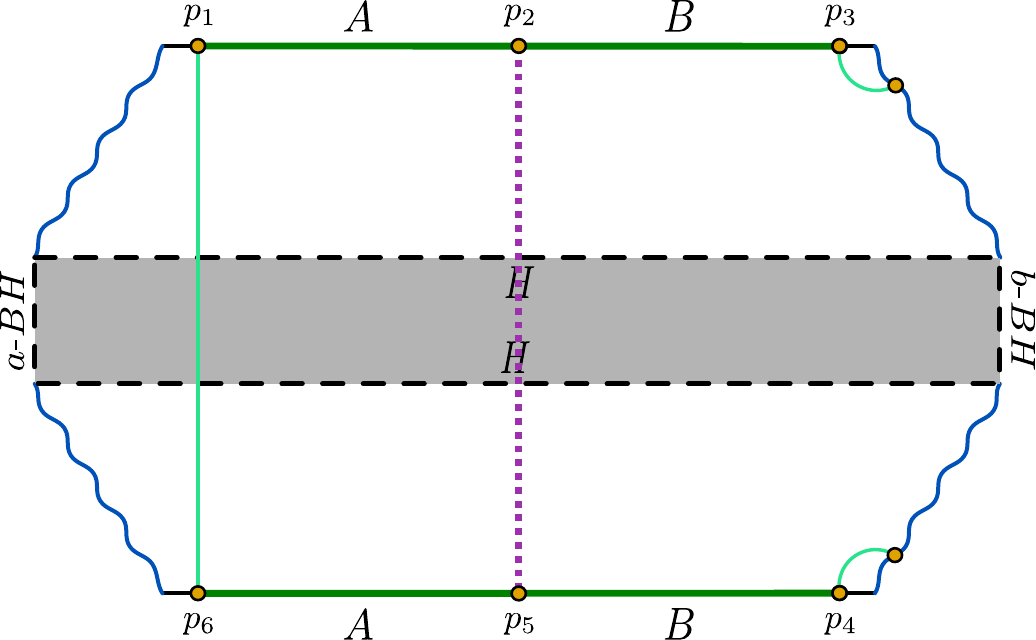}
		\caption{Configuration-11}
		\label{bulkbulk}
	\end{subfigure}
	\vspace{.5cm}
	\begin{subfigure}[b]{0.45\textwidth}
		\centering
		\includegraphics[width=\textwidth]{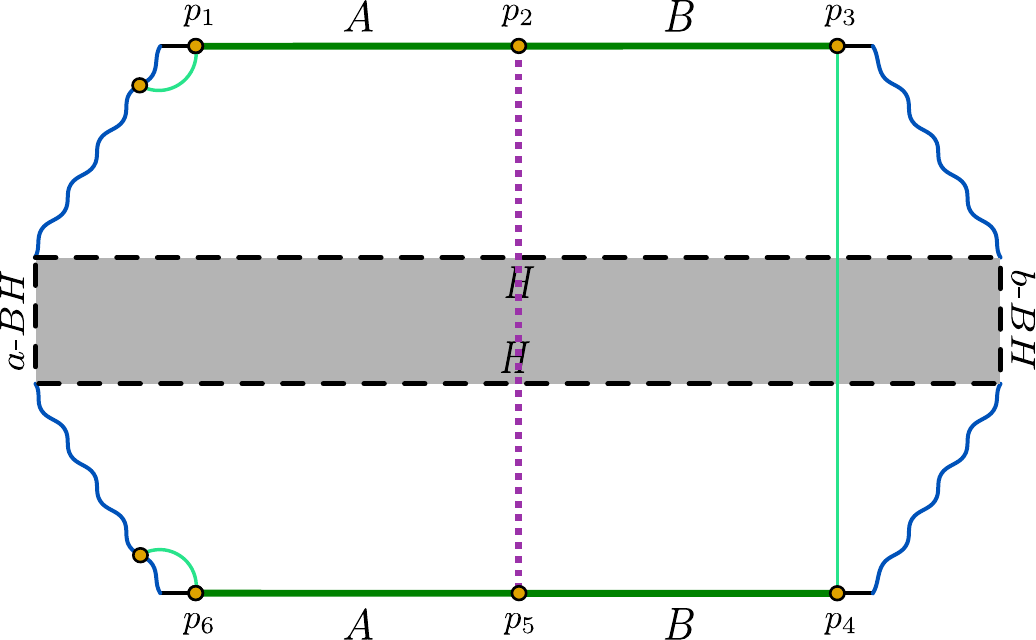}
		\caption{Configuration-12}
		\label{bulkbulk2}
	\end{subfigure}
	\caption{The corresponding figures depict the connected phase of the entanglement island with non-trivial cross section extending between the TFD copies of the radiation reservoirs. }\label{bulkresults}
\end{figure}

In these configurations, the computation of the reflected entropy for two adjacent subsystems $A$ and $B$ also follow the similar analysis described in configuration- 10, and the dominant correlators in these cases are given by the \cref{dominant10}. Therefore apart from the change in the factorization of the correlators in these configurations as depicted in \cref{bulkbulk,bulkbulk2}, the results of the reflected entropies remain identical, which is given by \cref{srbulk}. 

%%%%%%%%%%%%%%%%%%%%%%%%%%%%%%%%%%%%%%%%%%%%%%%%%%%%%%%%%%%%%%%%%%%%%%%%%%%%%%%%%

\subsubsection*{Configuration-13}
In this case, the R\'enyi reflected entropy of two adjacent subsystems $A$ and $B$ involves twist operators located on both the radiation reservoirs as depicted in \cref{bulkending} and the dominant contribution arises from the
the four point twist field correlator  in the computation for the reflected entropy. For this configuration, the R\'enyi reflected entropy may be obtained from the following expression
\begin{equation}
\begin{aligned}\label{sr-6}
	S^{R}_{n,m}(A:B) =\frac{1}{1-n}\log &\frac{\Big{<} \sigma_{g_A^{-1}}(a_1)\sigma_{g_A}(p_1)\sigma_{g_Bg_A^{-1}}(p_2)\sigma_{g_B^{-1}}(p_3)\sigma_{g_A^{-1}}(a_2)}{\Big{<}\sigma_{g^{-1}_m}(a_1)\sigma_{g_m}(p_1)\sigma_{g_m^{-1}}(p_3)}\\ 
	&\times\frac{\sigma_{g_A}(p_4)\sigma_{g_Bg_A^{-1}}(p_5)\sigma_{g_B^{-1}}(p_6)\Big{>}_{\mathrm{CFT}^{\bigotimes mn}}}{\sigma_{g^{-1}_m}(a_2)\sigma_{g_m}(p_4)\sigma_{g_m^{-1}}(p_6)\Big{>}^n_{\mathrm{CFT}^{\bigotimes m}} }\,,
\end{aligned}
\end{equation}
where the correlator does not involve any QES point on the $b$-brane. The factorization of the above correlator in the large central charge limit implies the following expression of the R\'enyi reflected entropy
\begin{equation}
\begin{aligned}\label{srbulkending13}
	S^{R}_{n,m}(A:B) =\frac{1}{1-n}\log &\frac{\Big{<} \sigma_{g_A^{-1}}(a_1)\sigma_{g_A}(p_1)\Big{>}\Big{<}\sigma_{g_Bg_A^{-1}}(p_2)\sigma_{g_B^{-1}}(p_3)\sigma_{g_Bg_A^{-1}}(p_5)\sigma_{g_B^{-1}}(p_6)\Big{>} }{\left(\Big{<}\sigma_{g^{-1}_m}(a_1)\sigma_{g_m}(p_1)\Big{>}\Big{<}\sigma_{g_m^{-1}}(p_3)\sigma_{g_m^{-1}}(p_6)\Big{>}\right)^n}\\ 
	&\times\frac{\Big{<}\sigma_{g_A^{-1}}(a_2)\sigma_{g_A}(p_4)\Big{>}}{\left(\Big{<}\sigma_{g^{-1}_m}(a_2)\sigma_{g_m}(p_4)\Big{>}\right)^n}\,.
\end{aligned}
\end{equation}
In the above equation, the dominant correlator involves twist operators located at the points $p_2$, $p_3$, $p_4$ and $p_5$ while the other contributions cancel from the nominator and denominator in \cref{srbulkending13}. Hence the R\'enyi reflected entropy reduced to the following form
\begin{equation}
	\begin{aligned}\label{srbulkendingdominant13}
		S^{R}_{n,m}(A:B) =\frac{1}{1-n}\log &\frac{\Big{<}\sigma_{g_Bg_A^{-1}}(p_2)\sigma_{g_B^{-1}}(p_3)\sigma_{g_Bg_A^{-1}}(p_5)\sigma_{g_B^{-1}}(p_6)\Big{>} }{\left(\sigma_{g_m^{-1}}(p_3)\sigma_{g_m^{-1}}(p_6)\Big{>}\right)^n}\,.
	\end{aligned}
\end{equation}
Now, we utilize a technique termed as inverse doubling trick described in \cite{Shao:2022gpg} to compute the reflected entropy from \cref{srbulkendingdominant13}. In this trick, the four point twist correlator is reduced to the two point function in the $BCFT_2$, and we can obtain the expression of this two point twist correlator in the OPE channel by following the analysis described in \cite{Shao:2022gpg}. Finally, the reflected entropy in the replica limit may be computed as
\begin{align}\label{sr13}
	S^{R}_{\text{eff}}(A:B)= \frac{2c}{3} \log \left[\frac{ (r_3-r_2)~ \text{sech}\left(\frac{2 \pi  t}{\beta }\right) \sqrt{r_3^2+2 r_3 r_2 \cosh \left(\frac{4 \pi  t}{\beta }\right)+r_2^2}}{r_3 \epsilon }\right]\,,
\end{align}
where $r_3$ and $r_2$ are the location of the points $p_3$ and $p_2$ respectively. 
\subsubsection*{Configuration-14}
This case is similar to the configuration-13 as shown in \cref{bulkending2}, and the twist field correlator for the R\'enyi reflected entropy factorizes in the following contraction in the large central charge limit
\begin{equation}
\begin{aligned}\label{srbulkending14}
	S^{R}_{n,m}(A:B) =\frac{1}{1-n}\log &\frac{\Big{<} \sigma_{g_A^{-1}}(b_1)\sigma_{g_B^{-1}}(p_3)\Big{>}\Big{<}\sigma_{g_A}(p_1)\sigma_{g_Bg_A^{-1}}(p_2)\sigma_{g_Bg_A^{-1}}(p_4)\sigma_{g_B^{-1}}(p_5)\Big{>} }{\left(\Big{<}\sigma_{g^{-1}_m}(b_1)\sigma_{g_m}(p_3)\Big{>}\Big{<}\sigma_{g_m^{-1}}(p_1)\sigma_{g_m^{-1}}(p_5)\Big{>}\right)^n}\\ 
	&\times\frac{\Big{<}\sigma_{g_A^{-1}}(b_2)\sigma_{g_A}(p_6)\Big{>}}{\left(\Big{<}\sigma_{g^{-1}_m}(b_2)\sigma_{g_m}(p_6)\Big{>}\right)^n}\,.
\end{aligned}
\end{equation}
The dominant correlator in \cref{srbulkending14} involves twist operators located at the points $p_1$, $p_2$, $p_4$ and $p_5$ while the other contributions cancel from the nominator and denominator in the replica limit. Hence the R\'enyi reflected entropy may be reduced as
\begin{equation}
	\begin{aligned}\label{srbulkendingdominant14}
		S^{R}_{n,m}(A:B) =\frac{1}{1-n}\log &\frac{\Big{<}\sigma_{g_A}(p_1)\sigma_{g_Bg_A^{-1}}(p_2)\sigma_{g_Bg_A^{-1}}(p_4)\sigma_{g_B^{-1}}(p_5)\Big{>} }{\left(\Big{<}\sigma_{g_m^{-1}}(p_1)\sigma_{g_m^{-1}}(p_5)\Big{>}\right)^n}\,.
	\end{aligned}
\end{equation}
On utilization of the techniques discussed in \cite{Dutta:2019gen,Shao:2022gpg}, the reflected entropy may be obtained as follows
\begin{align}\label{sr14}
	S^{R}_{\text{eff}}(A:B)= \frac{2c}{3} \log\left[\frac{(r_1-r_2)~ \text{sech}\left(\frac{2 \pi  t}{\beta }\right) \sqrt{r_1^2+2 r_1 r_2 \cosh \left(\frac{4 \pi  t}{\beta }\right)+r_2^2}}{r_1 \epsilon }\right] \,,
\end{align}
where $r_1$ and $r_2$ are location of the points $p_1$ and $p_2$ respectively.
\subsubsection*{Configuration-15 and 16}
\begin{figure}[H]
	\centering
	\begin{subfigure}[b]{0.45\textwidth}
		\centering
		\includegraphics[width=\textwidth]{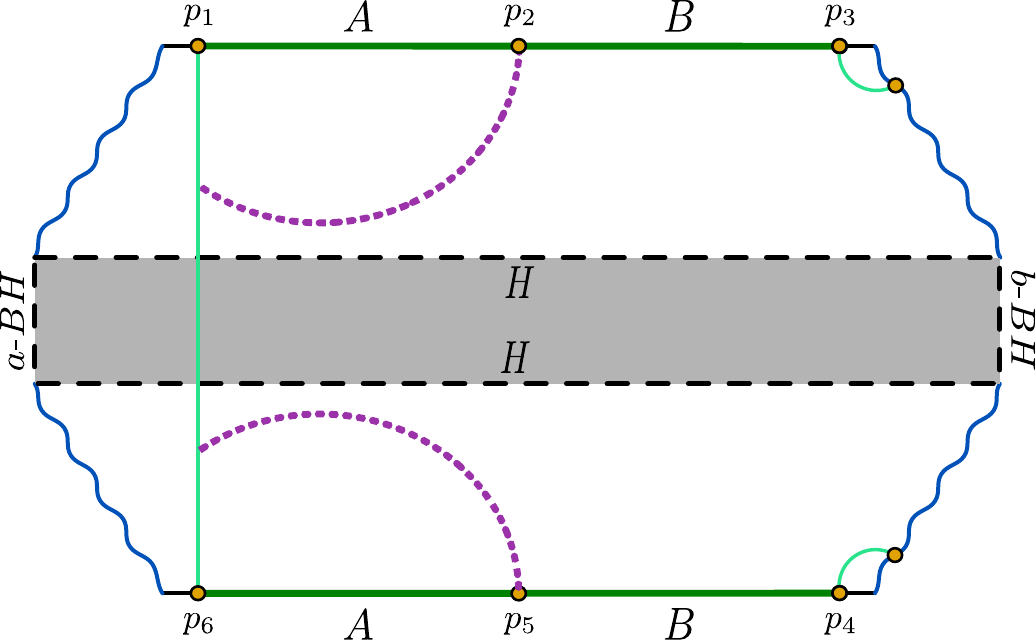}
		\caption{Configuration-13}
		\label{bulkending}
	\end{subfigure}
	\hspace{.5cm}
	\begin{subfigure}[b]{0.45\textwidth}
		\centering
		\includegraphics[width=\textwidth]{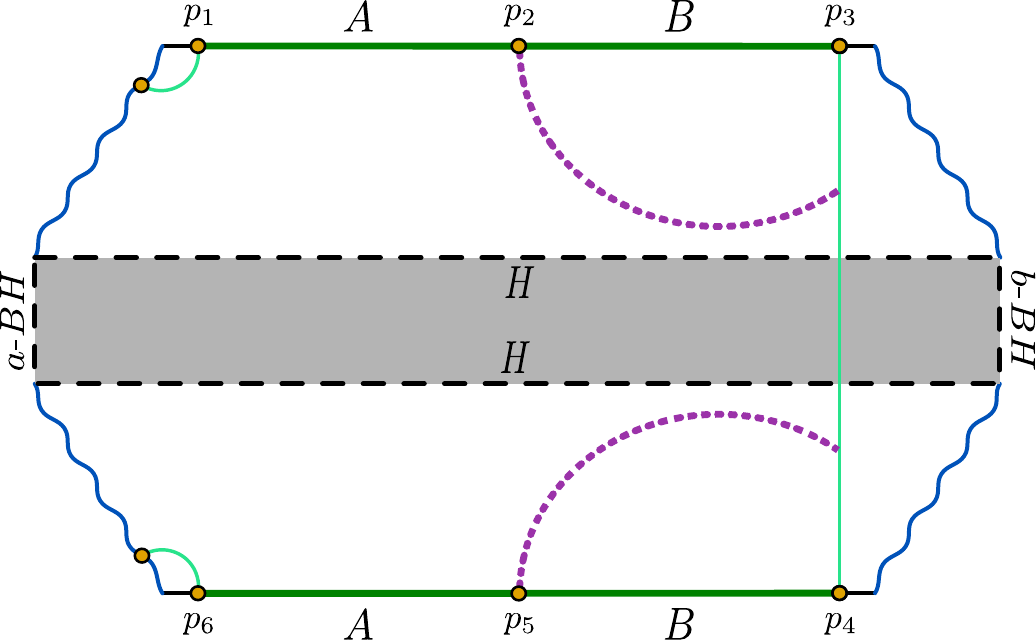}
		\caption{Configuration-14}
		\label{bulkending2}
	\end{subfigure}
	\hspace{.5cm}
	\begin{subfigure}[b]{0.45\textwidth}
		\centering
		\includegraphics[width=\textwidth]{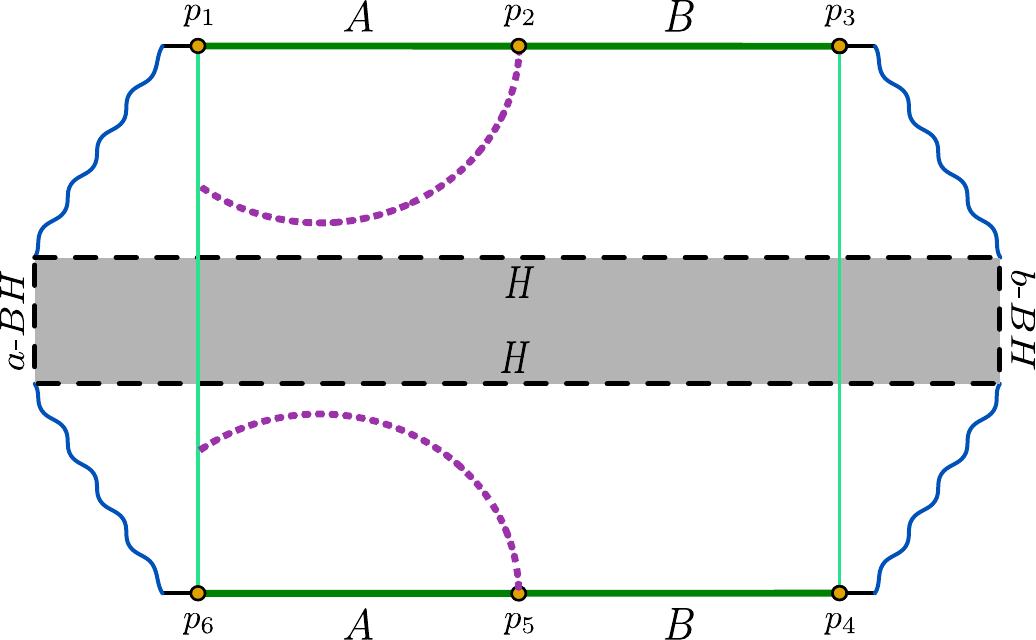}
		\caption{Configuration-15}
		\label{bulkendingbulk}
	\end{subfigure}
	\hspace{.5cm}
	\begin{subfigure}[b]{0.45\textwidth}
		\centering
		\includegraphics[width=\textwidth]{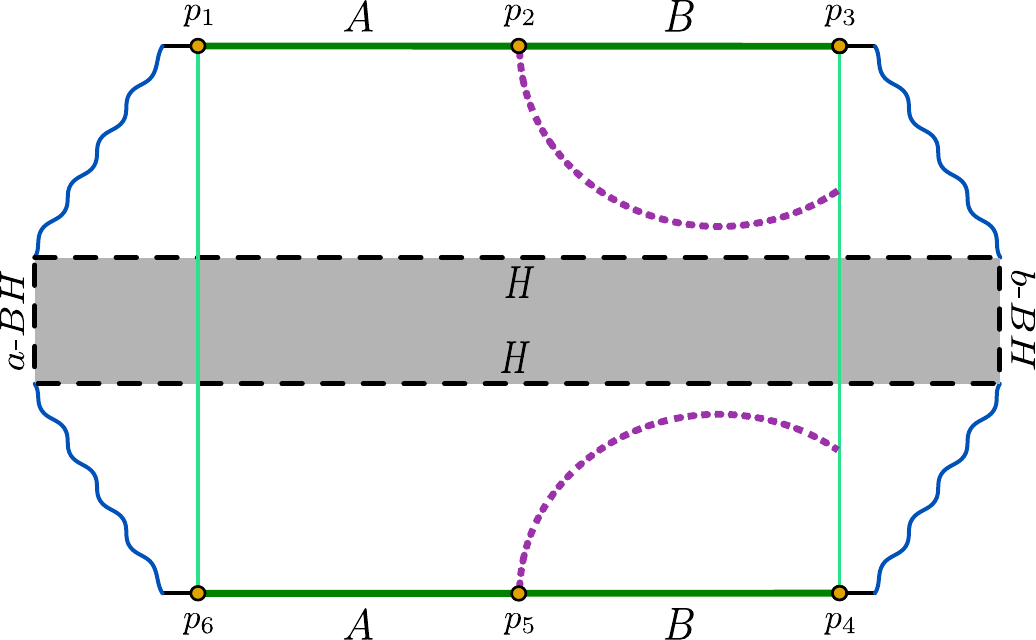}
		\caption{Configuration-16}
		\label{bulkendingbulk2}
	\end{subfigure}
	\caption{Schematics depicts the non-trivial cross section to the reflected entropy of the two adjacent subsystems $A$ and $B$ where the corresponding cross section ended of the Hartman-Maldacena surface.
	}\label{bulkendingresults}
\end{figure}

In these configurations, the computation of the reflected entropy for two adjacent subsystems $A$ and $B$ follows a similar analysis as described for the configurations-13 and 14. Here the dominant correlators for the configurations-15 and 16 are given by the \cref{srbulkendingdominant13,srbulkendingdominant14} respectively in the large central charge limit. The only difference in these configurations arise from the twist operators located at the points $p_1$ and $p_4$ as depicted in \cref{bulkendingbulk} for the first, while the other configuration \cref{bulkendingbulk2} involves twist operators located at the points $p_3$ and $p_6$. However the contributions from these two point twist correlators cancel from the numerator and denominator in the replica limit. Hence the expression for the reflected entropies in the configurations-15 and 16 are given by \cref{sr13,sr14} respectively.

%%%%%%%%%%%%%%%%%%%%%%%%%%%%%%%%%%%%%%%%%%%%%%%%%%%%%%%%%%%%%%%%%%%%%%%%%%%%%%%%%%%

\subsubsection{Disjoint subsystems}
In this subsection, we discuss the computation of the reflected entropy for two disjoint subsystems $A\equiv[p_1,p_2]\cup[p_5,p_6]$ and $B\equiv[p_3,p_4]\cup[p_7,p_8]$ located on both the radiation reservoirs described by $CFT_2$s in the communicating black hole setup as shown in \cref{penrose2}. In this context we obtain the reflected entropy for different configurations described by relative subsystem sizes utilizing the techniques developed in \cite{Dutta:2019gen}.
\subsubsection*{Configuration-1}
We first discuss the configuration which involves the twist operators located on the radiation reservoirs and the $a$, $b$-branes as depicted in \cref{distaka}. Since this configuration is symmetric on the reservoir copy of the double TFD states 
it suffices to compute the reflected entropy of the two disjoint subsystems in one of the reservoirs. Hence the R\'enyi reflected entropy may be obtained using the following twist correlator 
\begin{align}\label{srtaka}
	S^{R}_{n,m}(A:B) = 2\frac{1}{1-n}\log \frac{\left< \sigma_{g^{-1}_B}(a_1)\sigma_{g_A}(p_1)\sigma_{g_A^{-1}}(p_2)\sigma_{g_B}(p_3)\sigma_{g_B^{-1}}(p_4)\sigma_{g_B}(b_1) \right>_{\mathrm{CFT}^{\bigotimes mn}}}{\left<\sigma_{g^{-1}_m}(a_1)\sigma_{g_m}(p_1)\sigma_{g^{-1}_m}(p_2)\sigma_{g_m}(p_3)\sigma_{g_m^{-1}}(p_4)\sigma_{g_m}(b_1) \right>^n_{\mathrm{CFT}^{\bigotimes m}}}\,,	
\end{align}
where the factor 2 in the above equation incorporates the contribution to the reflected entropy from the other copy of the radiation reservoir. In the large central charge limit, the correlators in \cref{srtaka} factorize to the respective contractions
as follows 
\begin{equation}
\begin{aligned}	
	S^{R}_{n,m}(A:B) = 2\frac{1}{1-n}\log&\frac{\left<\sigma_{g^{-1}_B}(a_1)\sigma_{g_A}(p_1)\sigma_{g_A^{-1}}(p_2)\sigma_{g_B}(p_3)\right>_{\mathrm{CFT}^{\bigotimes mn}}}{\left(\left<\sigma_{g^{-1}_m}(a_1)\sigma_{g_m}(p_1)\sigma_{g^{-1}_m}(p_2)\sigma_{g_m}(p_3)\right> _{\mathrm{CFT}^{\bigotimes m}}\right)^n}\\
		&\frac{\left<\sigma_{g_B^{-1}}(p_4)\sigma_{g_B}(b_1)\right>_{\mathrm{CFT}^{\bigotimes mn}}}{\left(\left<\sigma_{g_m^{-1}}(p_4)\sigma_{g_m}(b_1) \right> _{\mathrm{CFT}^{\bigotimes m}}\right)^n}\,.
\end{aligned}
\end{equation}
Note that the two point function in the above equation cancel from the numerator and denominator in the replica limit. Thus the dominant correlator to the R\'enyi reflected entropy in this case may be expressed as
\begin{equation}\label{dis2dominant}
	\begin{aligned}	
		S^{R}_{n,m}(A:B) = 2\frac{1}{1-n}\log&\frac{\left<\sigma_{g^{-1}_B}(a_1)\sigma_{g_A}(p_1)\sigma_{g_A^{-1}}(p_2)\sigma_{g_B}(p_3)\right>_{\mathrm{CFT}^{\bigotimes mn}}}{\left<\sigma_{g^{-1}_m}(a_1)\sigma_{g_m}(p_1)\sigma_{g^{-1}_m}(p_2)\sigma_{g_m}(p_3)\right>_{\mathrm{CFT}^{\bigotimes m}}}\,.
	\end{aligned}
\end{equation}
Finally, the reflected entropy for the two disjoint subsystems for this configuration may be obtained as follows \cite{Dutta:2019gen,Chandrasekaran:2020qtn} 
\begin{equation}\label{srdistaka}
	S^{R}_{\text{eff}}(A:B)=\frac{2c}{3}\log\left(\frac{1+\sqrt{x}}{1-\sqrt{x}}\right)\,, \quad\quad x=\frac{\sinh{\frac{2 \pi (p_2-p_1)}{\beta}}\sinh{\frac{2 \pi (a_1+p_1)}{\beta}}}{\sinh{\frac{2 \pi (p_3-p_1)}{\beta}}\sinh{\frac{2 \pi (a_1+p_2)}{\beta}}}\,,
\end{equation} 
%%%%%%%%%%%%%%%%%%%%%%%%%%%%%%%%%%%%%%%%%%%%%%%%%%%%%%%%%%%%%%%%%%%%%%%%%%%%%%%%%%%%

\subsubsection*{Configuration-2}
This configuration \cref{distaka} is similar to the previous case and the computation of the reflected entropy for the two disjoint subsystems in this case follow the same analysis. However the twist field correlator in \cref{srtaka} factorizes to the respective contraction in the large central charge limit 
\begin{equation}
\begin{aligned}\label{srtak2}	
	S^{R}_{n,m}(A:B) = 2\frac{1}{1-n}\log&\frac{\left< \sigma_{g^{-1}_A}(a_1)\sigma_{g_A}(p_1)\right>_{\mathrm{CFT}^{\bigotimes mn}}}{\left<\sigma_{g^{-1}_m}(a_1)\sigma_{g_m}(p_1)\right>_{\mathrm{CFT}^{\bigotimes m}}}\\
	&\frac{\left<\sigma_{g_A^{-1}}(p_2)\sigma_{g_B}(p_3)\sigma_{g_B^{-1}}(p_4)\sigma_{g_A}(b_1)\right>_{\mathrm{CFT}^{\bigotimes mn}}}{\left(\left<\sigma_{g^{-1}_m}(p_2)\sigma_{g_m}(p_3)\sigma_{g_m^{-1}}(p_4)\sigma_{g_m}(b_1) \right>_{\mathrm{CFT}^{\bigotimes m}}\right)^n} \,.
\end{aligned}
\end{equation}
From the above equation, the dominant contribution to the reflected entropy arises from the four point twist correlator. Therefore we may obtain the reflected entropy for this configuration as \cite{Dutta:2019gen,Chandrasekaran:2020qtn}
\begin{equation}\label{srdistaka2}
	S^{R}_{\text{eff}}(A:B)=\frac{2c}{3}\log\left(\frac{1+\sqrt{x}}{1-\sqrt{x}}\right)\,, \quad\quad x=\frac{\sinh{\frac{2 \pi (p_4-p_3)}{\beta}}\sinh{\frac{2 \pi (b_1+p_2)}{\beta}}}{\sinh{\frac{2 \pi (p_3-p_2)}{\beta}}\sinh{\frac{2 \pi (b_1+p_4)}{\beta}}}\,,
\end{equation} 
%%%%%%%%%%%%%%%%%%%%%%%%%%%%%%%%%%%%%%%%%%%%%%%%%%%%%%%%%%%%%%%%%%%%%%%%%%%%%%%%%%%

\subsubsection*{Configuration- 3 and 4}
It may be observed from \cref{distakabulk} that this case is similar to the configuration-3 and the dominant contribution to the reflected entropy is given by \cref{dis2dominant}. The only difference in this case arises from the two point correlator involving the twist operators located on the points $p_4$ and $p_8$ which is obtained after the factorization in the large central charge limit. However this does not contribute to the reflected entropy in the replica limit as it cancels out from the numerator and the denominator as earlier. Hence the reflected entropy in this case is given by \cref{srdistaka}.

For the configuration-4 as depicted in \cref{takabulk2}, we follow arguments similar to those described above for the
configuration-3 and obtain the reflected entropy from \cref{srdistaka2}.
%%%%%%%%%%%%%%%%%%%%%%%%%%%%%%%%%%%%%%%%%%%%%%%%%%%%%%%%%%%%%%%%%%%%%%%%%%%%%%%%%%%%

\subsubsection*{Configuration- 5}
In this case, the reflected entropy for the two disjoint subsystems may be obtained by utilizing the procedure described in \cite{Dutta:2019gen}. Since this case involves four point twist correlator as the dominant contribution to the reflected entropy this may be given as

\begin{equation}\label{srdisRTending}
	S^{R}_{\text{eff}}(A:B)=\frac{2c}{3}\log\left(\frac{1+\sqrt{x}}{1-\sqrt{x}}\right)\,, \quad\quad x=\frac{\sinh{\frac{2 \pi (p_2-p_1)}{\beta}}\sinh{\frac{2 \pi (p_4-p_3)}{\beta}}}{\sinh{\frac{2 \pi (p_3-p_2)}{\beta}}\sinh{\frac{2 \pi (p_4-p_2)}{\beta}}}\,,
\end{equation} 
%%%%%%%%%%%%%%%%%%%%%%%%%%%%%%%%%%%%%%%%%%%%%%%%%%%%%%%%%%%%%%%%%%%%%%%%%%%%%%%%%%

\subsubsection*{Configuration-6}
In this case, the computation of the reflected entropy involves a nontrivial contribution from the twist operators located at the QES point $a_2$ on the $a$-brane and the points $p_2$ and $p_3$ in the radiation reservoir as depicted in \cref{disbraneending}. Here the reflected entropy also incorporates an additional contribution from the Weyl factor associated with the QES point $a_2$. Thus the R\'enyi reflected entropy in this case may be obtained as
\begin{align}\label{dissrbraneending}
	S^{R}_{n,m}(A:B) = 2\frac{1}{1-n}\log \frac{\left< \sigma_{g_B^{-1}g_A}(a_2)\sigma_{g_A^{-1}}(a_1)\sigma_{g_A}(p_1)\sigma_{g_A^{-1}}(p_2)\sigma_{g_B}(p_3)\sigma_{g_B^{-1}}(p_4)\sigma_{g_B}(b_1) \right>_{\mathrm{CFT}^{\bigotimes mn}}}{\left(\left<\sigma_{g^{-1}_m}(a_1)\sigma_{g_m}(p_1)\sigma_{g_m^{-1}}(p_2)\sigma_{g_m}(p_3)\sigma_{g_m^{-1}}(p_4)\sigma_{g_m}(b_1) \right>_{\mathrm{CFT}^{\bigotimes m}}\right)^n}\,,
\end{align}
where the factor 2 involves the contribution to the reflected entropy from the other copy of the radiation reservoir. In the large central charge limit, the correlator in \cref{dissrbraneending} factorizes to the following contraction
\begin{align}\label{dissrbraneendingnext}
	S^{R}_{n,m}(A:B)= 2\frac{1}{1-n}\log \frac{\Big< \sigma_{g_A^{-1}}(a_1)\sigma_{g_A}(p_1)\Big>\left<\sigma_{g_B^{-1}g_A}(a_2)\sigma_{g_A^{-1}}(p_2)\sigma_{g_B}(p_3)\right>\left<\sigma_{g_B^{-1}}(p_4)\sigma_{g_B}(b_1)\right>}{\left(\Big<\sigma_{g^{-1}_m}(a_1)\sigma_{g_m}(p_1)\Big>\left<\sigma_{g_m^{-1}}(p_2)\sigma_{g_m}(p_3)\right>\left<\sigma_{g_m^{-1}}(p_4)\sigma_{g_m}(b_1) \right>\right)^n}\,. 
\end{align}
From the above equation, the dominant correlator for this configuration arises from the three point twist correlator while the other contributions to the reflected entropy cancel from the numerator and denominator in the replica limit. Hence the R\'enyi reflected entropy may be expressed as follows
\begin{align}\label{dissrdominant6}
	S^{R}_{n,m}(A:B)= 2\frac{1}{1-n}\log \frac{\left<\sigma_{g_B^{-1}g_A}(a_2)\sigma_{g_A^{-1}}(p_2)\sigma_{g_B}(p_3)\right>}{\left(\left<\sigma_{g_m^{-1}}(p_2)\sigma_{g_m}(p_3)\right>\right)^n}\,.
\end{align}
Finally in the replica limit, the reflected entropy of the two disjoint subsystems for this case may be given by
\begin{align}\label{srconfig6dis}
	S^{R}_{\text{eff}}(A:B)= \frac{2c}{3} \log\left[\frac{\beta \left(\cosh \left(\frac{2 \pi  (R-a_2)}{\beta 
		}\right)-1\right)}{ \pi  r \sinh\left(\frac{2 \pi  a_2}{\beta}\right)}\right]\,.
\end{align}
where $R$ and $r$ are related to the points $p_2$ and $p_3$ as $R=(p_3+p_2)/2$ and $r=(p_3-p_2)/2$. Now we may obtain the generalized reflected entropy using the island formula described in \cref{srew} as
\begin{align}\label{srdisbraneending}
	S_{\text{gen}}^{R}(A:B)= 2\Phi_0+\frac{4 \pi  \Phi_r}{\beta } \coth \left(\frac{2 \pi a_2}{\beta }\right)+\frac{2c}{3} \log\left[\frac{\beta \left(\cosh \left(\frac{2 \pi  (R-a_2)}{\beta 
		}\right)-1\right)}{ \pi  r \sinh\left(\frac{2 \pi  a_2}{\beta}\right)}\right]\,.
\end{align}
\begin{figure}[H]
	\centering
	\begin{subfigure}[b]{0.45\textwidth}
		\centering
		\includegraphics[width=\textwidth]{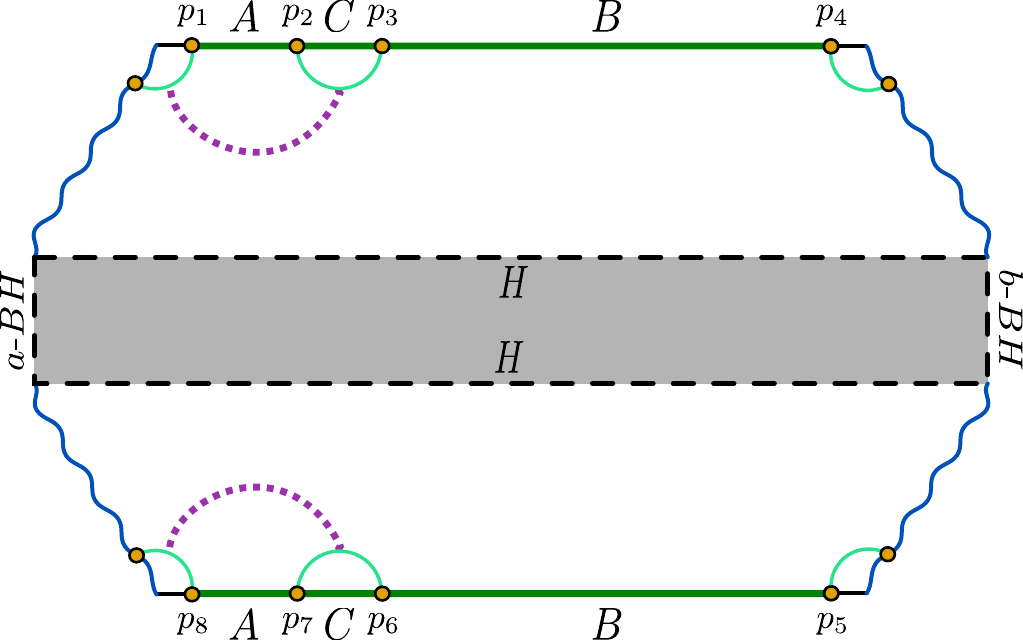}
		\caption{Configuration-1}
		\label{distaka}
	\end{subfigure}
	\hspace{.5cm}
	\begin{subfigure}[b]{0.45\textwidth}
		\centering
		\includegraphics[width=\textwidth]{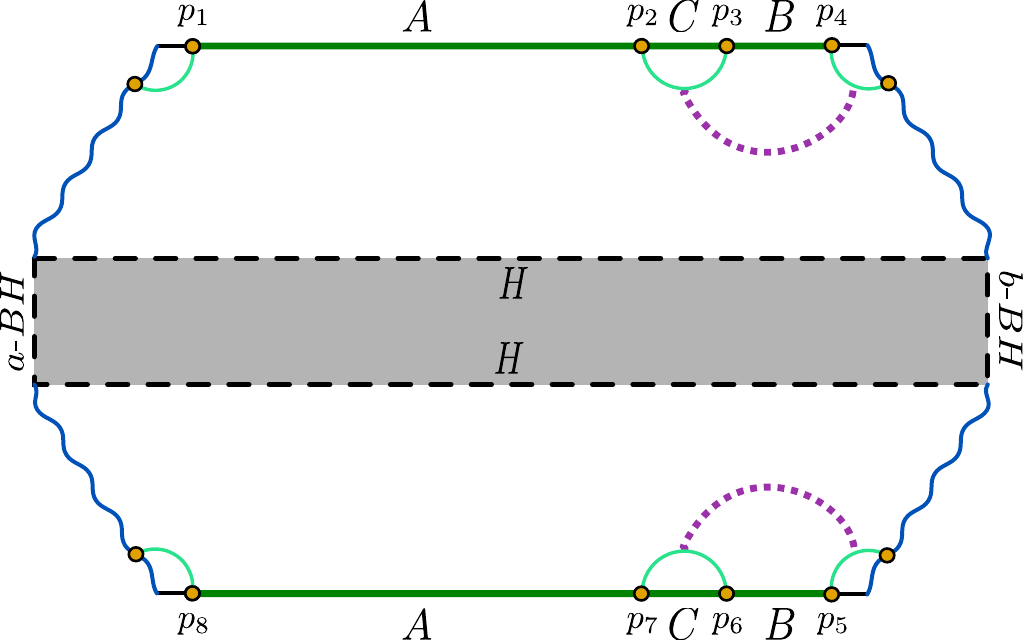}
		\caption{Configuration-2}
		\label{distaka2}
	\end{subfigure}
	\hspace{.5cm}
	\begin{subfigure}[b]{0.45\textwidth}
		\centering
		\includegraphics[width=\textwidth]{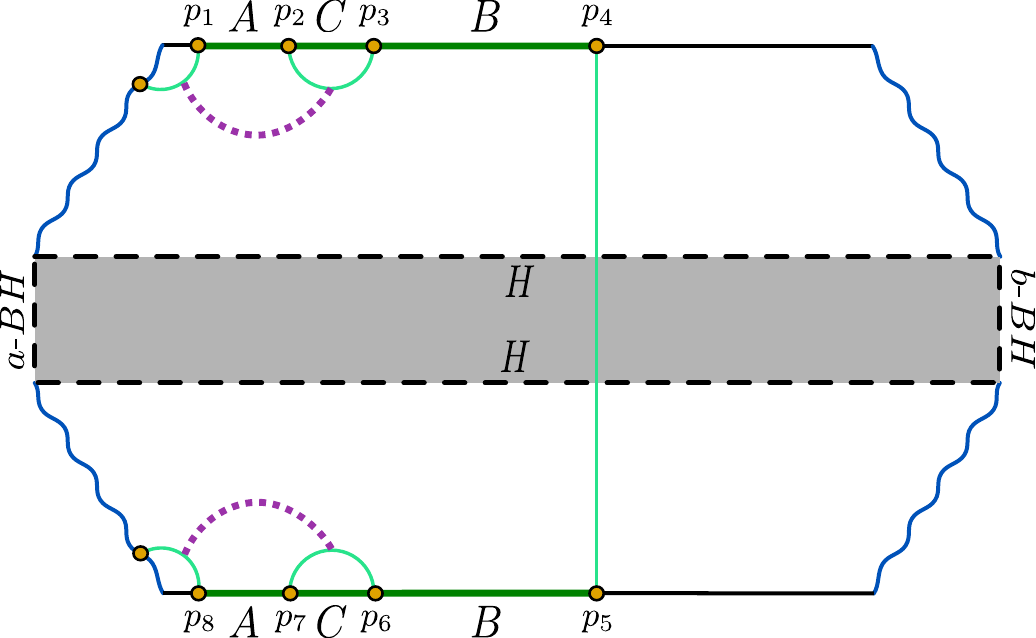}
		\caption{Configuration-3}
		\label{distakabulk}
	\end{subfigure}
	\hspace{.5cm}
	\begin{subfigure}[b]{0.45\textwidth}
		\centering
		\includegraphics[width=\textwidth]{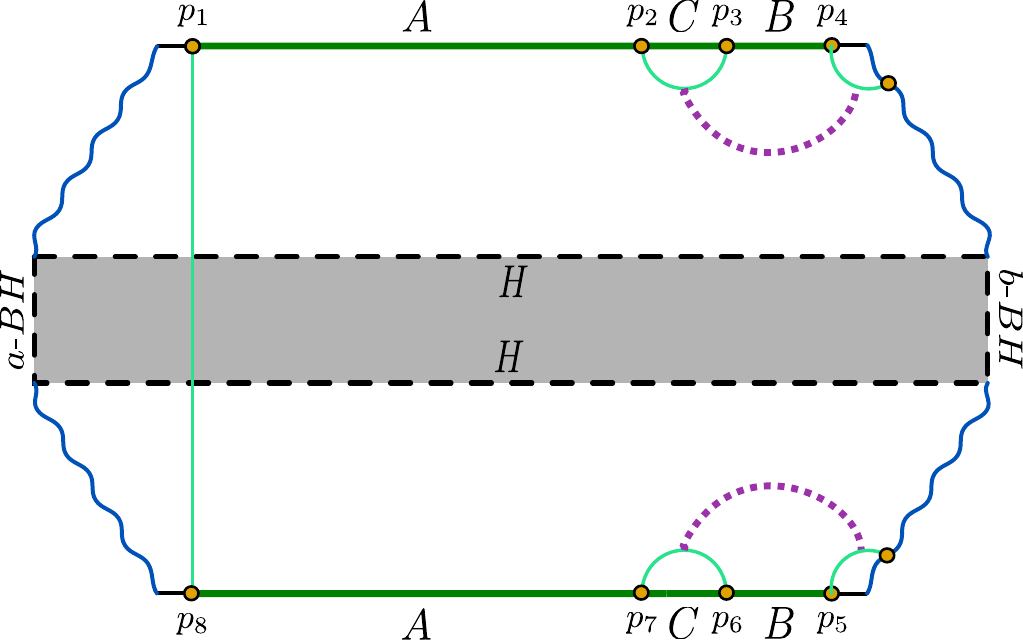}
		\caption{Configuration-4}
		\label{distakabulk2}
	\end{subfigure}
	\vspace{.5cm}
	\begin{subfigure}[b]{0.45\textwidth}
		\centering
		\includegraphics[width=\textwidth]{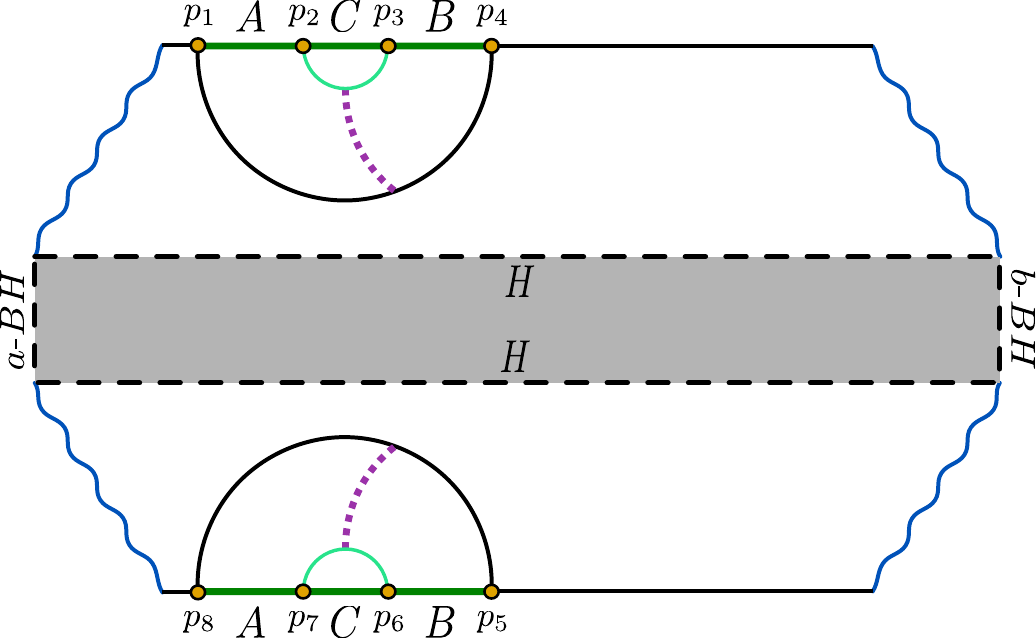}
		\caption{Configuration-5}
		\label{disrtending}
	\end{subfigure}
	\caption{Schematics depicts all the possible contributions to the reflected entropy for the case of two disjoint subsystems $A$ and $B$ where the  contraction of the correlator leads to the similar result. }\label{distakaresults}
\end{figure}

%%%%%%%%%%%%%%%%%%%%%%%%%%%%%%%%%%%%%%%%%%%%%%%%%%%%%%%%%%%%%%%%%%%%%%%%%%%%%%%%%%

\subsubsection*{Configuration-7}
\begin{figure}[H]
	\centering
	\begin{subfigure}[b]{0.45\textwidth}
		\centering
		\includegraphics[width=\textwidth]{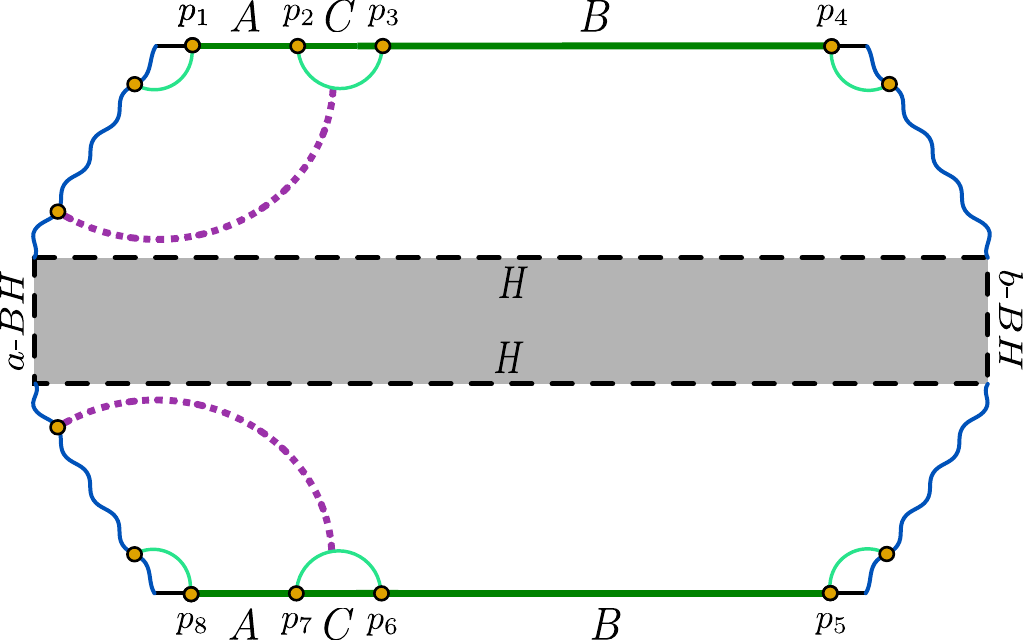}
		\caption{Configuration-6}
		\label{disbraneending}
	\end{subfigure}
	\hspace{.5cm}
	\begin{subfigure}[b]{0.45\textwidth}
		\centering
		\includegraphics[width=\textwidth]{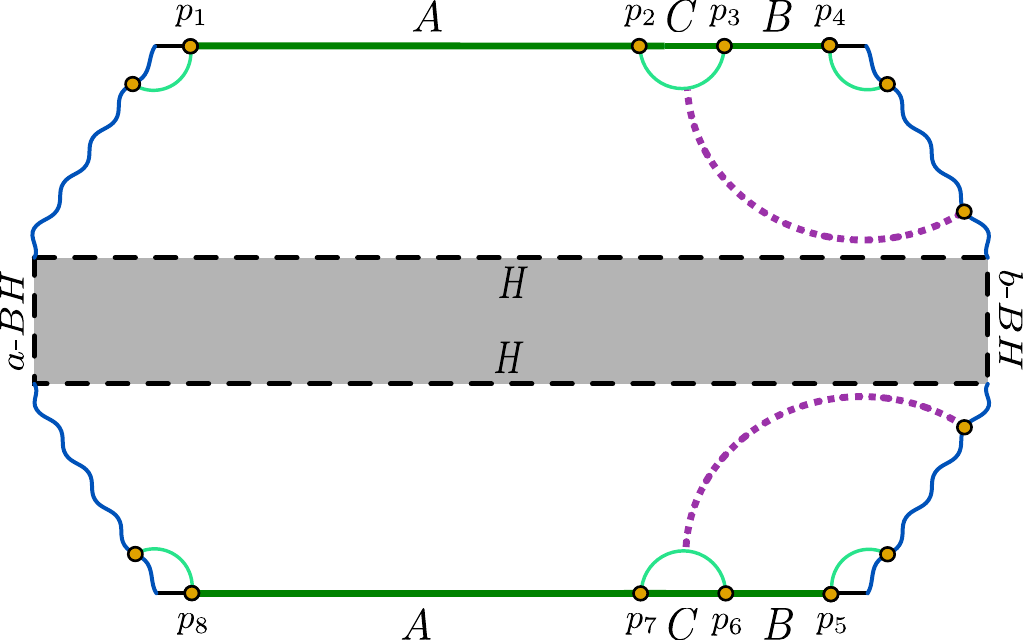}
		\caption{Configuration-7}
		\label{disbraneending2}
	\end{subfigure}
	\hspace{.5cm}
	\begin{subfigure}[b]{0.45\textwidth}
		\centering
		\includegraphics[width=\textwidth]{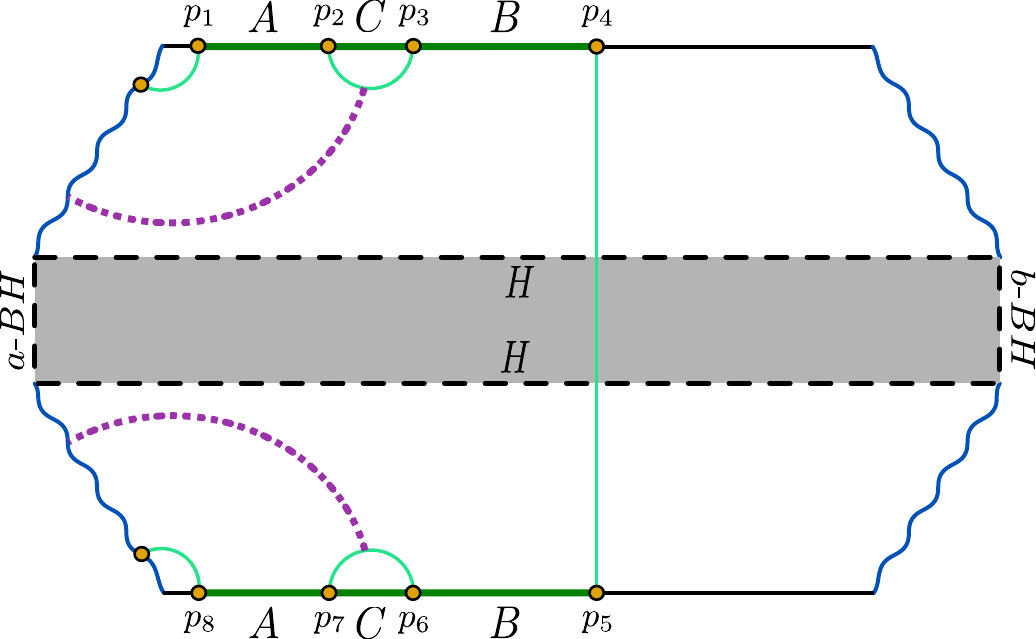}
		\caption{Configuration-8}
		\label{disbraneendingbulk}
	\end{subfigure}
	\hspace{.5cm}
	\begin{subfigure}[b]{0.45\textwidth}
		\centering
		\includegraphics[width=\textwidth]{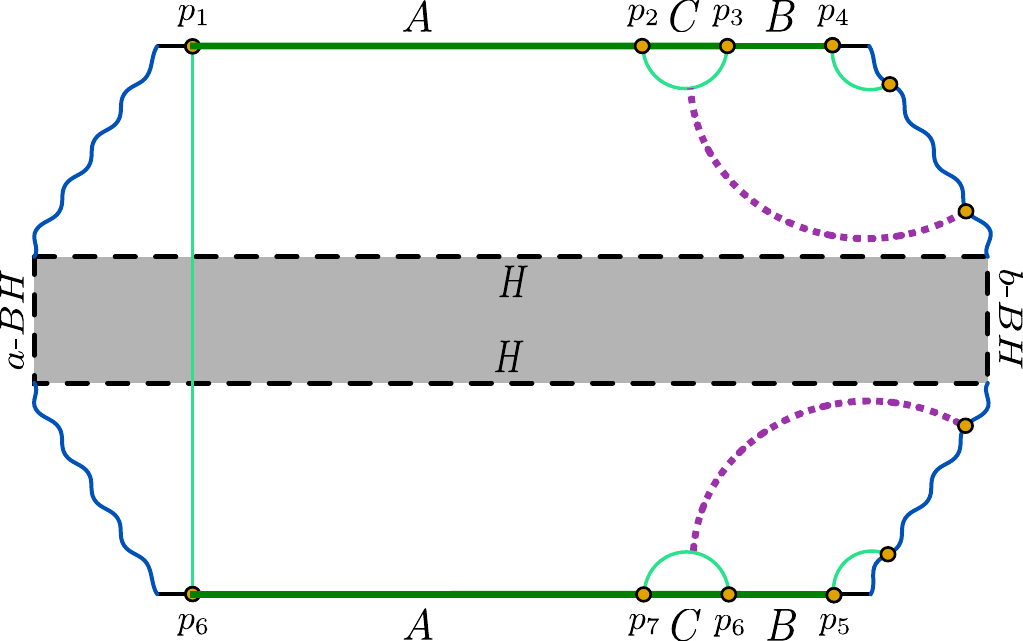}
		\caption{Configuration-9}
		\label{disbraneendingbulk2}
	\end{subfigure}
	\caption{Schematics depicts all the possible contributions to the reflected entropy for the case of two disjoint subsystems $A$ and $B$ where we get non trivial cross section from the entanglement island. }\label{disbraneendingresults}
\end{figure}

This configuration is similar to the previous case however the dominant correlator for the reflected entropy for the two disjoint subsystems now involves a QES point $b_2$ located on the $b$-brane as shown in \cref{disbraneending2}. Thus the factorization of the twist correlator for the R\'enyi reflected entropy in the large central charge limit may be given by
\begin{align}\label{dissrbraneending7}
	S^{R}_{n,m}(A:B)= 2\frac{1}{1-n}\log \frac{\Big< \sigma_{g_A^{-1}}(a_1)\sigma_{g_A}(p_1)\Big>\left<\sigma_{g_B^{-1}g_A}(b_2)\sigma_{g_A^{-1}}(p_2)\sigma_{g_B}(p_3)\right>\left<\sigma_{g_B^{-1}}(p_4)\sigma_{g_B}(b_1)\right>}{\left(\Big<\sigma_{g^{-1}_m}(a_1)\sigma_{g_m}(p_1)\Big>\left<\sigma_{g_m^{-1}}(p_2)\sigma_{g_m}(p_3)\right>\left<\sigma_{g_m^{-1}}(p_4)\sigma_{g_m}(b_1) \right>\right)^n}\,.
\end{align}
Note that the dominant correlator from the above equation arises from the three point twist correlator involving the points $p_2$, $p_3$ and $b_2$ while the other contributions to the reflected entropy cancel from the numerator and denominator in the replica limit. Hence the R\'enyi reflected entropy may be obtained as follows
\begin{align}\label{dissrdominant7}
	S^{R}_{n,m}(A:B)= 2\frac{1}{1-n}\log \frac{\left<\sigma_{g_B^{-1}g_A}(b_2)\sigma_{g_A^{-1}}(p_2)\sigma_{g_B}(p_3)\right>}{\left(\left<\sigma_{g_m^{-1}}(p_2)\sigma_{g_m}(p_3)\right>\right)^n}\,.
\end{align}
Finally in the replica limit the expression for the reflected entropy in this case is given by the following
\begin{align}\label{srconfig7dis}
	S^{R}_{\text{eff}}(A:B)= \frac{2c}{3} \log\left[\frac{\beta \left(\cosh \left(\frac{2 \pi  (R-b_2)}{\beta 
	}\right)-1\right)}{ \pi  r \sinh\left(\frac{2 \pi  b_2}{\beta}\right)}\right]\,,
\end{align}
where $R$ and $r$ are related to the points $p_2$ and $p_3$ as $R=(p_3+p_2)/2$ and $r=(p_3-p_2)/2$. Once again we may obtain the generalized reflected entropy using the island formula in \cref{srew} as follow
\begin{align}\label{srdisbraneending2}
	S_{\text{gen}}^{R}(A:B)= 2\Phi_0+\frac{4 \pi  \Phi_r}{\beta } \coth \left(\frac{2 \pi b_2}{\beta }\right)+\frac{2c}{3} \log\left[\frac{\beta \left(\cosh \left(\frac{2 \pi  (R-b_2)}{\beta 
		}\right)-1\right)}{ \pi  r \sinh\left(\frac{2 \pi  b_2}{\beta}\right)}\right]\,.
\end{align}
%%%%%%%%%%%%%%%%%%%%%%%%%%%%%%%%%%%%%%%%%%%%%%%%%%%%%%%%%%%%%%

%%%%%%%%%%%%%%%%%%%%%%%%%%%%%%%%%%%%%%%%%%%%%%%%%%%%%%%%%%%%%%

\subsubsection*{Configuration-8 and 9}
In these configurations, the computation of the reflected entropy follows a similar analysis to that described in configuration-3 and 4. The dominant twist correlators in these cases are given by the \cref{dissrdominant6,dissrdominant7} in the large central charge limit. The only difference in configuration-8 as depicted in \cref{disbraneendingbulk} arises from the two point correlator involving the twist operators located at the points $p_4$ and $p_8$ however this contribution to the reflected entropy cancel from the numerator and denominator in the replica limit. Hence the expression for the reflected entropy in the above configuration is given by \cref{srdisbraneending} in the replica limit.

For the configuration-9 (\cref{disbraneendingbulk2}), we may employ similar arguments as described for the configuration-8 to compute the reflected entropy of two disjoint subsystems. Thus the reflected entropy in this case is given by \cref{srdisbraneending2} in the replica limit.
\subsubsection*{Configuration-10}
The reflected entropy in this configuration involves twist operators located on both the radiation reservoirs and the $a$ and $b$-branes as depicted in \cref{disbulk}. The R\'enyi reflected entropy in this case may be obtained as
\begin{equation}
\begin{aligned}\label{dissrbulk}
	S^{R}_{n,m}(A:B) =\frac{1}{1-n}\log &\frac{\Big{<} \sigma_{g_A^{-1}}(a_1)\sigma_{g_A}(p_1)\sigma_{g_A^{-1}}(p_2)\sigma_{g_B}(p_3)\sigma_{g_B^{-1}}(p_4)\sigma_{g_B}(b_1) }{\Big{<}\sigma_{g^{-1}_m}(a_1)\sigma_{g_m}(p_1)\sigma_{g_m^{-1}}(p_2)\sigma_{g_m}(p_3)\sigma_{g^{-1}_m}(p_4)\sigma_{g_m}(b_1) }\\
	&\times\frac{\sigma_{g_A}(a_2)\sigma_{g_A^{-1}}(p_5)\sigma_{g_A}(p_6)\sigma_{g_B^{-1}}(p_7)\sigma_{g_B}(p_8)\sigma_{g_B^{-1}}(b_2)\Big{>}_{\mathrm{CFT}^{\bigotimes mn}}}{\sigma_{g_m}(a_2)\sigma_{g_m^{-1}}(p_5)\sigma_{g_m}(p_6)\sigma_{g_m^{-1}}(p_7)\sigma_{g_m}(p_8)\sigma_{g_m^{-1}}(b_2)\Big{>}^n_{\mathrm{CFT}^{\bigotimes m}}}\,.
\end{aligned}
\end{equation}
The factorization of the above correlator to the respective contraction in the large central charge limit is given by the following equation
\begin{equation}
\begin{aligned}\label{dissrbulknext}
	S^{R}_{n,m}(A:B) =\frac{1}{1-n}\log &\frac{\Big{<} \sigma_{g_A^{-1}}(a_1)\sigma_{g_A}(p_1)\Big{>}\Big{<}\sigma_{g_A^{-1}}(p_2)\sigma_{g_B}(p_3)\sigma_{g_A}(p_6)\sigma_{g_B^{-1}}(p_7)\Big{>}}{\left(\Big{<}\sigma_{g^{-1}_m}(a_1)\sigma_{g_m}(p_1)\Big{>}\Big{<}\sigma_{g_m^{-1}}(p_2)\sigma_{g_m}(p_3)\sigma_{g_m}(p_6)\sigma_{g_m^{-1}}(p_7)\Big{>}\right)^n}\\
	&\times\frac{\Big{<}\sigma_{g_B^{-1}}(p_4)\sigma_{g_B}(b_1)\Big{>}\Big{<}\sigma_{g_A}(a_2)\sigma_{g_A^{-1}}(p_5)\Big{>}\Big{<}\sigma_{g_B}(p_8)\sigma_{g_B^{-1}}(b_2)\Big{>}}{\left(\Big{<}\sigma_{g_m^{-1}}(p_4)\sigma_{g_m}(b_1)\Big{>}\Big{<}\sigma_{g_m}(a_2)\sigma_{g_m^{-1}}(p_5)\Big{>}\Big{<}\sigma_{g_m}(p_8)\sigma_{g_m^{-1}}(b_2)\Big{>}\right)^n}\,,
\end{aligned}
\end{equation}
where the Weyl factors associated with the points on the branes cancel from the numerator and denominator in the replica limit. The dominant twist correlator in this case for the reflected entropy involves the four point correlator with the twist operators located on both the radiation reservoirs. Hence the R\'enyi reflected entropy is given by
\begin{align}\label{dissrbulkdominant10}
	S^{R}_{n,m}(A:B) =\frac{1}{1-n}\log \frac{\Big{<}\sigma_{g_A^{-1}}(p_2)\sigma_{g_B}(p_3)\sigma_{g_A}(p_6)\sigma_{g_B^{-1}}(p_7)\Big{>}}{\left(\Big{<}\sigma_{g_m^{-1}}(p_2)\sigma_{g_m}(p_3)\sigma_{g_m}(p_6)\sigma_{g_m^{-1}}(p_7)\Big{>}\right)^n}\,,
\end{align}
where the other two point twist correlators in the above equation cancel from the numerator and denominator in the replica limit. Finally, the reflected entropy for this case in the replica limit may be obtained as follows  
\begin{align}\label{disjointsrbulk}
	S^{R}_{\text{eff}}(A:B) =\frac{2c}{3}\log \left[\frac{\beta  \cosh \left(\frac{2 \pi  t}{\beta}\right)}{\pi  r}\right]\,.
\end{align}
In \cref{disjointsrbulk}, $r$ is related to the points $p_2$ and $p_3$ as $r=(p_3-p_2)/2$.
\subsubsection*{Configuration-11 and 12}
\begin{figure}[H]
	\centering
	\begin{subfigure}[b]{0.45\textwidth}
		\centering
		\includegraphics[width=\textwidth]{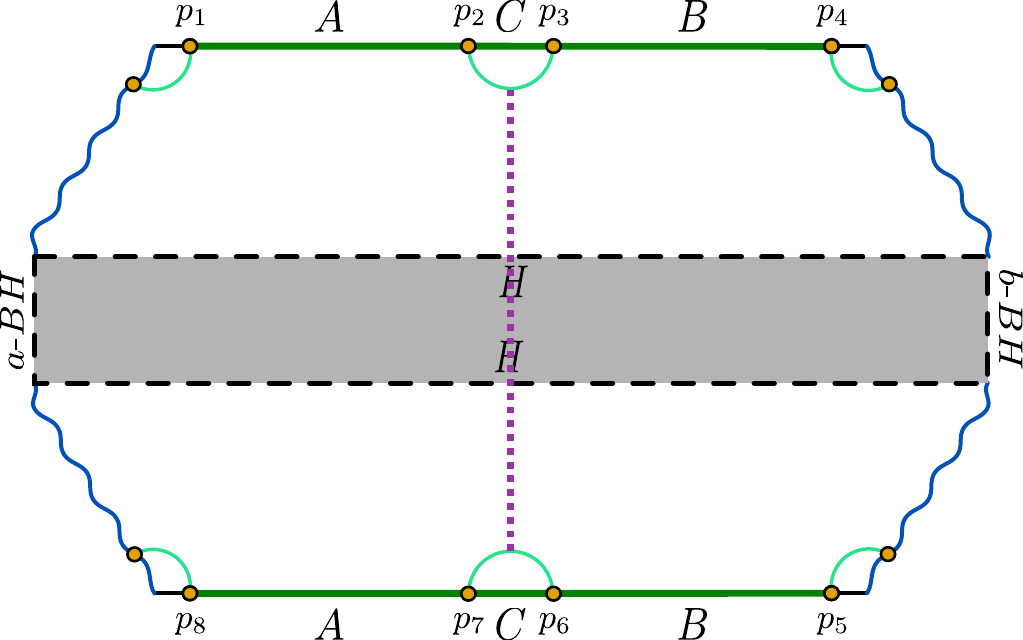}
		\caption{Configuration-10}
		\label{disbulk}
	\end{subfigure}
	\hspace{.5cm}
	\begin{subfigure}[b]{0.45\textwidth}
		\centering
		\includegraphics[width=\textwidth]{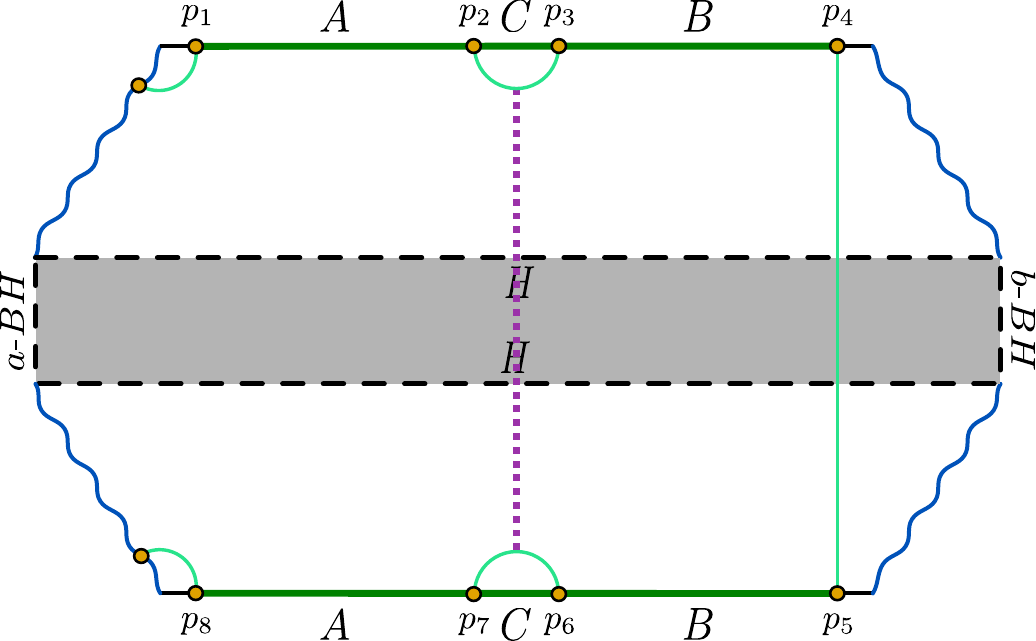}
		\caption{Configuration-11}
		\label{disbulkbulk}
	\end{subfigure}
	\vspace{.5cm}
	\begin{subfigure}[b]{0.45\textwidth}
		\centering
		\includegraphics[width=\textwidth]{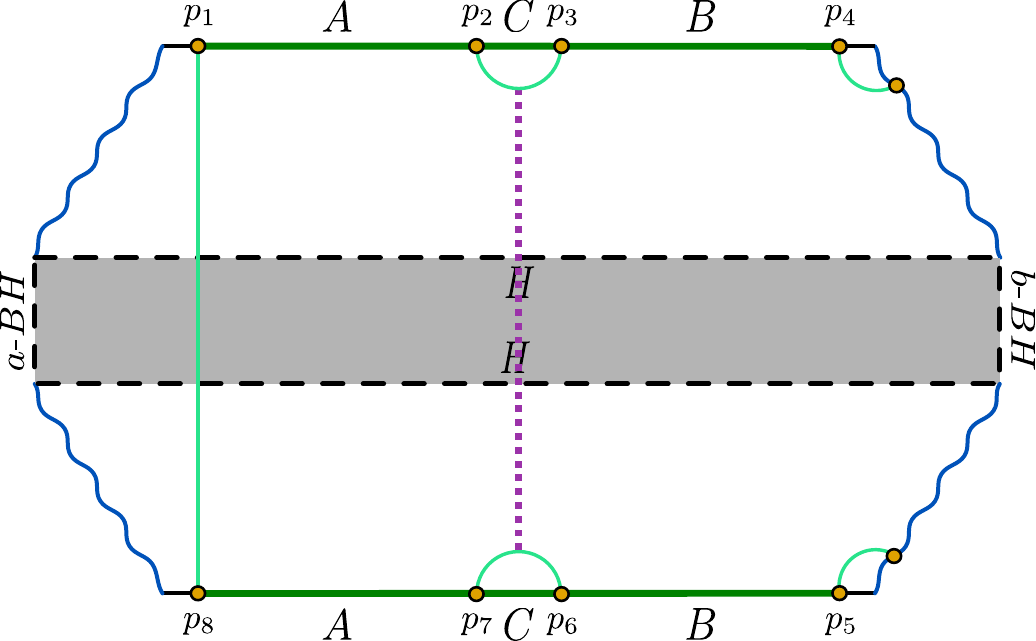}
		\caption{Configuration-12}
		\label{disbulkbulk2}
	\end{subfigure}
	\caption{ The corresponding diagrams depict non-trivial cross section extending between the TFD copies of the radiation reservoirs.}\label{disbulkresults}
\end{figure}

In these configurations (\cref{disbulkbulk,disbulkbulk2}), the computation of the reflected entropy follows similar analysis as described in the configuration-10. The only difference in these cases arise from the two point correlators with the twist operators located at the points $p_4$ and $p_8$ in the radiation reservoirs for the first while the other configuration-12 (\cref{disbulkbulk2}) incorporates the twist operators located at the points $p_1$ and $p_5$. However these contributions to the reflected entropy for these cases cancel from the numerator and denominator in the replica limit. Thus the expression of the reflected entropies in these configurations are given by \cref{disjointsrbulk}.

%%%%%%%%%%%%%%%%%%%%%%%%%%%%%%%%%%%%%%%%%%%%%%%%%%%%%%%%%%%%%%%%%%%%%%%%%%%%%%%

\subsubsection*{Configuration-13}
In this case, the reflected entropy for the two disjoint subsystems $A$ and $B$ involves the twist operators located on both the radiation reservoirs as depicted in \cref{disbulkending}. The R\'enyi reflected entropy in this scenario may be expressed as follows   
\begin{equation}
\begin{aligned}\label{dissrbulkending}
	S^{R}_{n,m}(A:B) =\frac{1}{1-n}\log &\frac{\Big{<} \sigma_{g_A^{-1}}(a_1)\sigma_{g_A}(p_1)\sigma_{g_A^{-1}}(p_2)\sigma_{g_B}(p_3)\sigma_{g_B^{-1}}(p_4)}{\Big{(}\Big{<}\sigma_{g^{-1}_m}(a_1)\sigma_{g_m}(p_1)\sigma_{g_m^{-1}}(p_2)\sigma_{g_m}(p_3)\sigma_{g_m^{-1}}(p_4)}\\ 
	&\times\frac{\sigma_{g_A}(a_2)\sigma_{g_A^{-1}}(p_5)\sigma_{g_A}(p_6)\sigma_{g_B^{-1}}(p_7)\sigma_{g_B}(p_8)\Big{>}_{\mathrm{CFT}^{\bigotimes mn}}}{\sigma_{g_m}(a_2)\sigma_{g^{-1}_m}(p_5)\sigma_{g_m}(p_6)\sigma_{g_m^{-1}}(p_7)\sigma_{g_m}(p_8)\Big{>}_{\mathrm{CFT}^{\bigotimes m}}\Big{)}^n}\,.
\end{aligned}
\end{equation}
 In the large central charge limit, the correlator in \cref{dissrbulkending} factorizes to the respective contraction as follows
 
 \begin{equation}
\begin{aligned}\label{srbulkending}
	S^{R}_{n,m}(A:B) =\frac{1}{1-n}\log &\frac{\Big{<}\sigma_{g_A^{-1}}(p_2)\sigma_{g_B}(p_3)\sigma_{g_B^{-1}}(p_4)\sigma_{g_A}(p_6)\sigma_{g_B^{-1}}(p_7)\sigma_{g_B}(p_8)\Big{>} }{\left(\Big{<}\sigma_{g_m^{-1}}(p_2)\sigma_{g_m}(p_3)\sigma_{g_m^{-1}}(p_4)\sigma_{g_m}(p_6)\sigma_{g_m^{-1}}(p_7)\sigma_{g_m}(p_8)\Big{>}\right)^n}\\ 
	&\times\frac{\Big{<} \sigma_{g_A^{-1}}(a_1)\sigma_{g_A}(p_1)\Big{>}\Big{<}\sigma_{g_A^{-1}}(a_2)\sigma_{g_A}(p_5)\Big{>}}{\left(\Big{<}\sigma_{g^{-1}_m}(a_1)\sigma_{g_m}(p_1)\Big{>}\Big{<}\sigma_{g^{-1}_m}(a_2)\sigma_{g_m}(p_5)\Big{>}\right)^n}\,.
\end{aligned}
\end{equation}
 The dominant twist correlator in this case for the reflected entropy involves the six point function with the twist operators located on both the radiation reservoirs, however the other two point twist correlators in \cref{srbulkending} cancel from the numerator and denominator in the replica limit. Hence the R\'enyi reflected entropy is given as
  \begin{equation}
 	\begin{aligned}\label{dominantsrbulkending}
 		S^{R}_{n,m}(A:B) =\frac{1}{1-n}\log \frac{\Big{<}\sigma_{g_A^{-1}}(p_2)\sigma_{g_B}(p_3)\sigma_{g_B^{-1}}(p_4)\sigma_{g_A}(p_6)\sigma_{g_B^{-1}}(p_7)\sigma_{g_B}(p_8)\Big{>} }{\left(\Big{<}\sigma_{g_m^{-1}}(p_2)\sigma_{g_m}(p_3)\sigma_{g_m^{-1}}(p_4)\sigma_{g_m}(p_6)\sigma_{g_m^{-1}}(p_7)\sigma_{g_m}(p_8)\Big{>}\right)^n}\,.
 	\end{aligned}
 \end{equation}
We utilize a technique termed as inverse doubling trick to compute the above dominant correlator \cite{Shao:2022gpg}. In this context, the six point function in the $CFT_2$ reduces to a three point function in the $BCFT_2$ and the expression for the reflected entropy may be obtained in the appropriate OPE channel by following a similar analysis to that described in \cite{Shao:2022gpg}
\begin{align}\label{srbulkendingfinal}
	S^{R}_{\text{eff}}(A:B)=\frac{c}{3} \log \left[\frac{\sech(\frac{2 \pi t}{\beta}) \sqrt{r^2+(R-r_2)^2} \sqrt{r^2+R^2+2 R r_2 \cosh (\frac{4 \pi t}{\beta})+r_2^2}}{r_2 \,\,r}\right]\,,
\end{align}
where $R$ and $r$ are related to the points $p_2$ and $p_3$ as $R=(p_3+p_2)/2$ and $r=(p_3-p_2)/2$.
\subsubsection*{Configuration-14}
This configuration is similar to above case
and the twist correlator for the R\'enyi reflected entropy factorizes to the respective contraction in the large central charge limit as follows
\begin{equation}
\begin{aligned}	
	S^{R}_{n,m}(A:B) =\frac{1}{1-n}\log &\frac{\Big{<}\sigma_{g_A}(p_1)\sigma_{g_A^{-1}}(p_2)\sigma_{g_B}(p_3)\sigma_{g_A^{-1}}(p_5)\sigma_{g_B}(p_6)\sigma_{g_B^{-1}}(p_7)\Big{>} }{\left(\Big{<}\sigma_{g_m}(p_1)\sigma_{g_m^{-1}}(p_2)\sigma_{g_m}(p_3)\sigma_{g_m^{-1}}(p_5)\sigma_{g_m}(p_6)\sigma_{g_m^{-1}}(p_7)\Big{>}\right)^n}\\
	&\times\frac{\Big{<} \sigma_{g_B}(b_1)\sigma_{g_B^{-1}}(p_4)\Big{>}\Big{<}\sigma_{g_B^{-1}}(b_2)\sigma_{g_B}(p_8)\Big{>}}{\left(\Big{<}\sigma_{g^{-1}_m}(b_1)\sigma_{g_m}(p_3)\Big{>}\Big{<}\sigma_{g^{-1}_m}(b_2)\sigma_{g_m}(p_8)\Big{>}\right)^n}\,.
\end{aligned}
\end{equation}
In this case, the dominant correlator in the above equation involves a six point twist correlator, however the other two point twist correlators cancel from the numerator and the denominator. Thus the R\'enyi reflected entropy may be obtained as
\begin{equation}\label{dominantconfig14}
	\begin{aligned}	
		S^{R}_{n,m}(A:B) =\frac{1}{1-n}\log \frac{\Big{<}\sigma_{g_A}(p_1)\sigma_{g_A^{-1}}(p_2)\sigma_{g_B}(p_3)\sigma_{g_A^{-1}}(p_5)\sigma_{g_B}(p_6)\sigma_{g_B^{-1}}(p_7)\Big{>} }{\left(\Big{<}\sigma_{g_m}(p_1)\sigma_{g_m^{-1}}(p_2)\sigma_{g_m}(p_3)\sigma_{g_m^{-1}}(p_5)\sigma_{g_m}(p_6)\sigma_{g_m^{-1}}(p_7)\Big{>}\right)^n}\,.
	\end{aligned}
\end{equation}
On utilization of the techniques discussed in \cite{Shao:2022gpg}, the reflected entropy for the two disjoint subsystems is given by the following equation
\begin{align}\label{srbulkendingdis14}
	S^{R}_{\text{eff}}(A:B)=\frac{c}{3} \log \left[\frac{\sech(\frac{2 \pi t}{\beta}) \sqrt{r^2+(R-r_2)^2} \sqrt{r^2+R^2+2 R r_2 \cosh (\frac{4 \pi t}{\beta})+r_2^2}}{r_2 \,\,r}\right]\,,
\end{align}
where $R$ and $r$ are related to the points $p_2$ and $p_3$ as $R=(p_3+p_2)/2$ and $r=(p_3-p_2)/2$.
\subsubsection*{Configuration-15 and 16}
\begin{figure}[H]
	\centering
	\begin{subfigure}[b]{0.45\textwidth}
		\centering
		\includegraphics[width=\textwidth]{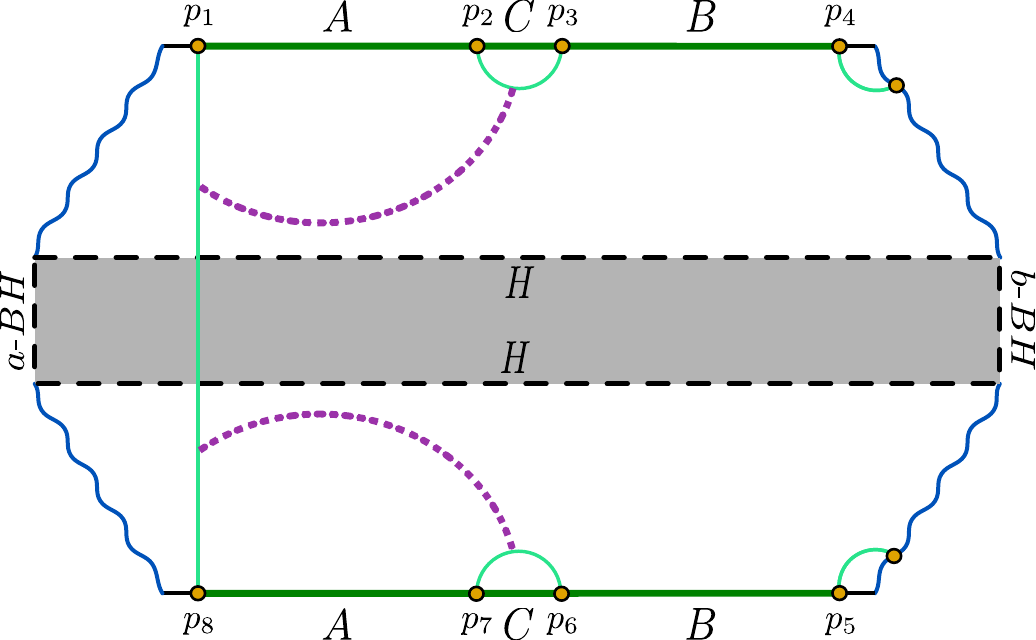}
		\caption{Configuration-13}
		\label{disbulkending}
	\end{subfigure}
	\hspace{.5cm}\vspace{.5cm}
	\begin{subfigure}[b]{0.45\textwidth}
		\centering
		\includegraphics[width=\textwidth]{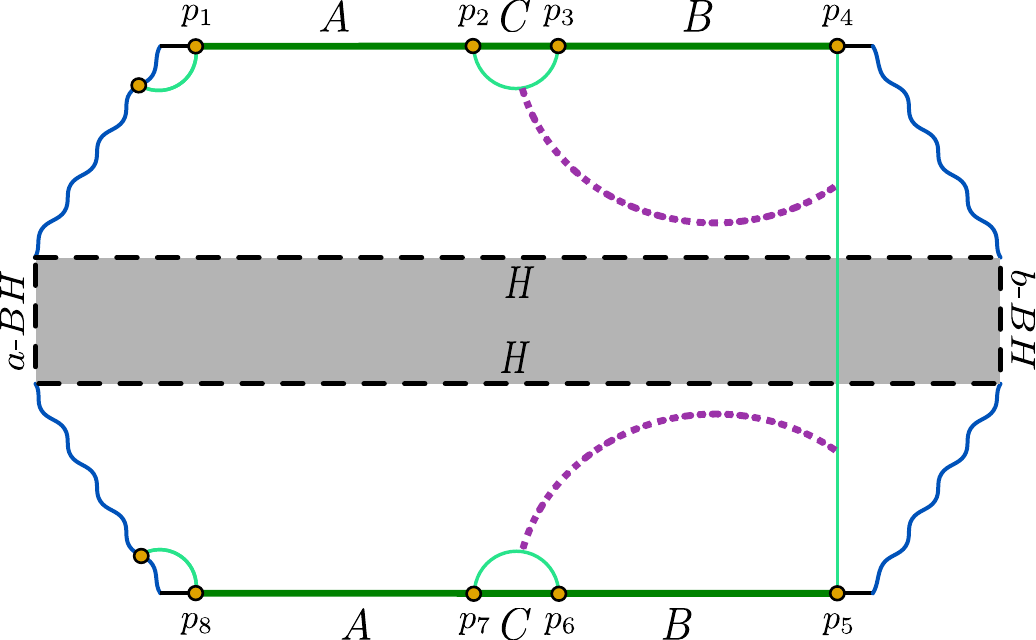}
		\caption{Configuration-14}
		\label{disbulkending2}
	\end{subfigure}
	\hspace{.5cm}
	\begin{subfigure}[b]{0.45\textwidth}
		\centering
		\includegraphics[width=\textwidth]{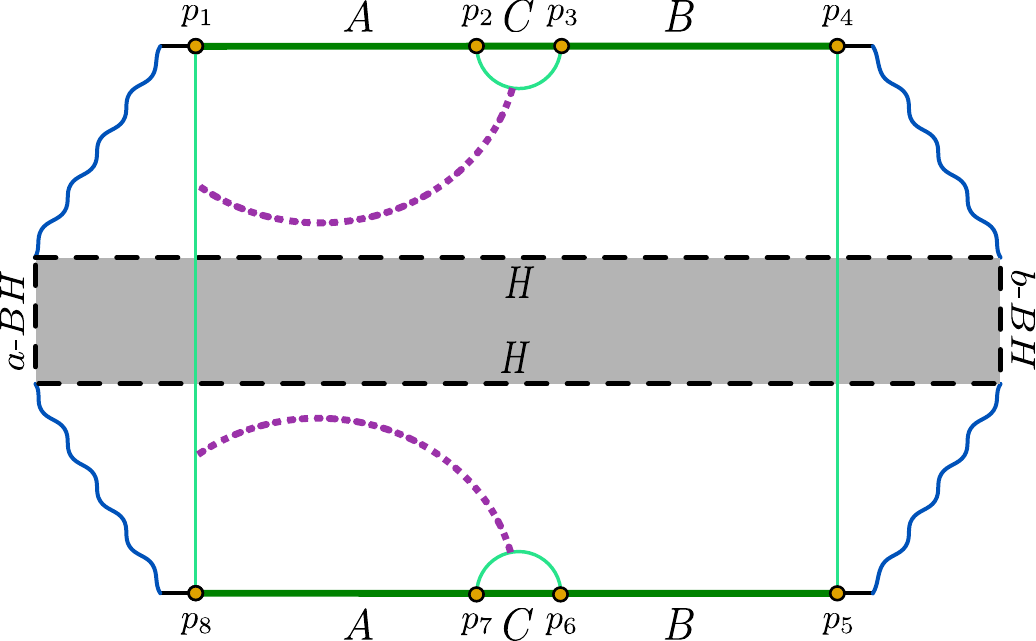}
		\caption{Configuration-15}
		\label{disbulkendingbulk}
	\end{subfigure}
	\hspace{.5cm}
	\begin{subfigure}[b]{0.45\textwidth}
		\centering
		\includegraphics[width=\textwidth]{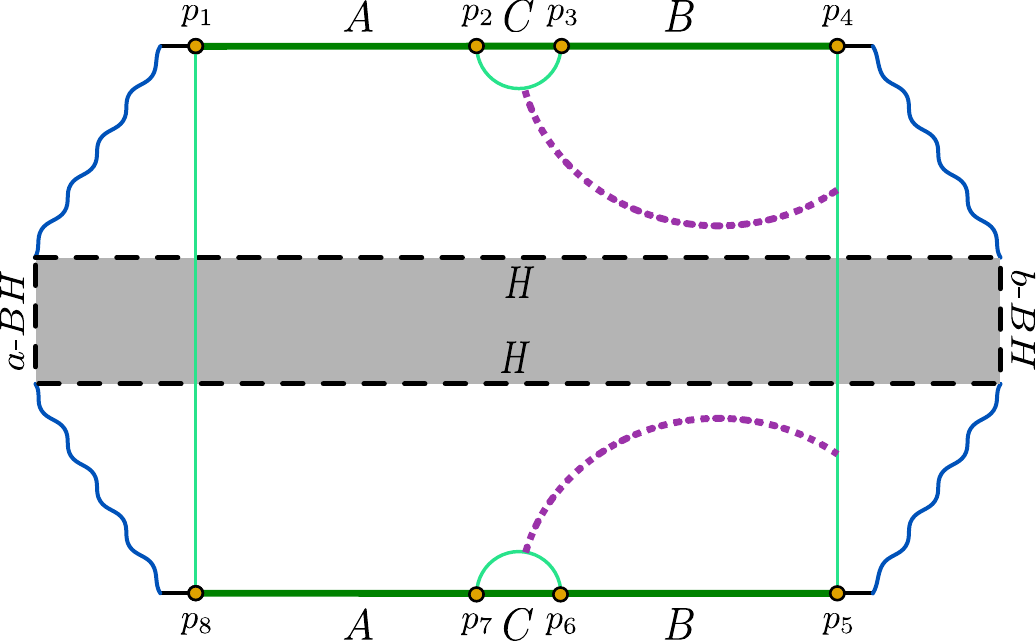}
		\caption{Configuration-16}
		\label{disbulkendingbulk2}
	\end{subfigure}
	\caption{ Starting two figures indicate connected phase of the entanglement island with non-trivial cross section however later diagrams depict disconnected phase of the entanglement island with cross section ending the Hartman-Maldacena surface.}\label{disbulkendingresults}
\end{figure}

In these configurations, the computation of the reflected entropies for the two disjoint subsystems $A$ and $B$ follow a similar analysis as described for the configurations-13 and 14. Here the dominant correlators for these configurations are given by the \cref{dominantsrbulkending,dominantconfig14}. The only difference in these cases arise from the two point correlators with the twist operators located at the points $p_4$ and $p_8$ for the first (\cref{disbulkendingbulk}) while the other configuration-16 (\cref{disbulkendingbulk2}) incorporates the twist operators located at the points $p_1$ and $p_5$ in the radiation reservoirs. However these contributions to the reflected entropy for these cases cancel from the numerator and denominator in the replica limit.
Finally, the reflected entropy for these configurations are given by the \cref{srbulkendingfinal,srbulkendingdis14}. 
\subsection{Entanglement wedge cross section}\label{subsec:EWCScomputation}
In this subsection, we compute the bulk EWCS for the various mixed states described by the two adjacent and disjoint subsystems located in both the radiation reservoirs for which the field theory computations were described in the earlier sections. In this context, we observed a rich phase structure for the EWCS arising from the various contributions from
different relative sizes of the two adjacent and disjoint subsystems as depicted in the \cref{takaresults,braneendingresults,bulkresults,bulkendingresults,distakaresults,disbraneendingresults,disbulkresults,disbulkendingresults}.  In particular, we utilized the embedding space formalism to compute the bulk EWCS for different configurations.  

\subsubsection{Adjacent subsystems}
We first consider the two adjacent subsystems $A$ and $B$ in both the copies of the radiation reservoirs and compute the EWCS for all the possible contributions while considering different sizes of the subsystems.
\subsubsection*{Configuration-1}
The EWCS for this configuration is shown in \cref{taka} as a dotted line and this may be computed using the result obtained in \cite{Takayanagi:2017knl} in the adjacent limit as follows
\begin{equation}\label{ewcstaka}
	E_W=\frac{c}{3}\log\left(4 z\right)\,, \quad\quad z=\frac{\sinh{\frac{2 \pi z_{12}}{\beta}}\sinh{\frac{2 \pi z_{34}}{\beta}}}{\sinh{\frac{2 \pi z_{23}}{\beta}}\sinh{\frac{2 \pi z_{14}}{\beta}}}\,,
\end{equation} 
where $z$ is the cross-ratio at a finite temperature and we have utilized the Brown-Henneaux formula $c=\frac{3}{2 G_N}$ \cite{Brown:1986nw}. The above result may be expressed in terms of the boundary coordinates of the corresponding subsystems shown in \cref{taka} as follows
\begin{align}\label{EWCStaka}
	E_W= \frac{c}{3}\log\left[\frac{\sinh\frac{2\pi(p_2-p_1)}{\beta}\sinh\frac{2\pi(a_1+p_2)}{\beta}}{\sinh\frac{2\pi(a_1+p_1)}{\beta}\sinh\frac{2\pi(\epsilon)}{\beta}}\right].
\end{align}
Interestingly the field theory replica technique result for the the reflected entropy described in \cref{reflectedtaka}
matches exactly with twice the bulk EWCS in \cref{ewcstaka} for the two adjacent subsystems in accordance with the
proposed holographic duality in \cite{Dutta:2019gen,Chandrasekaran:2020qtn}.

%%%%%%%%%%%%%%%%%%%%%%%%%%%%%%%%%%%%%%%%%%%%%%%%%%%%%%%%%%%%%%%%%%%%

\subsubsection*{Configuration-2}
As depicted in \cref{taka2}, the bulk EWCS for this configuration may be obtained by following a similar analysis as discussed in the previous configuration. Hence the expression for the EWCS is given by   
%and the using \cref{ewcstaka}
\begin{align}\label{EWCStaka2}
	E_W= \frac{c}{3}\log\left[\frac{\sinh\frac{2\pi(p_3-p_2)}{\beta}\sinh\frac{2\pi(b_1+p_2)}{\beta}}{\sinh\frac{2\pi(b_1+p_3)}{\beta}\sinh\frac{2\pi(\epsilon)}{\beta}}\right]\,,
\end{align}
where $\epsilon$ is the UV cutoff. Once again the replica technique result in \cref{reflectedtaka2} matches with twice the bulk EWCS in \cref{EWCStaka2} in agreement with the proposed holographic duality in \cite{Dutta:2019gen,Chandrasekaran:2020qtn}.

%%%%%%%%%%%%%%%%%%%%%%%%%%%%%%%%%%%%%%%%%%%%%%%%%%%%%%%%%%%%%%%%%%%%
\subsubsection*{Configuration-3 and 4}
The configuration-3 is similar to the case-1. However the only difference in this configuration arises from the entanglement wedge of the subsystem $A\cup B$ as described in \cref{takabulk}. Thus the bulk EWCS for the configuration-3 may be obtained from \cref{EWCStaka}.

On utilization of similar argument for the configuration-4  which is analogous to case-2 as shown in \cref{takabulk2}. Therefore, the bulk EWCS for the configuration-4 is given by \cref{EWCStaka2}.
%%%%%%%%%%%%%%%%%%%%%%%%%%%%%%%%%%%%%%%%%%%%%%%%%%%%%%%%%%%%%%%%%%%%
\subsubsection*{Configuration-5}
The computation of the bulk EWCS for this configuration as shown in \cref{RTending} is trivial since it reduces to the usual expression of the EWCS in the context of the $AdS/CFT$ scenario. Therefore the bulk EWCS is given by
\begin{align}\label{EWCS}
	E_W= \frac{c}{3}\log\left[\frac{\sinh\frac{2\pi(p_3-p_2)}{\beta}\sinh\frac{2\pi(p_3-p_2)}{\beta}}{\sinh\frac{2\pi(p_3-p_1)}{\beta}\sinh\frac{2\pi(\epsilon)}{\beta}}\right].
\end{align}
Once again twice of the above bulk EWCS exactly matches with the replica technique results in \cref{srRTending} which is consistent with the holographic duality described in \cite{Dutta:2019gen,Chandrasekaran:2020qtn}.
%%%%%%%%%%%%%%%%%%%%%%%%%%%%%%%%%%%%%%%%%%%%%%%%%%%%%%%%%%%%%%%%%%%%

\subsubsection*{Configuration-6}

The bulk EWCS in this configuration involves QES point located on the $a$-brane and the point $p_2$ situated in the radiation reservoir as shown in the \cref{braneending}. To compute the bulk EWCS for this configuration, we utilize the embedding space formalism which involves following coordinate transformations \cite{Grimaldi:2022suv,Geng:2021iyq}  

\begin{equation}\label{embeddingtransformation}
\begin{aligned}
	T_1(z,t)&=z_h \sqrt{\frac{1}{z^2}-\frac{1}{z_h^2}} \cosh \left(\frac{t}{z_h}\right)\,,\\ 
	T_2(z,t)&=\frac{z_h \sinh \left(\frac{x}{z_h}\right)}{z}\,,\\ 
	X_1(z,t)&=z_h \sqrt{\frac{1}{z^2}-\frac{1}{z_h^2}} \sinh \left(\frac{t}{z_h}\right)\,,\\ 
	X_2(z,t)&=\frac{z_h \cosh \left(\frac{x}{z_h}\right)}{z}\,. 
\end{aligned}
\end{equation}
where $z_h$ is related to the inverse temperature as $z_h=\frac{\beta}{2 \pi}$. Note that, the $AdS_3$ BTZ black hole metric may be reduced to the embedding metric using the above coordinate transformations as follows \cite{Grimaldi:2022suv,Geng:2021iyq}   
\begin{align}\label{embeddingmetric}
	ds^2= dX^2_1 +dX^2_2-dT^2_1-dT^2_2\,.
\end{align}
 In the above background, the expression of the geodesic length connecting two arbitrary points is given by
\begin{align}\label{embeddinglength}
	L= \cosh^{-1}\left(X_1 X'_1+ X_2 X'_2- T_1 T'_1 - T_2 T'_2\right)\,,
\end{align}
where the unprimed and the primed coordinates are the location of the arbitrary points in term of the embedding coordinates. For this configuration, the endpoints of the bulk EWCS in the BTZ coordinates are defined as $(p_2,\epsilon,t)$ and $(a_2, zQES ,t)$. Finally, we may obtain the bulk EWCS utilizing the \cref{embeddinglength,embeddingtransformation} as 
\begin{equation}
	E_W=\frac{c}{3}\cosh ^{-1}\left[\frac{z_h ^2 \left(\cosh \left(\frac{a_2-p_2}{z_h}\right)-1\right)}{zQES \epsilon }\right]\,,
\end{equation} 
where $zQES$ may be obtained using the profile of the $a$-brane as $zQES \,k=z_h \sinh \left(\frac{a_2}{z_h}\right)$ with constant $k=1$ \cite{Grimaldi:2022suv}. Hence, the expression of the bulk EWCS becomes
\begin{equation}\label{EWbraneending}
	E_W=\frac{c}{3}\log \left[\frac{\beta}{ \pi}\frac{ \cosh \left(\frac{2 \pi(a_2-p_2)}{\beta}\right)-1}{ \sinh \left(\frac{2 \pi a_2}{\beta}\right) }\right]\,,
\end{equation} 
where we have removed divergent term. In the above equation, we have utilized a relation between inverse temperature and $z_h$ as $z_h=\frac{\beta}{2 \pi}$. Once again twice of the above bulk EWCS exactly matches with the replica technique result in \cref{srconfig6} which is consistent with the proposed holographic duality in \cite{Dutta:2019gen,Chandrasekaran:2020qtn}.
%%%%%%%%%%%%%%%%%%%%%%%%%%%%%%%%%%%%%%%%%%%%%%%%%%%%%%%%%%%%%%%%%%%%

\subsubsection*{Configuration-7}
For this configuration, the computation of the bulk EWCS follows a similar analysis as described in the previous case.  However the bulk EWCS for this configuration involves a point $b_2$ located at the $b$-brane. Hence the expression of the bulk EWCS is given as 
\begin{equation}\label{EWbraneending2}
	E_W=\frac{c}{3}\log \left[\frac{\beta}{ \pi}\frac{ \cosh \left(\frac{2 \pi(b_2-p_2)}{\beta}\right)-1}{ \sinh \left(\frac{2 \pi b_2}{\beta}\right)  }\right]\,.
\end{equation} 
Once again the replica technique result in \cref{srconfig7} exactly matches with twice of the above bulk EWCS in accordance with the proposed holographic duality.

%%%%%%%%%%%%%%%%%%%%%%%%%%%%%%%%%%%%%%%%%%%%%%%%%%%%%%%%%%%%%%%%%%%%
\subsubsection*{Configuration-8 and 9}
For these configurations, the computation of the bulk EWCS follows a similar analysis described in the configurations 6 and 7. However the enclosed entanglement wedge regions for the subsystem $A\cup B$ in these two configurations are different as shown in the  \cref{braneendingbulk,braneendingbulk2}. Thus the expressions of the bulk EWCS for the configurations 8 and 9  are given by the \cref{EWbraneending,EWbraneending2} respectively.      

%%%%%%%%%%%%%%%%%%%%%%%%%%%%%%%%%%%%%%%%%%%%%%%%%%%%%%%%%%%%%%%%%%%%

\subsubsection*{Configuration-10}
The computation of the bulk EWCS for this configuration involves the endpoints ($p_2,\epsilon,t$) and ($p_2,\epsilon,-t+i\beta/2$) which are located in both the radiation reservoirs as depicted in \cref{bulkewcs}. Utilizing the \cref{embeddinglength,embeddingtransformation}, we may obtain the expression of the bulk EWCS as follows
   \begin{align}\label{EWbulk}
   	E_W =\frac{c}{3} \log \left[\frac{\beta}{ \pi}
   	\cosh \left(\frac{2 \pi  t}{\beta }\right)\right]\,,
   \end{align}
where we have removed divergent term. Note that twice of above the bulk EWCS exactly matches with the replica technique result for the reflected entropy described in \cref{srbulk} which is consistent with the holographic duality in \cite{Dutta:2019gen,Chandrasekaran:2020qtn}. 

%%%%%%%%%%%%%%%%%%%%%%%%%%%%%%%%%%%%%%%%%%%%%%%%%%%%%%%%%%%%%%%%%%%%

\subsubsection*{Configuration-11 and 12}
The computation of the bulk EWCS for these configurations follow a similar analysis described in the configuration-10. The only difference in these configurations arises from the enclosed entanglement wedge regions for the subsystem $A\cup B$ as shown in the  \cref{bulkbulk,bulkbulk2}.  Thus the expressions of the bulk EWCS for the configurations 11 and 12 are given by \cref{EWbulk}. 

%%%%%%%%%%%%%%%%%%%%%%%%%%%%%%%%%%%%%%%%%%%%%%%%%%%%%%%%%%%%%%%%%%%%

\subsubsection*{Configuration-13}

For the computation of the bulk EWCS in this configuration, we may utilize a coordinate transformation which relates the $AdS_3$ BTZ metric described in \cref{BTZmetric} to Poincare metric as follows \footnote{This kind of bulk EWCS was also studied in the article \cite{Shao:2022gpg} in a different toy model of the brane world geometry.}
\begin{align}
	\nonumber
	T=&\sqrt{1-\frac{z^2}{z_h^2}} e^{\frac{x}{z_h}} \sinh \left(\frac{t}{z_h}\right)\\ 
	X=&\sqrt{1-\frac{z^2}{z_h^2}} e^{\frac{x}{z_h}} \cosh \left(\frac{t}{z_h}\right)\\ \nonumber
	Z=&\frac{z e^{\frac{x}{z_h}}}{z_h}\,,
\end{align}
where $T,X,Z$ are the Poincare coordinates. The bulk EWCS may then be obtained from the following geodesic length formula 
\begin{align}\label{poincaregeodesic}
	L=\cosh^{-1}\left[\frac{(X_2-X_1)^2-(T_2-T_1)^2+Z_1 ^2+Z_2 ^2 }{2 Z_1 Z_2 }\right]\,.
\end{align}  
As described in \cref{bulkending}, the endpoints of the bulk EWCS for this case are given by  $\left(r_1 \cosh\frac{2 \pi t}{\beta},r_1 \sinh\frac{2 \pi t}{\beta},\epsilon\right)$ and $\left(x_2, r_2 \sinh\frac{2 \pi t}{\beta},z_2\right)$. Note that the bulk EWCS for this configuration describes the following geodesic equation $z_2 = \sqrt{r_2^2 (\cosh\frac{2 \pi t}{\beta})^2-x_2 ^2}$, which is also similar to the Hartman-Maldacena surface discussed in \cite{Hartman:2013qma}.
Thus the expression of the bulk EWCS may be obtained as
\begin{align}\label{EW13}
	E_W= \frac{c}{3} \log \left[\frac{(r_1-r_2)~ \text{sech}\left(\frac{2 \pi  t}{\beta }\right) \sqrt{r_1^2+2 r_1 r_2 \cosh \left(\frac{4 \pi  t}{\beta }\right)+r_2^2}}{r_1 \epsilon }\right]\,,
\end{align}
where we have utilized the relation between inverse temperature and $z_h$. Once again the replica technique result in \cref{sr13} exactly matches with twice the bulk EWCS which is consistent with the proposed holographic duality described in \cite{Dutta:2019gen,Chandrasekaran:2020qtn}.

%%%%%%%%%%%%%%%%%%%%%%%%%%%%%%%%%%%%%%%%%%%%%%%%%%%%%%%%%%%%%%%%%%%%

\subsubsection*{Configuration-14}
The computation of the bulk EWCS for this configuration follows a similar analysis described in the previous case. Thus, the expression of the bulk EWCS may be obtained using \cref{poincaregeodesic} as 
\begin{align}\label{EW14}
	E_W= \frac{c}{3} \log \left[\frac{(r_1-r_2)~ \text{sech}\left(\frac{2 \pi  t}{\beta }\right) \sqrt{r_1^2+2 r_1 r_2 \cosh \left(\frac{4 \pi  t}{\beta }\right)+r_2^2}}{r_1 \epsilon }\right]\,.
\end{align}
Interestingly the the replica technique result for the reflected entropy in \cref{sr14} exactly matches with twice the above bulk EWCS in accordance with holographic duality.
%%%%%%%%%%%%%%%%%%%%%%%%%%%%%%%%%%%%%%%%%%%%%%%%%%%%%%%%%%%%%%%%%%%%

\subsubsection*{Configuration-15 and 16}
The computation of the bulk EWCS for these configurations as show in the \cref{bulkendingbulk,bulkendingbulk2} follow a similar analysis described in the configuration 13. Hence the expressions of the bulk EWCS for the configurations 15 and 16 are given by the \cref{EW13,EW14} respectively.
%%%%%%%%%%%%%%%%%%%%%%%%%%%%%%%%%%%%%%%%%%%%%%%%%%%%%%%%%%%%%%%%%%%%
  
\subsubsection{Disjoint subsystems}
In this subsection, we discuss the computation of the bulk EWCS for the case of two disjoint subsystems $A$ and $B$ which are sandwiched with an auxiliary subsystem $C$. These subsystems are located in both the copies of the radiation reservoirs described by $CFT_2$s. In this context, we obtain different possible contributions to the bulk EWCS as depicted in \cref{disbraneendingresults,disbulkendingresults,disbulkresults,distakaresults} with different sizes of the subsystems.    

%%%%%%%%%%%%%%%%%%%%%%%%%%%%%%%%%%%%%%%%%%%%%%%%%%%%%%%%%%%%%%%%%%%%

\subsubsection*{Configuration-1}
To compute the bulk EWCS for this configuration, we utilize the result of the bulk EWCS obtained in \cite{Takayanagi:2017knl} for two disjoint subsystems $A$ and $B$ at finite temperature in the context of $AdS_3/CFT_2$ scenario. The corresponding expression of the EWCS may be obtained in terms of the finite temperature cross ratio $z$ as follows
\begin{equation}\label{disewcs}
	E_W=\frac{c}{6}\log\left[1+2 z +2\sqrt{z(z+1)}\right]\,, \quad\quad z=\frac{\sinh{\frac{2 \pi z_{12}}{\beta}}\sinh{\frac{2 \pi z_{34}}{\beta}}}{\sinh{\frac{2 \pi z_{23}}{\beta}}\sinh{\frac{2 \pi z_{14}}{\beta}}}\,,
\end{equation} 
where we have used Brown-Henneaux formula $c=3/2 G_N$ \cite{Brown:1986nw}. For the configuration described in \cref{distaka}, the bulk EWCS may be obtained using the coordinate of the boundary subsystems $A=[p_1,p_2]$ and $B=[p_3,a_1]$ as
\begin{equation}\label{disewcstaka}
	E_W=2 \frac{c}{6}\log\left[1+2 z +2\sqrt{z(z+1)}\right]\,, \quad\quad z=\frac{\sinh{\frac{2 \pi (p_2-p_1)}{\beta}}\sinh{\frac{2 \pi (a_1-p_1)}{\beta}}}{\sinh{\frac{2 \pi (p_3-p_2)}{\beta}}\sinh{\frac{2 \pi (a_1-p_1)}{\beta}}}\,,
\end{equation} 
The above bulk EWCS is consistent with the replica technique result of the reflected entropy in \cref{srdistaka} in accordance with the holographic duality as described in \cite{Dutta:2019gen,Chandrasekaran:2020qtn}. 

%%%%%%%%%%%%%%%%%%%%%%%%%%%%%%%%%%%%%%%%%%%%%%%%%%%%%%%%%%%%%%%%%%%%

\subsubsection*{Configuration-2}
From the \cref{distaka2}, we can observe that the bulk EWCS in this configuration may be obtained using a similar analysis followed in the configuration-1. Hence the expression of the bulk EWCS is given by
\begin{equation}\label{disewcstaka2}
	E_W=2 \frac{c}{6}\log\left[1+2 z +2\sqrt{z(z+1)}\right]\,, \quad\quad z=\frac{\sinh{\frac{2 \pi (p_4-p_3)}{\beta}}\sinh{\frac{2 \pi (b_1+p_2)}{\beta}}}{\sinh{\frac{2 \pi (p_4-p_2)}{\beta}}\sinh{\frac{2 \pi (b_1+p_2)}{\beta}}}\,.
\end{equation}    
Once again the replica technique result in \cref{srdistaka2} exactly matches with twice the bulk EWCS which is consistent with the proposed holographic duality described in \cite{Dutta:2019gen,Chandrasekaran:2020qtn}.

%%%%%%%%%%%%%%%%%%%%%%%%%%%%%%%%%%%%%%%%%%%%%%%%%%%%%%%%%%%%%%%%%%%%

\subsubsection*{Configuration-3 and 4}
The computation of the bulk EWCS for these configurations are similar to the cases discussed earlier.  However the only difference in these configurations arises from the enclosed entanglement wedge regions for the subsystem $A\cup B$ as shown in \cref{distakabulk,distakabulk2}. Thus the expression of the bulk EWCS in the configurations 3 and 4 are given by the \cref{disewcstaka,disewcstaka2} respectively.

%%%%%%%%%%%%%%%%%%%%%%%%%%%%%%%%%%%%%%%%%%%%%%%%%%%%%%%%%%%%%%%%%%%%

\subsubsection*{Configuration-5}
The computation of the EWCS follows a similar analysis described in \cite{Takayanagi:2017knl} in the context of the $AdS/CFT$ scenario. At a finite temperature, the expression of the EWCS in this configuration may be given by the following equation
\begin{equation}\label{disEWCS}
	E_W=2 \frac{c}{6}\log\left[1+2 z +2\sqrt{z(z+1)}\right]\,, \quad\quad z=\frac{\sinh{\frac{2 \pi (p_2-p_1)}{\beta}}\sinh{\frac{2 \pi (p_4-p_3)}{\beta}}}{\sinh{\frac{2 \pi (p_3-p_1)}{\beta}}\sinh{\frac{2 \pi (p_4-p_1)}{\beta}}}\,.
\end{equation}    
Once again twice of the above bulk EWCS is consistent with the replica technique expression in \cref{srdisRTending} in accordance with the proposed holographic duality. 

%%%%%%%%%%%%%%%%%%%%%%%%%%%%%%%%%%%%%%%%%%%%%%%%%%%%%%%%%%%%%%%%%%%%

\subsubsection*{Configuration-6}
In this configuration, we follow a similar analysis described in the configuration-6 of the adjacent subsystems to obtain the expression of the bulk EWCS. In particular, we utilize the embedding space formalism to obtain the expression of the bulk EWCS which is given by the following geodesic length \cref{embeddinglength}. As depicted in \cref{disbraneending}, the endpoints of the bulk EWCS in this case involves a point on the $a$-brane and an arbitrary point on the RT surface homologous to the subsystem $C$. We consider these endpoints of the bulk EWCS as $(a_2,z_{QES},t)$ on the $a$-brane and  $(R+r \sin(\phi),r \cos(\phi),t)$. Here $\phi$ is the angle 
defined to get the coordinates of the endpoint on the RT surface and $R, r$ are defined in terms of the boundary coordinate as $\frac{p_2+p_3}{2},\frac{p_3-p_2}{2}$ respectively. Therefore, the expression of the bulk EWCS can be obtained in this configuration using the \cref{embeddinglength} as
\begin{align}
	E_W= \frac{c}{3} \cosh^{-1} \left[\frac{z_h ^2 \left(\cosh \left(\frac{ (R-a_2)}{z_h 
		}\right)-1\right)}{ r\,\,\,z_{QES} }\right]\,.
\end{align}
where we considered that the bulk EWCS becomes minimum for $\phi=0$. In the above equation, the $zQES$ may be obtained using the profile of the $a$-brane as $zQES=z_h \sinh \left(\frac{a_2}{z_h}\right)$ with $k=1$ \cite{Grimaldi:2022suv}.  Therefore, the expression of the EWCS becomes
\begin{align}\label{disewcsbraneending}
	E_W= \frac{c}{3} \log\left[\frac{\beta \left(\cosh \left(\frac{2 \pi  (R-a_2)}{\beta 
		}\right)-1\right)}{ \pi  r \sinh\left(\frac{2 \pi  a_2}{\beta}\right)}\right]\,.
\end{align}
Note that the above result of the bulk EWCS exactly matches with the expression of the reflected entropy in \cref{srconfig6dis} using \cref{srew2}.
%%%%%%%%%%%%%%%%%%%%%%%%%%%%%%%%%%%%%%%%%%%%%%%%%%%%%%%%%%%%%%

\subsubsection*{Configuration-7}
The contribution of the bulk EWCS in this configuration may be obtained by following the analogous analysis described in the previous case. As shown in \cref{disbraneending2}, the bulk EWCS now obtains contribution from the $b$-brane. However the computation of the bulk EWCS remains same as configuration-6, thus the expression of the EWCS may be given by 
\begin{align}\label{disewcsbraneending2}
	E_W= \frac{c}{3} \log\left[\frac{\beta \left(\cosh \left(\frac{2 \pi  (R-b_2)}{\beta 
		}\right)-1\right)}{ \pi  r \sinh\left(\frac{2 \pi  b_2}{\beta}\right)}\right]\,.
\end{align}
Using \cref{srew2}, the above result matches with the replica technique result described in \cref{srconfig7dis}.

%%%%%%%%%%%%%%%%%%%%%%%%%%%%%%%%%%%%%%%%%%%%%%%%%%%%%%%%%%%%%%%%%%%%

\subsubsection*{Configuration-8 and 9}
In these configurations, the bulk EWCS may be obtained by following a similar procedure provided in the configurations- 6. The only  difference may be observed in the corresponding configurations arise from the enclosed entanglement wedge regions as depicted in \cref{disbraneendingbulk,disbraneendingbulk2}. Thus the expressions of the bulk EWCS in these cases are described by \cref{disewcsbraneending,disewcsbraneending2} for the configurations 8 and 9 respectively.

%%%%%%%%%%%%%%%%%%%%%%%%%%%%%%%%%%%%%%%%%%%%%%%%%%%%%%%%%%%%%%%%%%%%

\subsubsection*{Configuration-10}
For the computation of the bulk EWCS in this configuration as shown in \cref{disbulk}, we utilize the embedding space formalism to express the length of the EWCS which is given by \cref{embeddinglength}. Consequently, we obtain the corresponding length using the endpoints $(R+r \sin(\phi),r \cos(\phi),t)$ and $(R+r \sin(\phi),r \cos(\phi),-t+i\beta/2)$ of the bulk EWCS. Note that these endpoints of the EWCS are located on the RT surface homologous to subsystem $C$. Thus, The expression of the bulk EWCS may then be obtained as follows
\begin{align}\label{disewcsbulk}
	E_W =\frac{c}{3}\log \left[\frac{\beta  \cosh \left(\frac{2 \pi  t}{\beta}\right)}{\pi  r}\right]\,,
\end{align}
where $r$ is related to the points $p_2$ and $p_3$ as $r=(p_3-p_2)/2$. Above result matches with the replica technique result \cref{disjointsrbulk} of the reflected entropy by utilizing the proposal \cref{srew2}.

%%%%%%%%%%%%%%%%%%%%%%%%%%%%%%%%%%%%%%%%%%%%%%%%%%%%%%%%%%%%%%%%%%%%

\subsubsection*{Configuration-11 and 12}
As we can notice from the \cref{disbulkbulk,disbulkbulk2} that the bulk EWCS in these configurations is same as above case. However, the only observed difference in these configuration can be indicated from the enclosed entanglement wedge regions of the subsystem $A\cup B$. Therefore the expression of the bulk EWCS may be described by  \cref{disewcsbulk}.

%%%%%%%%%%%%%%%%%%%%%%%%%%%%%%%%%%%%%%%%%%%%%%%%%%%%%%%%%%%%%%%%%%%%

\subsubsection*{Configuration-13}
We utilize the method discussed in the configuration-13 of the adjacent subsystems to compute the EWCS in this configuration described in \cref{disbulkending} for the disjoint subsystems $A$ and $B$. Consequently, we use \cref{poincaregeodesic} to obtain the expression of the EWCS for two disjoint subsystems. As depicted in \cref{disbulkending} that the endpoints of the bulk EWCS are given as  ($(R+r \sin \phi) \cosh(\frac{2 \pi t}{\beta}),(R+r \sin \phi) \sinh(\frac{2 \pi t}{\beta}),r \cos \phi$) and ($x_2,r_3 \sinh(\frac{2 \pi t}{\beta}),\sqrt{\left(r_3 \cosh(\frac{2 \pi t}{\beta})\right) ^2-x_2 ^2}$). Here $\phi$ is the angle defined to get the coordinates of the endpoint on the RT surface homologous to the subsystem $C$. Finally the expression of the bulk EWCS may be given by  
\begin{align}\label{disEW13}
	E_W=\frac{c}{3} \log \left[\frac{\sech(\frac{2 \pi t}{\beta}) \sqrt{r^2+(R-r_2)^2} \sqrt{r^2+R^2+2 R r_2 \cosh (\frac{4 \pi t}{\beta})+r_2^2}}{r_2 \,\,r}\right]\,,
\end{align}
where  $R\,,\,r$ are defined in terms of the boundary coordinates as $\frac{p_3+p_2}{2}\,,\,\frac{p_3-p_2}{2}$ respectively. In the above equation, we obtained the minimum length of the bulk EWCS for $\phi=0$. Thus the corresponding result matches with the replica technique expression using the proposal \cref{srew2}.

%%%%%%%%%%%%%%%%%%%%%%%%%%%%%%%%%%%%%%%%%%%%%%%%%%%%%%%%%%%%%%

\subsubsection*{Configuration-14}
We may obtain the expression of the bulk EWCS in this configuration as shown in \cref{disbulkending2} by following a similar analysis described in the earlier configuration. Thus the expression of the bulk EWCS is given by
\begin{align}\label{disEW14}
	E_W=\frac{c}{3} \log \left[\frac{\sech(\frac{2 \pi t}{\beta}) \sqrt{r^2+(R-r_2)^2} \sqrt{r^2+R^2+2 R r_2 \cosh (\frac{4 \pi t}{\beta})+r_2^2}}{r_2 \,\,r}\right]\,.
\end{align}
Above expression matches with the replica technique result using the proposal \cref{srew2}.
%%%%%%%%%%%%%%%%%%%%%%%%%%%%%%%%%%%%%%%%%%%%%%%%%%%%%%%%%%%%%%
\subsubsection*{Configuration-15 and 16}
In these configurations, the bulk EWCS as depicted in \cref{disbulkendingbulk,disbulkendingbulk2} may be obtained by following a similar analysis discussed in configuration-13 and 14. The only difference in these configuration may be observed from the enclosed entanglement wedge region for the subsystem $A\cup B$. Therefore the bulk EWCS is given by \cref{EW13,EW14} for the configurations 15 and 16 respectively.

%%%%%%%%%%%%%%%%%%%%%%%%%%%%%%%%%%%%%%%%%%%%%%%%%%%%%%%%%%%%%%%%%%%%

\section{Markov gap}\label{subsec:markovcomputaion}
In this section, we will compare the behaviour of the profiles for the reflected entropy with that of the mutual information for the bipartite mixed states in three different cases where we will vary the subsystem sizes and the time. As mentioned
in the introduction the difference between the holographic reflected entropy and the mutual information is
is described as the holographic Markov gap as defined in the context of quantum information theory and the Markov recovery process. To this end we first considered bipartite mixed states of adjacent subsystems in the reservoirs and obtained the profiles for the holographic reflected entropy by varying the size of the subsystem and time. Similarly, we 
described the corresponding holographic mutual information for these phases arising from the different structures of the RT surfaces supported by the subsystems. Subsequently we followed a similar analysis for the case of disjoint subsystems and in all the cases we demonstrated the behaviour of the holographic Markov gap described in  \cite{Hayden:2021gno}.

%%%%%%%%%%%%%%%%%%%%%%%%%%%%%%%%%%%%%%%%%%%%%%%%%%%%%%%%%%%%%%%%%%%%

\subsection{Adjacent subsystems}
We first consider adjacent subsystems $A$ and $B$ of finite lengths $l_1$ and $l_2$ respectively in both the TFD copies of the radiation reservoirs to describe the holographic Markov gap. In this context, the holographic mutual information for the adjacent subsystems $A$ and $B$ is defined as 
\begin{align}
	I(A:B)= S(A) +S(B)-S(A\cup B)\,,
\end{align} 
where $S(X)$ is the holographic entanglement entropy of a subsystem $X$. We will investigate the qualitative nature of the holographic Markov gap for three different scenarios involving the subsystem sizes and time. For this purpose we
will utilize the structure of the RT surfaces associated with the corresponding subsystems which was briefly described in \cite{Afrasiar:2022ebi} for the communicating black hole configurations.  

%%%%%%%%%%%%%%%%%%%%%%%%%%%%%%%%%%%%%%%%%%%%%%%%%%%%%%%%%%%%%%%%%%%%

\subsubsection*{$\bm{(i)}$ Full system ($\bm{A\cup B}$) fixed, common point varied}\label{adjcaseiEW}
\begin{figure}[h!]
	\centering
	\includegraphics[width=14cm]{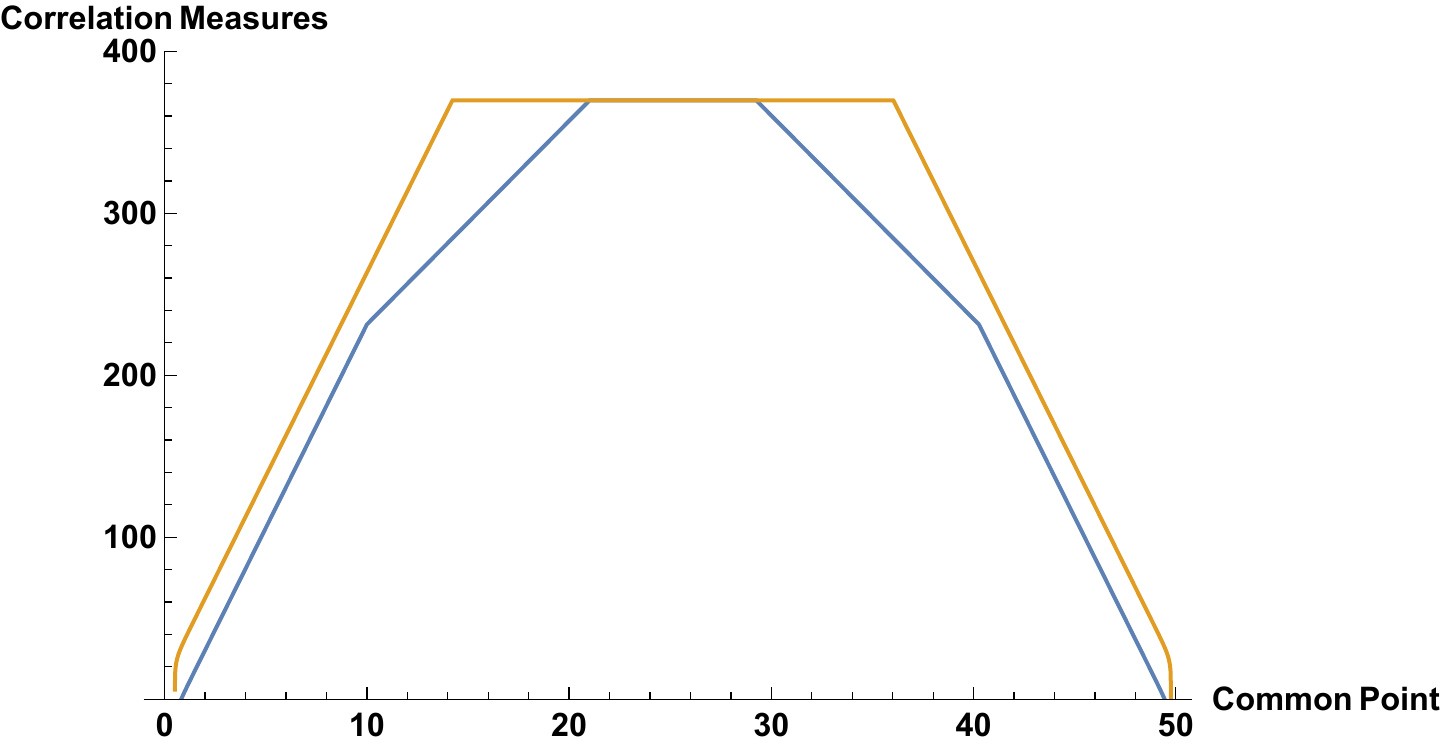}
	\caption{Schematics shows the variation of the correlation measures reflected entropy (yellow) and mutual information (blue) for two adjacent subsystems $A$ and $B$ while the common boundary between them is being shifted in the radiation reservoirs. In the y-axis, the correlation measures are plotted with a scale factor of $\frac{6}{c}$ where $c$ is the central charge of the $CFT_2$s located in the reservoirs and the Planck branes. Here $\beta=1$, $t=15$, $c=500$, $\phi_0= \frac{30c}{6}$, $\phi_r= \frac{30}{\pi}$, $L=\frac{16\pi}{\beta}$, $\epsilon=.001$, $A\cup B=[.01L,.99L]$. }\label{adjcase1}
\end{figure}

We first consider the case where the entire reservoir region is covered by the subsystem $A\cup B$ and the adjacent point is varied on a constant time slice. In this case, we obtained various phases of the holographic reflected entropy and the mutual information that characterizes the Markov gap as depicted in \cref{adjcase1}.

In the above figure, we observe different dominant contributions to the reflected entropy for two adjacent subsystems described in \cref{subsec:srcomputation}  while varying the common point. Specifically, the configurations (1),(10) and (2) for the two adjacent subsystems dominate in the consecutive phases of the reflected entropy in \cref{adjcase1}.

%%%%%%%%%%%%%%%%%%%%%%%%%%%%%%%%%%%%%%%%%%%%%%%%%%%%%%%%%%%%%%%%%%%%

\subsubsection*{$\bm{(ii)}$ Subsystem $\bm{A}$ fixed, $\bm{B}$ varied}\label{adjcaseiiEW}
\begin{figure}[h!]
	\centering
	\includegraphics[width=14cm]{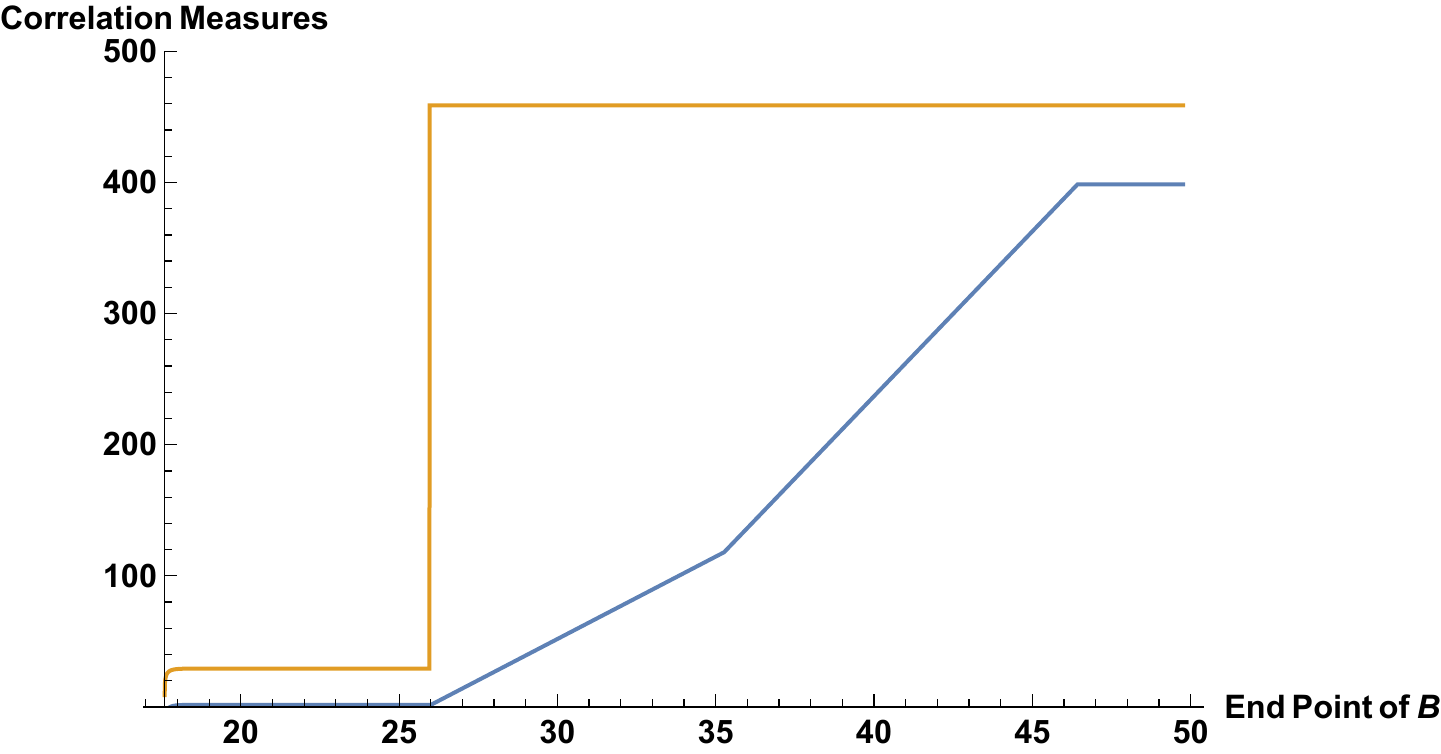}
	\caption{Schematics shows the variation of the correlation measures namely, reflected entropy (yellow) and mutual information (blue) for two adjacent subsystems $A$ and $B$ with increasing size of the subsystem $B$ in the radiation reservoirs. In the y-axis, the correlation measures are plotted with a scale factor of $\frac{6}{c}$ where $c$ is the central charge of the $CFT_2$s located in the reservoirs and the Planck branes. Here $\beta=1$, $t=20$, $c=500$, $\phi_0= \frac{30c}{6}$, $\phi_r= \frac{30}{\pi}$, $L=\frac{16\pi}{\beta}$, $\epsilon=.001$, $A=[.01L,.15L]$. }\label{adjcase2}
\end{figure}

Next we consider the size of the subsystem $A$ to be fixed on a constant time slice and analyzed the holographic Markov gap with the variation of the size of the subsystem $B$. We observed different phases of the reflected entropy and the mutual information to describe the Markov gap for each as shown in \cref{adjcase2}.

Similar to earlier case, we receive different dominant contributions corresponding to various configurations described in \cref{subsec:srcomputation} to the reflected entropy while varying the size of the subsystem $B$. In particular, we observe configurations (5),and (1) 
 dominate in the consecutive phase of the reflected entropy as shown in \cref{adjcase2}.

%%%%%%%%%%%%%%%%%%%%%%%%%%%%%%%%%%%%%%%%%%%%%%%%%%%%%%%%%%%%%%%%%%%%

\subsubsection*{$\bm{(iii)}$ Subsystems $\bm{A}$ and $\bm{B}$ fixed, time varied} \label{AdjtimeEW}
\begin{figure}[H]
	\centering
	\includegraphics[width=14cm]{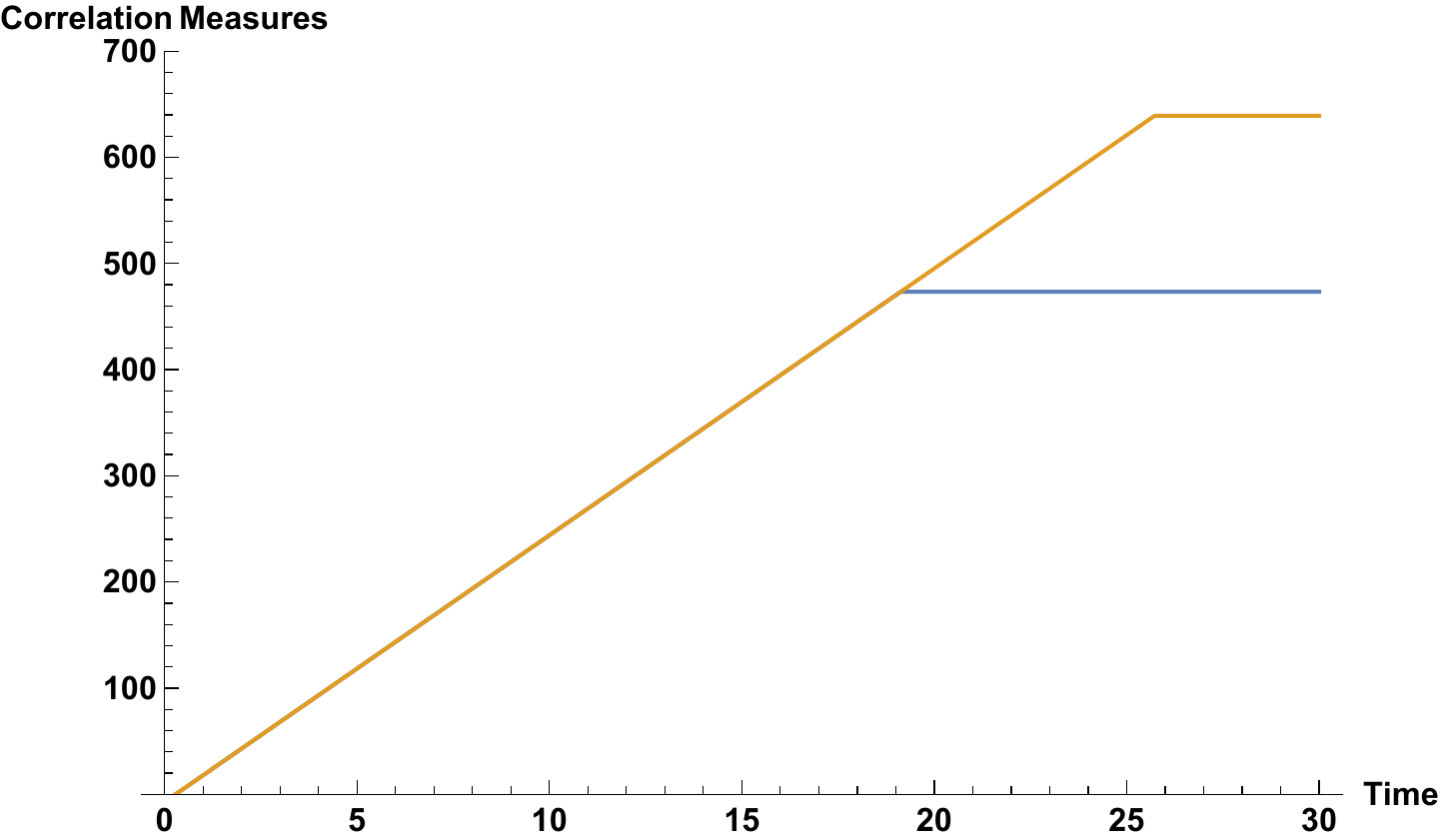}
	\caption{Schematics shows the variation of the correlation measures namely, reflected entropy (yellow) and mutual information (blue) for two adjacent subsystems $A$ and $B$ while increasing the time $t$. In the y-axis, the correlation measures are plotted with a scale factor of $\frac{6}{c}$ where $c$ is the central charge of the $CFT_2$s located in the reservoirs and the Planck branes. Here $\beta=1$, $c=500$, $\phi_0= \frac{30c}{6}$, $\phi_r= \frac{30}{\pi}$, $L=\frac{16\pi}{\beta}$, $\epsilon=.001$, $A=[.01L,.5L]$ and $B=[.5L,.99L]$. }\label{adjcase3}
\end{figure}

In this case, we fix both the size of the adjacent subsystems $A$ and $B$ and analyze the behaviour of the holographic Markov gap with time. In particular, we consider the equal lengths of the subsystems $A$ and $B$ \cref{adjcase3}.

Finally in the last case, we observe different dominant contributions (10) and (2) dominate consecutively in the reflected entropy for two adjacent subsystems as shown in \cref{adjcase3}.
%%%%%%%%%%%%%%%%%%%%%%%%%%%%%%%%%%%%%%%%%%%%%%%%%%%%%%%%%%%%%%

\subsection{Disjoint subsystems}
We now consider a mixed state configuration of disjoint subsystems $A$ and $B$ where a subsystem $C$ is sandwiched between them in both the TFD copies of the radiation reservoirs. Here we describe the holographic Markov gap in three different scenarios involving the subsystem sizes and the time. For the disjoint subsystems the mutual information is defined as
\begin{align}\label{dismutual}
	I(A:B)=S(A)+S(B)-S(A\cup B\cup C)-S(C)\,,
\end{align}  
where $S(X)$ is the holographic entanglement entropy of a subsystem $X$. Here we will utilize the structure of the RT surfaces associated with the subsystems in question which was again briefly described in \cite{Afrasiar:2022ebi} for the communicating black hole scenarios.

%%%%%%%%%%%%%%%%%%%%%%%%%%%%%%%%%%%%%%%%%%%%%%%%%%%%%%%%%%%%%%

\subsubsection*{$\bm{(i)}$ Subsystem $\bm{A}$ fixed, $\bm{C}$ varied}\label{discaseiEW}
\begin{figure}[h!]
	\centering
	\includegraphics[width=14cm]{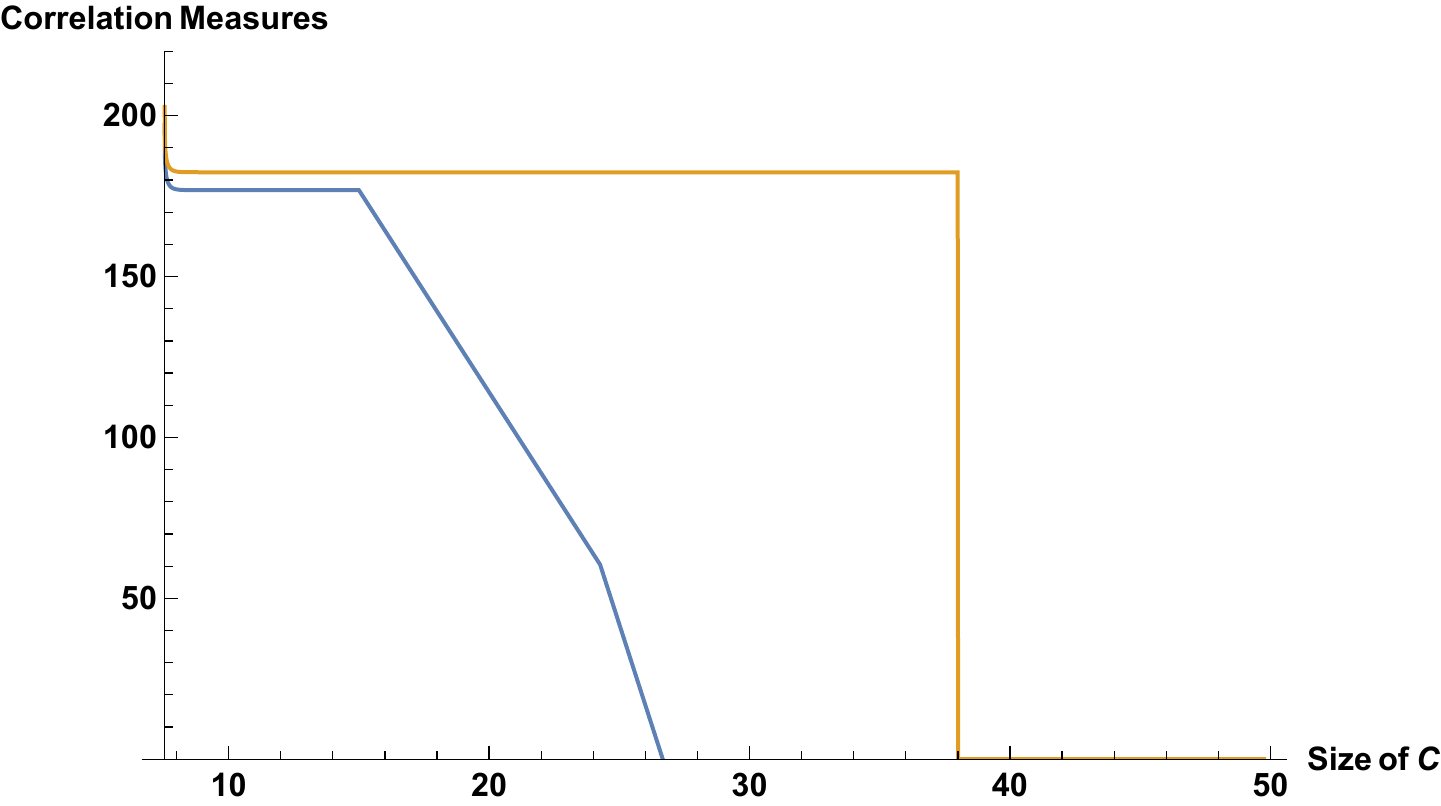}
	\caption{Schematics shows the variation of the correlation measures namely, reflected entropy (yellow) and mutual information (blue) for two disjoint subsystems $A$ and $B$ while increasing common point. In the y-axis, the correlation measures are plotted with a scale factor of $\frac{6}{c}$ where $c$ is the central charge of the $CFT_2$s located in the reservoirs and the Planck branes. Here $\beta=1$, $c=500$, $\phi_0= \frac{30c}{6}$, $\phi_r= \frac{30}{\pi}$, $L=\frac{16\pi}{\beta}$, $\epsilon=.001$ and $A=[.01L,.15L]$. }\label{discase1}
\end{figure}

In this case, we fix the size of the subsystem $A$ and gradually vary the size of the subsystem $C$ on a constant time slice. We analyze the behavior of the holographic Markov gap from the comparison between the reflected entropy described in \cref{subsec:srcomputation} and the mutual information in \cref{dismutual}. We observe different phases of the holographic Markov gap as shown in \cref{discase1}. 

In this case, we receive different dominant contributions for various configurations described in \cref{subsec:srcomputation} to the reflected entropy for two disjoint subsystems while varying the size of the subsystem $C$. We observe contribution (1) dominates in the starting phase of the reflected entropy as shown in \cref{discase1}.
%%%%%%%%%%%%%%%%%%%%%%%%%%%%%%%%%%%%%%%%%%%%%%%%%%%%%%%%%%%%%%

\subsubsection*{$\bm{(ii)}$ Subsystems $\bm{A}$ and $\bm{C}$ fixed, $\bm{B}$}\label{discaseiiEW}
\begin{figure}[h!]
	\centering
	\includegraphics[width=14cm]{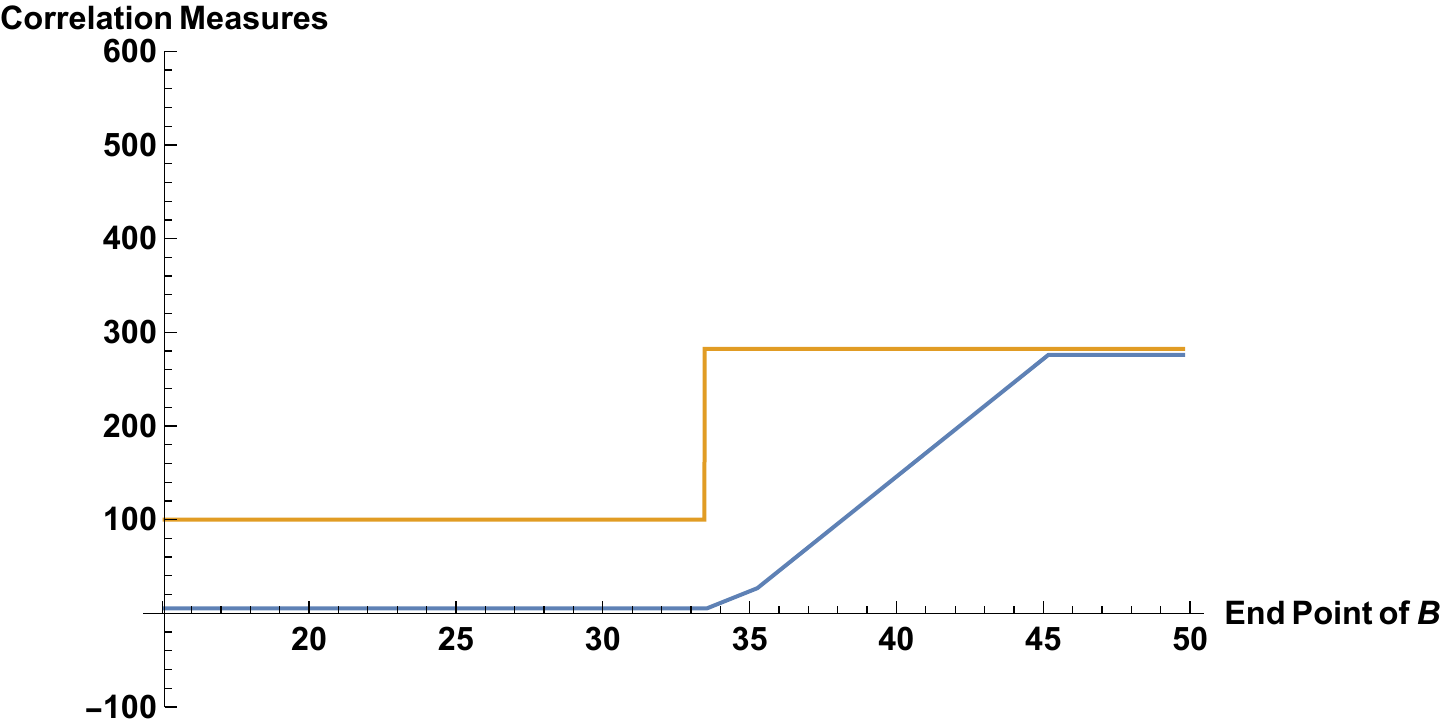}
	\caption{Schematics shows the variation of the correlation measures namely, reflected entropy (yellow) and mutual information (blue) for two disjoint subsystems $A$ and $B$ while increasing the size of the subsystem $B$. In the y-axis, the correlation measures are plotted with a scale factor of $\frac{6}{c}$ where $c$ is the central charge of the $CFT_2$s located in the reservoirs and the Planck branes. Here $\beta=1$, $c=500$, $\phi_0= \frac{30c}{6}$, $\phi_r= \frac{30}{\pi}$, $L=\frac{16\pi}{\beta}$, $\epsilon=.001$, $A=[.01L,.15L]$ and $C=[.15L,.30L]$. }\label{discase2}
\end{figure}

Next we consider the size of the subsystems $A$ and $C$ to be fixed on a constant time slice and analyze the behavior of the holographic Markov gap while varying the size of the subsystem $B$. We obtained various phases of the reflected entropy and the mutual information as described in \cref{discase2}

As earlier, we note that the reflected entropy curve in \cref{discase2} is dominated by various contributions from the configurations describe in \cref{subsec:srcomputation}. Particularly, configurations (5) and (1) of the reflected entropy for the disjoint subsystems dominate the corresponding phases shown in \cref{discase2}.
%%%%%%%%%%%%%%%%%%%%%%%%%%%%%%%%%%%%%%%%%%%%%%%%%%%%%%%%%%%%%%

\subsubsection*{$\bm{(iii)}$ Subsystems $\bm{A}$, $\bm{B}$ and $\bm{C}$ fixed, time varied}\label{distimeEW}
\begin{figure}[h!]
	\centering
	\includegraphics[width=14cm]{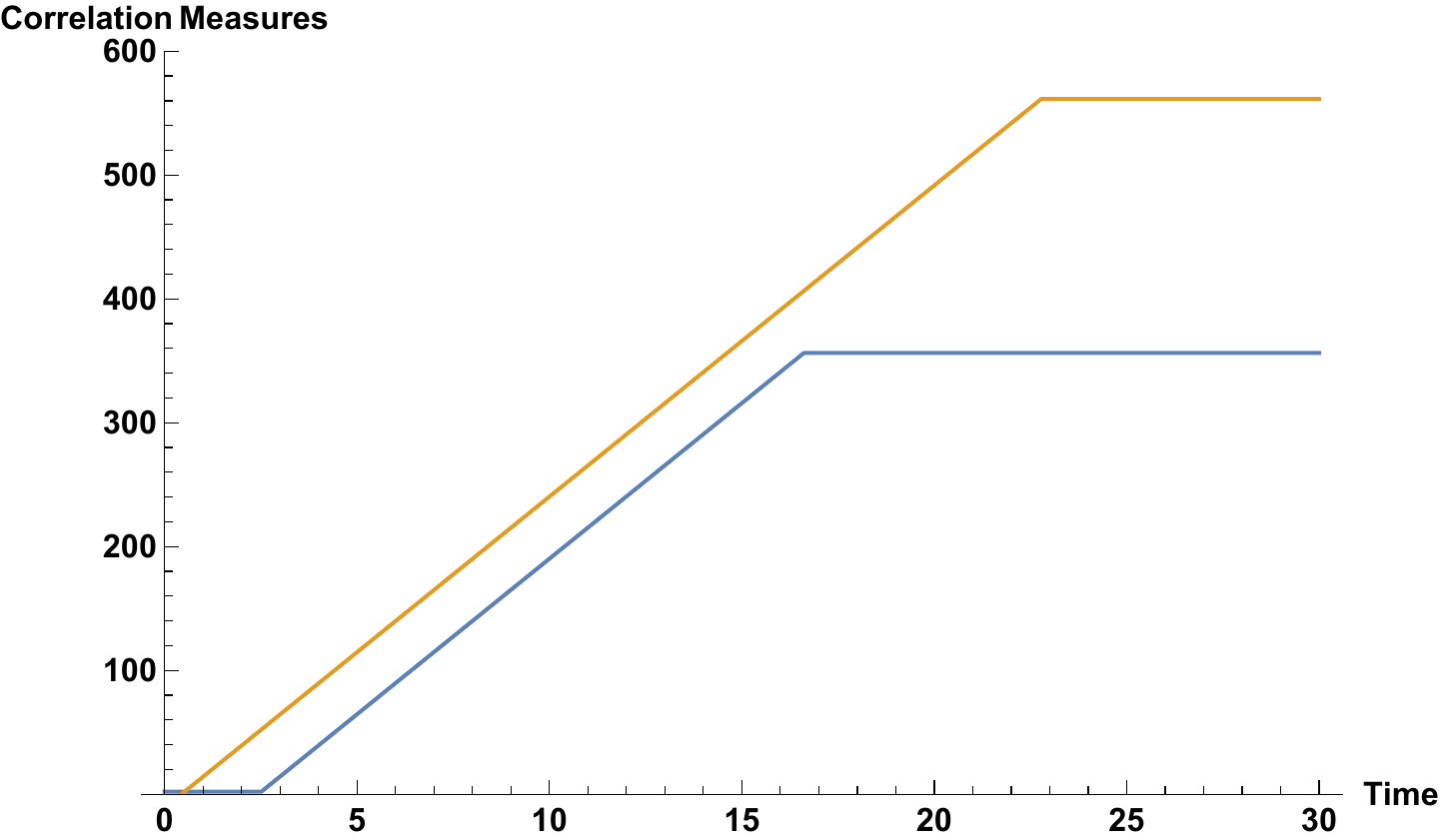}
	\caption{Schematics shows the variation of the correlation measures namely, reflected entropy (yellow) and mutual information (blue) for two disjoint subsystems $A$ and $B$ while varying the time $t$. In the y-axis, the correlation measures are plotted with a scale factor of $\frac{6}{c}$ where $c$ is the central charge of the $CFT_2$s located in the reservoirs and the Planck branes. Here $\beta=1$, $c=500$, $\phi_0= \frac{30c}{6}$, $\phi_r= \frac{30}{\pi}$, $L=\frac{16\pi}{\beta}$, $\epsilon=.001$, $A=[.01L,.45L]$, $C=[.45L,.55L]$ and $B=[.55L,.99L]$. }\label{discase3}
\end{figure}

Finally in the last case, we considered equal sizes of the subsystems $A$ and $B$ on a constant time slice and observed various phases of the holographic Markov gap through the analysis of the reflected entropy as obtained in \cref{subsec:srcomputation} and the mutual information in \cref{dismutual} with the variation of the time as shown in \cref{discase3}. 

Once again in the last scenario, the reflected entropy for the two disjoint subsystems receives different dominant contributions from the configurations described in \cref{subsec:srcomputation} with increasing time $t$. Specifically, the configurations (10) and (2) for the two disjoint subsystems dominate respectively in the consecutive phases of the reflected entropy depicted by the yellow curve in \cref{discase3}.

%%%%%%%%%%%%%%%%%%%%%%%%%%%%%%%%%%%%%%%%%%%%%%%%%%%%%%%%%%%%%%
\section{Summary and Discussion}\label{discussion}
To summarize, we have computed the reflected entropy for various bipartite mixed states described by adjacent and disjoint subsystems at a finite temperature for the communicating black hole configurations in a Planck brane world geometry. In this context, we considered a configuration involving two eternal JT black holes with two finite sized radiation reservoirs coupled to two quantum dots. The bulk dual of the corresponding configuration was described by an eternal $AdS_3$ BTZ black hole geometry truncated by two Planck branes. Interestingly, the gravitating nature of each of the radiation reservoirs corresponding to the two Planck branes could be demonstrated from the perspective of the other   
brane. Note that for this configuration the two dimensional eternal JT black holes are located on the Planck branes with
a single matter $CFT_2$ described on the entire geometry with transparent boundary conditions imposed at the interfaces of the radiation reservoirs and the Planck branes.

The reflected entropy for the mixed state configurations in question was computed for the communicating black hole configuration described above. We demonstrated the different possible dominant channels for the multipoint twist correlators involved  in the computation of the reflected entropy in the large central charge limit. In this context, we first investigated the reflected entropy in the large central charge limit for two adjacent subsystems located in the radiation reservoirs utilizing the replica technique developed in \cite{Dutta:2019gen}. Subsequently, we followed a similar analysis for the computation of the reflected entropy of two disjoint subsystems located in the radiation reservoirs. These field theory replica technique results were substantiated by explicit bulk holographic computation of the EWCS in the dual brane world geometry.   

Furthermore, we also analyzed the holographic Markov gap as described in \cite{Hayden:2021gno} between the reflected entropy and the mutual information for various bipartite mixed states in the context of $AdS_3/CFT_2$ correspondence. In this connection, we compared the holographic reflected entropy and the mutual information for various bipartite mixed state configurations under consideration. We obtained different profiles of the holographic Markov gap in various configurations related to the variation of size of the subsystems and time. 

There are various fascinating future directions which can be investigated to obtain a better understanding of the mixed state entanglement structure in Hawking radiation. Following some recent developments in \cite{Afrasiar:2022ebi,Lu:2022fxb,Afrasiar:2022fid}, it will be interesting to study other mixed state correlation measures and their corresponding Markov gaps. One may also extend the study of the present article to investigate the multipartite correlations where it is expected to observe the characteristics of the holographic Markov gap. We would like to address some these interesting issues in the near future.

\section{Acknowledgement}
The research work of JKB is supported by the grant 110-2636-M-110-008 by the National Science and Technology Council (NSTC) of Taiwan. The work of GS is partially supported by the Dr. Jagmohan Garg Chair Professor position at the Indian Institute of Technology, Kanpur.

\bibliographystyle{JHEP}
\bibliography{Citation}

%\printbibliography

%\bibliographystyle{habbrv}

%\bibliography{reference}

\end{document}